\documentclass[pdflatex,sn-mathphys-ay]{sn-jnl}% Math and Physical Sciences Author Year Reference Style

\usepackage{algorithm}%
\usepackage{algorithmicx}%
\usepackage{algpseudocode}%
\usepackage{amsmath,amssymb,amsfonts}%
\usepackage{accents}% to be loaded before 'amsmath' (\uvec)
\usepackage[title]{appendix}%
\usepackage{amsthm}%
\usepackage{booktabs}%
\usepackage{longtable}
\usepackage{dsfont}% to use \mathds
\usepackage{graphicx}%
\usepackage{listings}%
\usepackage{manyfoot}%
\usepackage{mathrsfs}%
\usepackage{mathtools}%
\usepackage{textcomp}%
\usepackage{xcolor}%
\usepackage{subcaption}%
\usepackage{multirow}%
\DeclareRobustCommand{\uvec}[1]{\underaccent{\tilde}{#1}}

\DeclareMathOperator{\SPD}{\mathrm{Sym}^{+}}
\DeclareMathOperator{\lat}{lat}
\DeclareMathOperator{\lon}{lon}
\DeclareMathOperator{\anc}{anc}
\DeclareMathOperator{\free}{free}
\DeclareMathOperator{\simu}{sim}
\DeclareMathOperator{\cov}{Cov}
\DeclareMathOperator{\cond}{\mathsf{cond}}

\DeclareMathOperator{\diag}{diag}

\DeclareMathOperator{\vect}{vec}
\DeclareMathOperator{\tr}{tr}

\DeclareMathOperator{\SMV}{partial}
\DeclareMathOperator{\OMV}{missing}
\DeclareMathOperator{\mis}{\mathsf{mis}}
\DeclareMathOperator{\obs}{\mathsf{obs}}
\DeclareMathOperator{\om}{\mathsf{o,m}}

\DeclareMathOperator{\gu}{\mathsf{g,u}}

\DeclareMathOperator{\interp}{\mathsf{int}}

\newcommand{\sfUnif}{\textrm{\sf Uniform}}
\newcommand{\sfNor}{\textrm{\sf Normal}}
\newcommand{\sfG}{\textrm{\sf Gamma}}
\newcommand{\sfIG}{\textrm{\sf Inverse-Gamma}}

\newcommand{\sfIW}{\textrm{\sf Inverse-Wishart}}

\theoremstyle{thmstyleone}%
\theoremstyle{thmstyletwo}%

\theoremstyle{thmstylethree}%

\begin{document}

\title[Article Title]{Spatial deformation in a Bayesian spatiotemporal model for incomplete matrix-variate responses}

%%=============================================================%%
%% GivenName	-> \fnm{Joergen W.}
%% Particle	-> \spfx{van der} -> surname prefix
%% FamilyName	-> \sur{Ploeg}
%% Suffix	-> \sfx{IV}
%% \author*[1,2]{\fnm{Joergen W.} \spfx{van der} \sur{Ploeg} 
%%  \sfx{IV}}\email{iauthor@gmail.com}
%%=============================================================%%

\author*[1]{\fnm{Rodrigo} \sur{de Souza Bulh\~{o}es}}\email{rbulhoes@ufba.br}

\author[2]{\fnm{Marina} \sur{Silva Paez}}\email{marina@im.ufrj.br}
%\equalcont{These authors contributed equally to this work.}

\author[2]{\fnm{Dani} \sur{Gamerman}}\email{dani@im.ufrj.br}
%\equalcont{These authors contributed equally to this work.}

\affil*[1]{\orgdiv{Department of Statistics, Institute of Mathematics and Statistics}, \orgname{Federal University of Bahia}, \orgaddress{\street{Av. Milton Santos, s/n}, \city{Salvador}, \postcode{40.170-110}, \state{Bahia}, \country{Brazil}}}

\affil[2]{\orgdiv{Department of Statistical Methods, Institute of Mathematics}, \orgname{Federal University of Rio de Janeiro}, \orgaddress{\street{Av. Athos da Silveira Ramos, 149}, \city{Rio de Janeiro}, \postcode{21.941-909}, \state{Rio de Janeiro}, \country{Brazil}}}

%%==================================%%
%% Sample for unstructured abstract %%
%%==================================%%

\abstract{
In this paper, we propose a Bayesian matrix-variate spatiotemporal modeling framework for jointly analyzing multiple response variables observed at spatial locations over time. The approach relaxes the standard assumption of spatial isotropy by incorporating a deformation-based mechanism, allowing the covariance structure to capture directional effects and nonstationary spatial dependence. Temporal dynamics are modeled through dynamic linear models, enabling coherent uncertainty propagation within a state-space formulation. Missing observations are handled via a data augmentation strategy that preserves the joint structure of the multivariate responses. The proposed methodology is evaluated through simulation studies and an application to air quality data. Results indicate that accounting for spatial deformation leads to substantial gains in predictive performance in anisotropic settings, while cross-variable dependence plays a secondary role in improving overall fit. The framework is computationally tractable for moderate numbers of spatial locations and responses, and provides a flexible basis for modeling multivariate spatiotemporal processes under incomplete data.
}


\maketitle

\section{Introduction}\label{sec:Introduction}

Environmental data sets are typically characterized by processes observed over a finite set of times at fixed monitoring sites within a geographic region. Over the past decades, several spatiotemporal models have been developed to analyze such processes. These models often assume that the process of interest follows a Gaussian process with a mean function and a valid covariance structure given by the product of a variance parameter and an isotropic correlation function, defined in terms of the Euclidean distance between locations \citep{schmidt2020flexible}. This implies translation and rotational invariance of the process, which are properties that characterize isotropy and stationarity, often referred to as spatial homogeneity.

However, the assumption of isotropy is frequently unrealistic for many environmental variables such as temperature, pollution, soil moisture, or rainfall. These processes are influenced by local factors that induce directional dependence or spatial heterogeneity, including landscape, topography, prevailing winds, and proximity to the ocean \citep{Morales2013, Morales2022}.

Spatial nonstationarity can be addressed through a variety of approaches. One important class is based on spatially varying covariance parameters, where the covariance function is allowed to change smoothly over space \citep{Paciorek2006, Stein2005}. Another line of work considers partition-based or locally stationary models, in which the spatial domain is divided into subregions with distinct dependence structures \citep{Konomi2014}. A third class includes convolution and basis-function approaches, which construct nonstationary covariance structures via spatially varying kernels or low-rank representations \citep{Lemos2009, Katzfuss2013}. Additional strategies include the incorporation of covariates directly into the covariance function \citep{Guttorp2013} and stochastic partial differential equation approaches \citep{Lindgren2011}.

Within this broad class of methods, the deformation-based approach provides a conceptually simple and interpretable mechanism for modeling anisotropy. Rather than directly specifying a nonstationary covariance function, the method maps the original geographic coordinates into a latent space where isotropy holds. In this transformed space, standard stationary models can be applied, while anisotropy and nonstationarity are captured through the deformation itself. Compared to other approaches, deformation models offer global coherence in the spatial dependence structure, avoiding abrupt transitions across subregions, although at the cost of increased computational complexity.

A seminal contribution in this direction is the framework introduced by \citet{Sampson1992}, which remains a cornerstone for modeling non-homogeneous spatial processes. The original proposal uses multidimensional scaling to obtain latent coordinates and thin-plate splines for interpolation at ungauged sites. In that formulation, temporal stationarity is assumed, with temporal dependence handled by detrending and analyzing residuals as repeated measurements. Later Bayesian extensions \citep{Damian2001, Schmidt2003} incorporated uncertainty in the latent coordinates, and subsequent works \citep{Damian2003, Bruno2009} relaxed the assumption of temporal stationarity by including temporal trends. However, in these approaches, inference was typically performed in separate stages.

Bayesian dynamic models \citep{West1997} have also been widely used to describe temporal evolution in spatiotemporal data \citep{Sanso1999, Stroud2001, Huerta2004}. These models relax temporal stationarity and provide a unified framework for uncertainty quantification and the inclusion of explanatory variables. \citet[Sec.~3]{Schmidt2011} and \citet{Morales2013} suggested combining dynamic models with spatial deformation, allowing temporal dynamics and spatial anisotropy to be handled simultaneously within a single inferential framework.

In this manuscript, we aim to analyze multidimensional spatiotemporal processes observed over a finite set of times at fixed monitoring stations. Building upon the matrix-normal Bayesian dynamic framework \citep{Quintana1987, Landim2000}, \citet{Paez2008} proposed modeling matrix-variate data using an isotropic Matérn covariance function to describe spatial dependence. Their results demonstrated that joint modeling of multiple response variables substantially improves interpolation accuracy compared with univariate approaches.

Several existing approaches address nonstationary spatial dependence, multivariate responses, or temporal dynamics separately. However, to the best of our knowledge, there are no methods that jointly incorporate spatial deformation, matrix-variate dependence, temporal dynamics, and missing data handling within a unified Bayesian framework. As a result, direct comparisons with existing methods are not straightforward. Instead, we focus on comparisons with nested versions of the proposed model, which allow isolating the contribution of spatial deformation and cross-variable dependence within a coherent inferential framework.

Here, we extend the matrix-normal Bayesian dynamic model of \citet{Paez2008} by incorporating spatial deformation, allowing spatial dependence to be represented in a latent isotropic space while preserving temporal dynamics through the Bayesian state-space formulation. This results in a unified model where deformation, covariance parameters, and temporal components are estimated jointly. The proposed formulation also leverages separability and the factorization of the dynamic regression parameters to enable computationally efficient inference. To perform inference, we develop an efficient Markov chain Monte Carlo algorithm for sampling from the posterior distribution of the model parameters.

The remainder of this paper is organized as follows. Section~\ref{sec:Complete} presents the proposed spatiotemporal model for complete matrices. Section~\ref{sec:Incomplete} extends the model to accommodate incomplete matrices, describes the imputation of missing values, and presents the inference procedure. Section~\ref{sec:Interpolation} discusses the interpolation problem. Section~\ref{sec:Model_Comparison-Checking} presents criteria for model comparison and checking. Simulation studies and an application are provided in Sections~\ref{sec:Simulation} and~\ref{sec:Application}, respectively. Finally, Section~\ref{sec:Conclusions} concludes the paper with a discussion of findings and future directions.

%--- Section ---%
\section{Spatiotemporal model for matrix-variate responses}
\label{sec:Complete}

\subsection{General framework}
\label{subsec:Framework}

We are interested in modeling multivariate spatiotemporal processes observed at a finite number of monitoring sites and time points. At each time $t$ and location $\uvec{\mathbf{s}}$, a $q$-dimensional response vector is recorded, representing multiple environmental variables measured simultaneously over space and time.

Formally, let $\mathcal{S} \subset \mathds{R}^{2}$ denote the spatial domain of interest and $\mathcal{T} = \{1, \ldots, T\}$ the discrete time index. We assume that $\{(Y_{1}(\uvec{\mathbf{s}}, t), \ldots, Y_{q}(\uvec{\mathbf{s}}, t)) \in \mathds{R}^{q}: \uvec{\mathbf{s}}\in \mathcal{S}, t\in\mathcal{T}\}$ is a Gaussian process (GP). Observations are available at $N$ fixed sites 
$\uvec{\mathbf{s}}_{1}, \ldots, \uvec{\mathbf{s}}_{N}$, with coordinates $\uvec{\mathbf{s}}_{n} = \left[\begin{smallmatrix}
  \lon_{n} & \lat_{n}
\end{smallmatrix}\right]^{\top}$ for $n \in \{1, \ldots, N\}$. 
The notation $Y_{n,i,t} = Y_{i}(\uvec{\mathbf{s}}_{n}, t)$ refers to the value of the $i$th response at site $\uvec{\mathbf{s}}_{n}$ and time $t$, with $i \in \{1, \ldots, q\}$.

For each $t \in \mathcal{T}$, the $N \times q$ response matrix is
\begin{equation}
\label{eq:Responses}
\mathbf{Y}_{t} =
  \begin{bmatrix}
      Y_{1,1,t} & \cdots & Y_{1,q,t} \\
      \vdots    & \ddots & \vdots    \\
      Y_{N,1,t} & \cdots & Y_{N,q,t}
  \end{bmatrix},
\end{equation}
whose $i$th column $\uvec{\mathbf{Y}}_{i,t} = \left[\begin{smallmatrix}
  Y_{1,i,t} & \cdots & Y_{N,i,t}
\end{smallmatrix}\right]^{\top}$ collects all $N$ spatial observations of variable $i$ at time $t$.

This matrix representation allows the model to capture both
(i) dependence over time, through the evolution of regression parameters, and 
(ii) dependence across space, through a covariance structure that accounts for anisotropy and deformation.

Throughout the paper, bold symbols denote matrices (e.g., $\mathbf{Y}_{t}$), bold symbols with an undertilde denote vectors (e.g., $\uvec{\mathbf{s}}$ and $\uvec{\mathbf{Y}}_{i,t}$), and non-bold symbols denote scalar quantities (e.g., $Y_{n,i,t}$).

\subsection{Modeling spatial dependence through deformation}
\label{subsec:Spatial_Structure}

Spatial correlation is rarely isotropic in environmental data. To flexibly describe anisotropy, we adopt the deformation approach of \citet{Sampson1992}, in which the observed geographic coordinates from $\mathcal{S}$ are mapped to a latent space $\mathcal{D} \subset \mathds{R}^{2}$ where isotropy approximately holds. This framework can be interpreted as a latent transformation of the input space, which connects spatial deformation conceptually to hierarchical or deep GP constructions \citep{damianou2013deep}.

Let $\{d(\uvec{\mathbf{s}}) \in \mathcal{D}: \uvec{\mathbf{s}} \in \mathcal{S})\}$ denote this mapping, with $d(\uvec{\mathbf{s}}_{n}) = \uvec{\mathbf{d}}_{n} = 
\left[\begin{smallmatrix}
  d_{1}(\uvec{\mathbf{s}}_{n}) & d_{2}(\uvec{\mathbf{s}}_{n})
\end{smallmatrix}\right]^{\top}$, and collect the latent coordinates in the $2 \times N$ matrix $\mathbf{D} = \left[\uvec{\mathbf{d}}_{1}, \ldots, \uvec{\mathbf{d}}_{N}\right]$. Based on \citet{Morales2013}, we model spatial dependence through an exponential kernel as a function of Euclidean distances between latent coordinates. Specifically, the $N \times N$ spatial dependence matrix $\mathbf{B}$ has entries
\begin{equation}
\label{eq:B}
B_{n,n'} = \exp\!\left\{-\phi\big\| d(\uvec{\mathbf{s}}_{n}) - d(\uvec{\mathbf{s}}_{n'}) \big\|\right\},
\end{equation}
where $\phi > 0$ controls the spatial range and $\|\cdot\|$ denotes the Euclidean distance.

Following \citet{Schmidt2003}, the deformation process $d(\cdot)$ is itself modeled as a bivariate GP, inducing the matrix-normal prior distribution
\begin{equation}
\label{eq:D_Prior}
[\mathbf{D} \mid \boldsymbol{\sigma}_{d}^{2}] \sim \sfNor_{2 \times N}(\mathbf{S}, \boldsymbol{\sigma}_{d}^{2}, \mathbf{R}_{d}),
\end{equation}
where $\mathbf{S} = \left[\uvec{\mathbf{s}}_{1}, \ldots, \uvec{\mathbf{s}}_{N}\right]$ is the $2 \times N$ matrix of observed coordinates, $\boldsymbol{\sigma}_{d}^{2}$ is a $2 \times 2$ covariance matrix controlling the magnitude of deformation along each coordinate, and the $N \times N$ matrix $\mathbf{R}_{d}$ encodes prior spatial correlation in $\mathcal{S}$.

In this work, $\sfNor_{p}(\uvec{\boldsymbol{\mu}}, \mathbf{U})$ denotes the multivariate normal distribution with mean vector $\uvec{\boldsymbol{\mu}} \in \mathds{R}^{p}$ and covariance matrix $\mathbf{U} \in \SPD(p)$, while $\sfNor_{p \times q}(\mathbf{M}, \mathbf{U}, \mathbf{V})$ denotes the matrix-normal distribution with mean matrix $\mathbf{M} \in \mathds{R}^{p \times q}$, row covariance matrix $\mathbf{U} \in \SPD(p)$, and column covariance matrix $\mathbf{V} \in \SPD(q)$, where $\SPD(\cdot)$ denotes the set of positive definite matrices. The matrix-normal distribution provides a convenient representation for matrix-valued random variables with separable covariance structure across rows and columns. In particular, \citet{Gupta2000} show the following equivalence between these distributions:
\begin{equation}
\label{eq:Vectorization}
\mathbf{X} \sim \sfNor_{p \times q}(\mathbf{M}, \mathbf{U}, \mathbf{V}) \iff
\vect(\mathbf{X}) \sim \sfNor_{pq}\!\left(\vect(\mathbf{M}), \mathbf{V} \otimes \mathbf{U}\right),
\end{equation}
where $\vect(\cdot)$ denotes the vectorization operator and $\otimes$ denotes the Kronecker product. Under this representation, the covariance between elements factorizes as $\cov(X_{i,j}, X_{i',j'}) = U_{i,i'}\, V_{j,j'}$, for $i,i' \in \{1, \ldots, p\}$ and $j,j' \in \{1, \ldots, q\}$.

\citet{Schmidt2003} adopt a Gaussian (squared-exponential) correlation function for $\mathbf{R}_{d}$, whose entries have the form
$R_{n,n'} = \exp\{-\psi \|\uvec{\mathbf{s}}_{n} - \uvec{\mathbf{s}}_{n'}\|^{2}\},$
with fixed $\psi > 0$, ensuring that geographically distant sites in $\mathcal{S}$ exhibit weaker prior dependence in $\mathcal{D}$. To avoid identifiability issues arising from invariances of the deformation map, these authors restrict $\boldsymbol{\sigma}_{d}^{2}$ to be diagonal, with elements $\sigma_{d_{1,1}}^{2}$ and $\sigma_{d_{2,2}}^{2}$ assigned independent inverse-gamma prior distributions.

An important practical consideration in deformation models is the possibility of ``folding'', where distinct locations in $\mathcal{S}$ are mapped to overlapping positions in $\mathcal{D}$. While not explicitly prohibited, folding is mitigated by the smoothness induced by the GP prior. As pointed out by \citet{Schmidt2003}, this formulation tends to discourage non-injective mappings and alleviates multimodality in posterior inference.

\subsection{Spatiotemporal hierarchical structure}
\label{subsec:ST_Modelling}

The temporal evolution of the process is modeled through a Bayesian dynamic structure applied to the matrix-variate responses defined in \eqref{eq:Responses}. 
At each time $t$, the $N \times q$ response matrix $\mathbf{Y}_{t}$ contains the $q$ variables measured simultaneously across the $N$ monitoring sites. 

The mean structure is defined through a design matrix $\mathbf{X}_{t}$ of dimension $N \times p$, where $p$ denotes the number of covariates. In particular, $p=1$ corresponds to a model with a time-varying intercept only, while $p>1$ allows for the inclusion of additional explanatory variables. The parameter $\boldsymbol{\Sigma} \in \SPD(q)$ accounts for contemporaneous dependence among the $q$ response variables. Spatial dependence is captured by the $N \times N$ matrix $\mathbf{B}$, a function of the spatial range $\phi$ and the deformations $\mathbf{D}$, whose elements are defined in \eqref{eq:B}, allowing for anisotropic spatial correlation among locations.

Associated with each time $t$, we define a latent parameter matrix $\boldsymbol{\beta}_{t} \in \mathbb{R}^{p \times q}$, whose entries represent regression effects driving the temporal evolution of the $q$ responses.

The hierarchical model consists of three stages:

\bmhead{Observation equation}
At each time $t$, the process $\mathbf{Y}_{t}$ is related to the current state $\boldsymbol{\beta}_{t}$ through the design matrix $\mathbf{X}_{t}$ and the spatial correlation structure $\mathbf{B}$ as
\begin{eqnarray}
\label{eq:Observation_Mat}
\bigl[\mathbf{Y}_{t} \mid \boldsymbol{\beta}_{t}, \phi,  \mathbf{D}, \boldsymbol{\Sigma} \bigr]
  &\sim& \sfNor_{N \times q}\bigl( \mathbf{X}_{t} \boldsymbol{\beta}_{t}, \mathbf{B},
  \boldsymbol{\Sigma} \bigr).
\end{eqnarray}

\bmhead{Evolution equation}
The latent states evolve dynamically over time according to
\begin{eqnarray}
\label{eq:Evolution_Mat}
\bigl[\boldsymbol{\beta}_{t} \mid \boldsymbol{\beta}_{t-1}, \mathbf{W}, \boldsymbol{\Sigma}\bigr]
  &\sim& \sfNor_{p \times q}\bigl( \mathbf{G}_{t} \boldsymbol{\beta}_{t-1}, \mathbf{W},
  \boldsymbol{\Sigma} \bigr),
\end{eqnarray}
where $\mathbf{G}_{t}$ is a known $p \times p$ state evolution matrix that governs the temporal propagation of the regression coefficients, typically specified to control their persistence over time (e.g., identity or other user-defined transition structures). 

The parameter $\mathbf{W} \in \SPD(p)$ controls the innovation variability of the latent states. In the proposed model, $\mathbf{W}$ is assumed to be diagonal, allowing for component-specific temporal variability while maintaining a parsimonious parameterization.

\bmhead{Initial information}
The initial state $\boldsymbol{\beta}_{0}$ follows
\begin{eqnarray}
\label{eq:InitialInformation_Mat}
\bigl[\boldsymbol{\beta}_{0} \mid \boldsymbol{\Sigma}\bigr]
  &\sim& \sfNor_{p \times q}\bigl( \mathbf{M}_{0}, \mathbf{C}_{0}, \boldsymbol{\Sigma} \bigr),
\end{eqnarray}
where $\mathbf{M}_{0} \in \mathbb{R}^{p \times q}$ and $\mathbf{C}_{0} \in \SPD(p)$ denote the prior mean and covariance matrices describing the initial uncertainty of the states.

More general specifications for $\mathbf{G}_t$ and $\mathbf{W}$, such as autoregressive or fully parameterized structures, could be considered. However, following common practice in Bayesian dynamic linear models \citep{West1997}, we adopt a parsimonious specification in which $\mathbf{G}_t$ is fixed and $\mathbf{W}$ is structured to avoid overparameterization. Jointly estimating more flexible temporal components alongside spatial deformation and cross-variable dependence would substantially increase the parameter space and may introduce confounding effects, potentially leading to reduced computational stability.

%--- Section ---%
\section{Model for incomplete matrix-variate responses}
\label{sec:Incomplete}

\subsection{General framework}
\label{subsec:General}

We allow each $N\times q$ response matrix $\mathbf{Y}_{t}$ to contain missing entries. 
Let $N_{t}^{\obs}$ and $N_{t}^{\mis}$ denote, respectively, the numbers of observed and missing elements at time $t$, with $N_{t}^{\obs}+N_{t}^{\mis}=Nq$. 
Depending on the pattern of availability, each matrix $\mathbf{Y}_{t}$ can fall into one of three categories: (i) fully observed ($N_{t}^{\obs}=Nq$), (ii) partially observed ($0<N_{t}^{\obs}<Nq$), and (iii) fully missing ($N_{t}^{\obs}=0$).

Let $\mathcal{T}_{\text{complete}}$, $\mathcal{T}_{\text{partial}}$, and $\mathcal{T}_{\text{missing}}$ denote the sets of time indices corresponding to these three cases. 
The complete set of time indices is then the disjoint union 
$\mathcal{T}=\mathcal{T}_{\text{complete}}\cup \mathcal{T}_{\text{partial}}\cup \mathcal{T}_{\text{missing}}$. 
Following \citet{Gamerman2022}, the missing-data mechanism is assumed to be Missing Completely at Random (MCAR) in the sense of \citet[Sec.~1.3]{Little2019}. Under this assumption, the probability of missingness does not depend on either observed or unobserved data values, which allows the missingness mechanism to be treated as ignorable for likelihood-based and Bayesian inference.

When $\mathbf{Y}_{t}$ is partially observed, missing entries may be arbitrarily located within the matrix. 
It is convenient to work with the vectorized form $\uvec{\mathbf{Y}}_{t}=\vect(\mathbf{Y}_{t})$, which allows us to separate observed and missing components using a permutation. 
For illustration, consider $N=4$ and $q=2$:
\[
\mathbf{Y}_{t}=
\begin{bmatrix}
Y_{1,1,t}^{\obs} & Y_{1,2,t}^{\mis}\\
Y_{2,1,t}^{\obs} & Y_{2,2,t}^{\mis}\\
Y_{3,1,t}^{\mis} & Y_{3,2,t}^{\obs}\\
Y_{4,1,t}^{\obs} & Y_{4,2,t}^{\obs}
\end{bmatrix},
\qquad
\uvec{\mathbf{Y}}_{t}=
\begin{bmatrix}
Y_{1,1,t}^{\obs} & Y_{2,1,t}^{\obs} & Y_{3,1,t}^{\mis} & Y_{4,1,t}^{\obs} &
Y_{1,2,t}^{\mis} & Y_{2,2,t}^{\mis} & Y_{3,2,t}^{\obs} & Y_{4,2,t}^{\obs}
\end{bmatrix}^{\top}.
\]

Let $\mathbf{P}_{t}$ be the $Nq\times Nq$ permutation matrix that rearranges $\uvec{\mathbf{Y}}_{t}$ so that observed elements come first:
\[
\mathbf{P}_{t}\uvec{\mathbf{Y}}_{t}=
\begin{bmatrix}
\uvec{\mathbf{Y}}_{t,\obs}\\[2pt]
\uvec{\mathbf{Y}}_{t,\mis}
\end{bmatrix},
\quad
\mathbf{L}_{t,\obs}=\begin{bmatrix}\mathbf{I}_{N_{t}^{\obs}} & \mathbf{0}\end{bmatrix},
\quad
\mathbf{L}_{t,\mis}=\begin{bmatrix}\mathbf{0} & \mathbf{I}_{N_{t}^{\mis}}\end{bmatrix},
\]
so that $\uvec{\mathbf{Y}}_{t,\obs}=\mathbf{L}_{t,\obs}\mathbf{P}_{t}\uvec{\mathbf{Y}}_{t}$ and $\uvec{\mathbf{Y}}_{t,\mis}=\mathbf{L}_{t,\mis}\mathbf{P}_{t}\uvec{\mathbf{Y}}_{t}$. 
Intuitively, $\mathbf{P}_{t}$ reorders the entries so that observed components appear first, while $\mathbf{L}_{t,\obs}$ and $\mathbf{L}_{t,\mis}$ extract the observed and missing subvectors, respectively. 
If $t\in\mathcal{T}_{\text{complete}}$, then $\mathbf{P}_{t}=\mathbf{I}_{Nq}$ and $\uvec{\mathbf{Y}}_{t,\obs}=\uvec{\mathbf{Y}}_{t}$.

To accommodate arbitrary missingness patterns within $\mathbf{Y}_{t}$, it is necessary to work with the vectorized form of the model. Write $\uvec{\mathbf{Y}}_{t} = \vect(\mathbf{Y}_{t})$, $\uvec{\boldsymbol{\beta}}_{t} = \vect(\boldsymbol{\beta}_{t})$, and $\uvec{\mathbf{M}}_{0} = \vect(\mathbf{M}_{0})$. Applying \eqref{eq:Vectorization}, then \eqref{eq:Observation_Mat}--\eqref{eq:InitialInformation_Mat} are respectively equivalent to
\begin{eqnarray}
\label{eq:Observation_Vec}
\left[\uvec{\mathbf{Y}}_{t} \mid \uvec{\boldsymbol{\beta}}_{t}, \phi, \mathbf{D}, \boldsymbol{\Sigma}\right] & \sim & \sfNor_{Nq}([\mathbf{I}_{q} \otimes \mathbf{X}_{t}]\uvec{\boldsymbol{\beta}}_{t}, \boldsymbol{\Sigma} \otimes \mathbf{B}), \\
\label{eq:Evolution_Vec}
\left[\uvec{\boldsymbol{\beta}}_{t} \mid \uvec{\boldsymbol{\beta}}_{t-1}, \mathbf{W}, \boldsymbol{\Sigma}\right] & \sim & \sfNor_{pq}([\mathbf{I}_{q} \otimes \mathbf{G}_{t}]\uvec{\boldsymbol{\beta}}_{t-1}, \boldsymbol{\Sigma} \otimes \mathbf{W}), \\
\label{eq:InitialInformation_Vec}
\left[\uvec{\boldsymbol{\beta}}_{0} \mid \boldsymbol{\Sigma}\right] & \sim & \sfNor_{pq}(\uvec{\mathbf{M}}_{0}, \boldsymbol{\Sigma} \otimes \mathbf{C}_{0}).
\end{eqnarray}

For $t \in \mathcal{T}_{\text{partial}}$, from \eqref{eq:Observation_Vec} we obtain
\begin{equation*}
\left[\begin{bmatrix}
\uvec{\mathbf{Y}}_{t,\obs}\\[2pt]
\uvec{\mathbf{Y}}_{t,\mis}
\end{bmatrix} \mid \boldsymbol{\theta}\right] \sim \sfNor_{Nq}(\uvec{\boldsymbol{\mu}}_{t}, \boldsymbol{\Delta}_{t}),
\end{equation*}
which leads to the usual marginal and conditional formulations:
\begin{equation}
\label{eq:Dist_Mis}  
[\uvec{\mathbf{Y}}_{t,\mis}\mid\boldsymbol{\theta}]
\sim \sfNor_{N_{t}^{\mis}}(\uvec{\boldsymbol{\mu}}_{t,\mis},  \boldsymbol{\Delta}_{t,\mis})
\end{equation}
and
\begin{equation}
\label{eq:Dist_Mis_Obs}
[\uvec{\mathbf{Y}}_{t,\mis}\mid \uvec{\mathbf{Y}}_{t,\obs}=\uvec{\mathbf{y}}_{t,\obs},\boldsymbol{\theta}]
\sim 
\sfNor_{N_{t}^{\mis}}\!\big(
\uvec{\boldsymbol{\mu}}_{t,\cond}, \boldsymbol{\Delta}_{t,\cond}\big),
\end{equation}
where
\begin{equation*}
\uvec{\boldsymbol{\mu}}_{t}
=\mathbf{P}_{t}[\mathbf{I}_{q}\otimes\mathbf{X}_{t}]\uvec{\boldsymbol{\beta}}_{t}
=\begin{bmatrix}\uvec{\boldsymbol{\mu}}_{t,\obs}\\ \uvec{\boldsymbol{\mu}}_{t,\mis}\end{bmatrix},
\qquad
\boldsymbol{\Delta}_{t}
=\mathbf{P}_{t}[\boldsymbol{\Sigma}\otimes\mathbf{B}]\mathbf{P}_{t}^{\top}
=\begin{bmatrix}
\boldsymbol{\Delta}_{t,\obs} & \boldsymbol{\Delta}_{t,\om}\\
\boldsymbol{\Delta}_{t,\om}^{\top} & \boldsymbol{\Delta}_{t,\mis}
\end{bmatrix},
\end{equation*}
\begin{align*}
\uvec{\boldsymbol{\mu}}_{t,\cond}
&= 
\uvec{\boldsymbol{\mu}}_{t,\mis}
+ \boldsymbol{\Delta}_{t,\om}^{\top}\boldsymbol{\Delta}_{t,\obs}^{-1}
  \big(\uvec{\mathbf{y}}_{t,\obs}-\uvec{\boldsymbol{\mu}}_{t,\obs}\big),
\\[2pt]
\boldsymbol{\Delta}_{t,\cond}
&=
\boldsymbol{\Delta}_{t,\mis}
- \boldsymbol{\Delta}_{t,\om}^{\top}\boldsymbol{\Delta}_{t,\obs}^{-1}\boldsymbol{\Delta}_{t,\om}.
\end{align*}

Collect all partially or fully missing vectors as
\[
\uvec{\mathbf{Y}}_{\mis}=\{\uvec{\mathbf{Y}}_{t,\mis}:0<N_{t}^{\mis}\le Nq\},
\qquad
\uvec{\mathbf{Y}}_{\obs}=\{\uvec{\mathbf{Y}}_{t,\obs}:N_{t}^{\obs}>0\}.
\]

Assuming conditional independence across time given the latent states and other model parameters,
\begin{equation}
\label{eq:Missing}
f(\uvec{\mathbf{y}}_{\mis}\mid \boldsymbol{\theta},\uvec{\mathbf{y}}_{\obs})
=
\prod_{t\in\mathcal{T}_{\text{partial}}}
f(\uvec{\mathbf{y}}_{t,\mis}\mid \uvec{\mathbf{y}}_{t,\obs},\boldsymbol{\theta})
\prod_{t\in\mathcal{T}_{\text{missing}}}
f(\uvec{\mathbf{y}}_{t,\mis}\mid \boldsymbol{\theta}),
\end{equation}
with support 
\[
\mathcal{Y}_{\mis}=
\left(\prod_{t\in\mathcal{T}_{\text{partial}}} \mathds{R}^{N_{t}^{\mis}}\right)
\times
\left(\prod_{t\in\mathcal{T}_{\text{missing}}} \mathds{R}^{Nq}\right).
\]

Algorithm~\ref{alg:Missing} samples from \eqref{eq:Missing} using Monte Carlo simulation. 
This step imputes missing values by sampling from their conditional Gaussian distributions, naturally incorporating spatial, temporal, and cross-variable dependence through $\boldsymbol{\Delta}_{t}$.

\subsection{Bayesian inference}

Assuming prior independence among the model components $\phi$, $\{\boldsymbol{\sigma}_{d}^{2}, \mathbf{D}\}$, $\mathbf{W}$, and $\boldsymbol{\Sigma}$, and applying the Markov property of the latent states $\uvec{\boldsymbol{\beta}}_{0:T} = \{\uvec{\boldsymbol{\beta}}_{0}, \uvec{\boldsymbol{\beta}}_{1}, \ldots, \uvec{\boldsymbol{\beta}}_{T}\}$ given in \eqref{eq:Evolution_Vec} and \eqref{eq:InitialInformation_Vec}, the joint prior distribution of the parameters can be written as
\begin{equation}
\label{eq:Prior}
f(\boldsymbol{\theta}) = f(\phi) f(\boldsymbol{\sigma}_{d}^{2}) f(\mathbf{D} \mid \boldsymbol{\sigma}_{d}^{2}) f(\mathbf{W}) f(\boldsymbol{\Sigma})
  f(\uvec{\boldsymbol{\beta}}_{0}\mid \boldsymbol{\Sigma})
  \prod_{t=1}^{T} f(\uvec{\boldsymbol{\beta}}_{t}\mid \uvec{\boldsymbol{\beta}}_{t-1}, \mathbf{W}, \boldsymbol{\Sigma}),
\end{equation}
where $\boldsymbol{\theta} = \{\uvec{\boldsymbol{\beta}}_{0:T}, \mathbf{W}, \boldsymbol{\Sigma}, \phi, \boldsymbol{\sigma}_{d}^{2}, \mathbf{D}\}$ denotes the complete set of model parameters. 

Given conditional independence of the observations across time given the latent states and other model parameters, the likelihood function is expressed as
\begin{equation}
\label{eq:Likelihood}
\ell(\boldsymbol{\theta}; \uvec{\mathbf{y}}) = \prod_{t=1}^{T}
f(\uvec{\mathbf{y}}_{t} \mid \uvec{\boldsymbol{\beta}}_{t}, \phi, \mathbf{D}, \boldsymbol{\Sigma}),
\end{equation}
and combining~\eqref{eq:Prior} and~\eqref{eq:Likelihood} through Bayes' theorem yields the posterior density
\begin{multline}
\label{eq:Posterior}
    f(\boldsymbol{\theta} \mid \uvec{\mathbf{y}}) \propto
    f(\phi) f(\boldsymbol{\sigma}_{d}^{2}) f(\mathbf{D} \mid \boldsymbol{\sigma}_{d}^{2}) f(\mathbf{W}) f(\boldsymbol{\Sigma})
    f(\uvec{\boldsymbol{\beta}}_{0} \mid \boldsymbol{\Sigma}) \\
    \times \prod_{t=1}^{T}
    f(\uvec{\boldsymbol{\beta}}_{t}\mid \uvec{\boldsymbol{\beta}}_{t-1}, \mathbf{W}, \boldsymbol{\Sigma})
    f(\uvec{\mathbf{y}}_{t} \mid \uvec{\boldsymbol{\beta}}_{t}, \phi, \mathbf{D}, \boldsymbol{\Sigma}).
\end{multline}

Given the prevalence of missing values in environmental data, we aim to sample from $f(\uvec{\mathbf{y}}_{\mis}, \boldsymbol{\theta} \mid \uvec{\mathbf{y}}_{\obs})$ via a data augmentation approach. Under the MCAR assumption described in Section~\ref{subsec:General}, the missingness mechanism is ignorable, and the likelihood can be expressed in terms of the observed and missing components. In particular, exploiting the invariance of the multivariate normal distribution under permutations of its components \citep[see, e.g.,][Sec.~2.4]{migon2014}, the joint density can be equivalently written in terms of $\{\uvec{\mathbf{y}}_{\obs}, \uvec{\mathbf{y}}_{\mis}\}$ without loss of generality. This implies that $\ell(\boldsymbol{\theta}; \uvec{\mathbf{y}}) = \ell(\boldsymbol{\theta}; \uvec{\mathbf{y}}_{\obs}, \uvec{\mathbf{y}}_{\mis})$, and consequently $f(\boldsymbol{\theta} \mid \uvec{\mathbf{y}}) = f(\boldsymbol{\theta} \mid \uvec{\mathbf{y}}_{\obs}, \uvec{\mathbf{y}}_{\mis})$. Posterior samples are obtained by alternating between draws of $\uvec{\mathbf{y}}_{\mis}$ and $\boldsymbol{\theta}$ from \eqref{eq:Missing} and \eqref{eq:Posterior}, respectively.

Details on the computational implementation are provided in Section~\ref{subsec:MCMC}.

\subsection{Markov chain Monte Carlo algorithm}
\label{subsec:MCMC}

Posterior inference is carried out via a hybrid Markov chain Monte Carlo (MCMC) algorithm. Missing responses are treated as additional unknowns and updated within a data augmentation framework. Since the posterior distribution in~\eqref{eq:Posterior} does not admit a closed-form expression, the scheme combines Gibbs sampling for conjugate components, Forward Filtering Backward Sampling (FFBS) for the latent dynamic coefficients, and simulation-based updates for non-conjugate parameters. These include Metropolis--Hastings steps for the spatial range parameter $\phi$, a marginal Metropolis--Hastings update for the state-evolution covariance matrix $\mathbf{W}$ based on an integrated likelihood, and a No-U-Turn Sampler (NUTS) step for the deformation matrix $\mathbf{D}$.

As the model admits equivalent matrix and vectorized representations, given in \eqref{eq:Observation_Mat}--\eqref{eq:InitialInformation_Mat} and \eqref{eq:Observation_Vec}--\eqref{eq:InitialInformation_Vec}, respectively, it induces two corresponding parameterizations of $\boldsymbol{\theta}$ with respect to the latent states. For computational purposes, we alternate between these representations. Missing values are imputed in vectorized form, using $\uvec{\mathbf{y}}_{t}$ as implemented in Algorithm~\ref{alg:Missing}, with latent states represented as $\uvec{\boldsymbol{\beta}}_{0:T}$. After imputation, the completed response vectors are reshaped into their corresponding $N \times q$ matrix forms $\mathbf{y}_{t}$, and the latent states are subsequently handled in matrix form as $\boldsymbol{\beta}_{0:T}$.

This choice is motivated by computational efficiency: working directly with matrix-normal distributions avoids repeated inversions of large Kronecker-structured covariance matrices, replacing them with operations on their lower-dimensional factors.

We next describe the sampling steps for each model component, followed by a description of the full MCMC scheme.

\subsubsection*{Contemporaneous covariance matrix}
\bmhead{Unrestricted case}
When the $q\times q$ contemporaneous covariance matrix $\boldsymbol{\Sigma}$ is unrestricted, we assign an inverse-Wishart prior
\[
\boldsymbol{\Sigma} \sim \sfIW_{q}(a_{\boldsymbol{\Sigma}}, \mathbf{b}_{\boldsymbol{\Sigma}}),
\]
where $\sfIW_{q}(a, \mathbf{b})$ denotes the inverse-Wishart distribution with $a > q - 1$ degrees of freedom and scale matrix $\mathbf{b}\in\SPD(q)$. 

One can show that its full conditional distribution remains inverse-Wishart (see Appendix~\ref{subsubsec:Sigma_Complete}), with
\begin{equation}
\label{eq:Sigma_Full}
[\boldsymbol{\Sigma} \mid \mathbf{Y} = \mathbf{y}, \boldsymbol{\beta}_{0:T}, \mathbf{W}, \phi, \mathbf{D}]
\sim \sfIW_{q}(a'_{\boldsymbol{\Sigma}}, \mathbf{b}'_{\boldsymbol{\Sigma}}),
\end{equation}
where
\begin{equation}
\label{eq:updated_a-Sigma}
a'_{\boldsymbol{\Sigma}} = a_{\boldsymbol{\Sigma}} + p + Tp + TN
\end{equation}
and
\begin{equation}
\label{eq:updated_b-Sigma}
\begin{array}{rcl}
\mathbf{b}'_{\boldsymbol{\Sigma}}
&=&
\mathbf{b}_{\boldsymbol{\Sigma}}
+ (\boldsymbol{\beta}_{0}-\mathbf{M}_{0})^{\top}\mathbf{C}_{0}^{-1}(\boldsymbol{\beta}_{0}-\mathbf{M}_{0})
\\[4pt]
&&
+ \sum\limits_{t=1}^{T}
(\boldsymbol{\beta}_{t}-\mathbf{G}_{t}\boldsymbol{\beta}_{t-1})^{\top}\mathbf{W}^{-1}(\boldsymbol{\beta}_{t}-\mathbf{G}_{t}\boldsymbol{\beta}_{t-1})
\\[4pt]
&&
+ \sum\limits_{t=1}^{T}
(\mathbf{y}_{t}-\mathbf{X}_{t}\boldsymbol{\beta}_{t})^{\top}\mathbf{B}^{-1}(\mathbf{y}_{t}-\mathbf{X}_{t}\boldsymbol{\beta}_{t}).
\end{array}
\end{equation}

\bmhead{Diagonal case}
When independence among response variables is assumed, the contemporaneous covariance matrix has the form $\boldsymbol{\Sigma} = \diag\{\Sigma_{1,1}, \ldots, \Sigma_{q,q}\}$. We assign independent inverse-gamma priors to its diagonal elements:
\[
\Sigma_{i,i} \sim \sfIG(a_{\Sigma_{i,i}}, b_{\Sigma_{i,i}}),
\qquad i \in \{1,\ldots,q\},
\]
where $\sfIG(a, b)$ denotes the inverse-gamma distribution with shape $a > 0$ and scale $b > 0$.

It can be shown that each full conditional distribution is also inverse-gamma (see Appendix~\ref{subsubsec:Sigma_Diagonal}), with
\begin{equation}
\label{eq:Sigma_Diag}
[\Sigma_{i,i} \mid \mathbf{Y} = \mathbf{y}, \boldsymbol{\beta}_{0:T}, \mathbf{W}, \phi, \mathbf{D}]
\sim \sfIG(a'_{\Sigma_{i,i}}, b'_{\Sigma_{i,i}}),
\end{equation}
where
\begin{equation}
\label{eq:updated_a-Sigma_i}
a'_{\Sigma_{i,i}} = a_{\Sigma_{i,i}} + \frac{p}{2} + \frac{Tp}{2} + \frac{TN}{2}
\end{equation}
and
\begin{equation}
\label{eq:updated_b-Sigma_i}
\begin{array}{rcl}
b'_{\Sigma_{i,i}}
&=&
b_{\Sigma_{i,i}}
+ \dfrac{1}{2}(\uvec{\boldsymbol{\beta}}_{i,0} - \uvec{\mathbf{M}}_{i,0})^{\top}\mathbf{C}_{0}^{-1}(\uvec{\boldsymbol{\beta}}_{i,0} - \uvec{\mathbf{M}}_{i,0})
\\[4pt]
&&
+ \dfrac{1}{2}\sum\limits_{t=1}^{T}(\uvec{\boldsymbol{\beta}}_{i,t}-\mathbf{G}_{t}\uvec{\boldsymbol{\beta}}_{i,t-1})^{\top}\mathbf{W}^{-1}(\uvec{\boldsymbol{\beta}}_{i,t}-\mathbf{G}_{t}\uvec{\boldsymbol{\beta}}_{i,t-1})
\\[4pt]
&&
+ \dfrac{1}{2}\sum\limits_{t=1}^{T}(\uvec{\mathbf{y}}_{i,t}-\mathbf{X}_{t}\uvec{\boldsymbol{\beta}}_{i,t})^{\top}\mathbf{B}^{-1}(\uvec{\mathbf{y}}_{i,t}-\mathbf{X}_{t}\uvec{\boldsymbol{\beta}}_{i,t}).
\end{array}
\end{equation}

This diagonal case serves as a benchmark model under conditional independence across response variables.

\subsubsection*{Dynamic coefficients} 
The full conditional distribution of the latent state trajectory $\boldsymbol{\beta}_{0:T} = \{\boldsymbol{\beta}_{0}, \boldsymbol{\beta}_{1}, \ldots, \boldsymbol{\beta}_{T}\}$ 
is given by
\begin{multline}
\label{eq:beta_FCD}
f(\boldsymbol{\beta}_{0:T} \mid \mathbf{y}, \mathbf{W}, \phi, \mathbf{D}, \boldsymbol{\Sigma}) \propto
f(\boldsymbol{\beta}_{0} \mid \boldsymbol{\Sigma}) \\
\times \prod\limits_{t=1}^{T}
f(\boldsymbol{\beta}_{t} \mid \boldsymbol{\beta}_{t-1}, \mathbf{W}, \boldsymbol{\Sigma})
f(\mathbf{y}_{t} \mid \boldsymbol{\beta}_{t}, \phi, \mathbf{D}, \boldsymbol{\Sigma}).
\end{multline}

Inference for the dynamic regression coefficients $\boldsymbol{\beta}_{0:T}$ is carried out using the Forward Filtering Backward Sampling (FFBS) algorithm. FFBS is a standard technique for Gaussian state-space models that enables efficient sampling of the entire latent state trajectory conditional on the model parameters. The forward filtering step recursively computes the filtering distributions using Kalman filter recursions, while the backward sampling step draws from the corresponding smoothing distribution. This procedure yields exact samples from the posterior distribution of the latent states and scales linearly with the number of time points \citep{fruhwirth1994data, carter1994gibbs, Chib1995}.

\subsubsection*{State-evolution covariance matrix}

We assume that the state-evolution covariance matrix has diagonal form,
\[
\mathbf{W}=\diag\{W_{1,1},\ldots,W_{p,p}\}.
\]

Rather than sampling $\mathbf{W}$ from its full conditional distribution given $\boldsymbol{\beta}_{0:T}$, we update it from a marginal posterior distribution obtained by integrating out the latent state trajectory. This strategy reduces posterior dependence between $\mathbf{W}$ and $\boldsymbol{\beta}_{0:T}$ and improves mixing in dynamic models \citep{fruhwirth1994data}.

For each diagonal element, we assume an independent Lomax (Pareto Type II) prior distribution with density
\[
f(W_{j,j})
=
\frac{\lambda}{\tau_j^2}
\left(1+\frac{W_{j,j}}{\tau_j^2}\right)^{-(\lambda+1)}
\mathds{1}_{(0,\infty)}(W_{j,j}),
\qquad j=1,\ldots,p,
\]
where $\tau_j^2>0$ is a scale parameter controlling the magnitude of the state innovations, and $\lambda>0$ governs the tail behavior. The Lomax distribution provides a heavy-tailed alternative to the inverse-gamma prior, allowing greater robustness with respect to prior specification of the innovation variances. In our implementation, we set $\lambda=1$, yielding a proper distribution with infinite variance and a weakly-informative specification that avoids excessive shrinkage while still penalizing large values of $W_{j,j}$.

To preserve positivity and improve numerical stability, the update is performed on the logarithmic scale. Let
\[
\eta_j=\ln W_{j,j},
\qquad
\uvec{\boldsymbol{\eta}}_W= \left[\eta_1 \; \cdots \; \eta_p\right]^\top.
\]

The marginal likelihood is obtained by integrating out the latent states,
\begin{equation*}
f(\mathbf{y}\mid \uvec{\boldsymbol{\eta}}_W,\boldsymbol{\Sigma},\phi,\mathbf{D})
=
\int\limits_{\prod\limits_{t=0}^{T} \mathds{R}^{p \times q}}
f(\mathbf{y} \mid \boldsymbol{\beta}_{0:T},\boldsymbol{\Sigma},\phi,\mathbf{D})
f(\boldsymbol{\beta}_{0:T} \mid \mathbf{W},\boldsymbol{\Sigma})
\, \partial \boldsymbol{\beta}_{0:T},
\end{equation*}
and is evaluated via the Kalman filter.

A random-walk Metropolis--Hastings step is then applied to $\mathbf{W}$, with proposals constructed on the logarithmic scale via the transformation $\eta_j = \ln W_{j,j}$. A detailed description of this update is provided in Algorithm~\ref{alg:Update_W}.

\subsubsection*{Spatial range parameter}  
The spatial range parameter $\phi$ controls the rate of decay of spatial dependence in the deformed space. We assign a gamma prior of the form
\begin{equation}
\label{eq:phi_Prior}
\phi \sim \sfG(1, 0.3 / \zeta),
\end{equation}
where $\sfG(a,b)$ denotes the Gamma distribution with shape $a>0$ and rate $b>0$, and $\zeta$ is the median Euclidean distance between all pairs of gauged sites in $\mathcal{S}$.

The prior specification in \eqref{eq:phi_Prior} follows the scale-adaptive construction proposed by \citet[Sec.~4]{Fonseca2011} and yields a prior that reflects the spatial configuration of the monitoring network while remaining weakly informative. In particular, scaling the rate parameter by $\zeta$ ensures that the prior is properly calibrated to the spatial scale of the domain and invariant to changes in measurement units.

Since the full conditional distribution is not available in closed form, $\phi$ is updated by a Metropolis--Hastings step using the log-target density
\begin{multline}
\label{eq:phi_FCD}
\ln f(\phi\mid \mathbf{y}, \boldsymbol{\beta}_{0:T},\mathbf{D},\boldsymbol{\Sigma})
=
\textnormal{Constant}
-\frac{0.3}{\zeta}\phi
-\frac{Tq}{2}\ln\det\mathbf{B}
\\
-\frac{1}{2}
\sum_{t=1}^{T}
\tr\left[\Bigl(
\mathbf{Y}_{t}-\mathbf{X}_{t}\boldsymbol{\beta}_t
\Bigr)^\top \mathbf{B}^{-1}
\Bigl(
\mathbf{Y}_{t}-\mathbf{X}_{t}\boldsymbol{\beta}_t
\Bigr) \boldsymbol{\Sigma}^{-1} \right],
\end{multline}
and log-normal random-walk proposals,
\[
\phi^{\textnormal{prop}}=\phi^{\textnormal{curr}}\exp(\varepsilon_\phi),
\qquad
\varepsilon_\phi\sim \sfNor(0,\delta_\phi^2),
\]
which automatically preserve positivity.

\subsubsection*{Deformation variance parameters}

The left covariance matrix in \eqref{eq:D_Prior} is assumed to have diagonal form,
\[
\boldsymbol{\sigma}_d^2=\diag\{\sigma_{d_{1,1}}^2,\sigma_{d_{2,2}}^2\}.
\]

Suppose that $\sigma_{d_{1,1}}^2$ and $\sigma_{d_{2,2}}^2$ have independent prior distributions. For $m \in \{1, 2\}$, assigning  
\begin{equation}
\label{eq:sigma2d_Prior}
\sigma_{d_{m,m}}^2 \sim \sfIG(a_{\sigma_{d,m}},b_{\sigma_{d,m}}),
\end{equation}
it follows by conjugacy that the full conditional distributions are also inverse-gamma (see Appendix~\ref{subsec:sigma2d}), given by
\begin{equation}
\label{eq:sigma2d_m-FCD}
[\sigma_{d_{m,m}}^2\mid \mathbf{D}]
\sim
\sfIG\!\left(
a_{\sigma_{d,m}}+\frac{N}{2},
\;
b_{\sigma_{d,m}}+\frac{1}{2}\mathbf{\Delta}_{m,\cdot}\mathbf{R}_d^{-1}\mathbf{\Delta}_{m,\cdot}^{\top}
\right),
\end{equation}
where $\mathbf{\Delta}_{m,\cdot}$ ($1 \times N$) denotes the $m$th row of $\mathbf{\Delta} = \mathbf{D}-\mathbf{S}$.

\subsubsection*{Deformation matrix}

\bmhead{General case}
The latent deformation matrix $\mathbf{D}$ is assigned a matrix-normal prior given in \eqref{eq:D_Prior}. Its full conditional distribution has no closed form, and its log-target density is
\begin{multline}
\label{eq:D_FCD}
\ln f(\mathbf{D} \mid \mathbf{y}, \boldsymbol{\beta}_{0:T},
\phi,\boldsymbol{\sigma}_d^2,\boldsymbol{\Sigma})
= \textnormal{Constant}
-\frac{1}{2}
\tr\!\left[
(\mathbf{D}-\mathbf{S})^\top
\boldsymbol{\sigma}_d^{-2}
(\mathbf{D}-\mathbf{S})
\mathbf{R}_d^{-1}
\right]
\\
-\frac{Tq}{2}\ln\det\mathbf{B}
-\frac{1}{2}
\tr\!\left(\mathbf{B}^{-1}\mathbf{Q}\right),
\end{multline}
where
\[
\mathbf{Q} = \sum_{t=1}^{T}
(\mathbf{Y}_{t}-\mathbf{X}_t\boldsymbol{\beta}_t)
\boldsymbol{\Sigma}^{-1}
(\mathbf{Y}_{t}-\mathbf{X}_t\boldsymbol{\beta}_t)^\top
\]
collects the aggregated cross-products of the residuals.

Regarding the spatial component, note that \eqref{eq:B} is a strictly decreasing function of the distance in the deformed space. As a result, the covariance structure defined by $\mathbf{B}$ is unique up to a scaling factor, and the deformation $d(\cdot)$ is unique up to homothetic transformations \citep{perrin1999}. Following standard practice in Bayesian spatial deformation models \citep{Damian2001, Morales2013, Gamerman2024, sampson2014}, the first two columns of $\mathbf{D}$ are fixed at the corresponding observed spatial coordinates to serve as anchor points. That is, by setting $\uvec{\mathbf{d}}_1=\uvec{\mathbf{s}}_1$ and $\uvec{\mathbf{d}}_2=\uvec{\mathbf{s}}_2$, we ensure identifiability of the deformation. Accordingly, only the remaining $N-2$ columns are sampled, and we write $\mathbf{D} = \left[\uvec{\mathbf{s}}_{1}, \uvec{\mathbf{s}}_{2}, \mathbf{D}_{\free}\right]$.

The deformation matrix $\mathbf{D}$ is the most computationally demanding component of the posterior simulation due to the nonlinear dependence of $\mathbf{B}$ on all pairwise distances in the deformed space. In this work, $\mathbf{D}$ is sampled jointly using the No-U-Turn Sampler (NUTS) \citep[see, e.g.,][Chap.~12]{gelman2013}, which exploits gradient information from the log-target density \eqref{eq:D_FCD}. Details of the gradient computation are provided in Appendix~\ref{subsec:D_gradient}.

At each iteration, NUTS introduces an auxiliary Gaussian momentum vector and simulates Hamiltonian trajectories using the leapfrog integrator. The algorithm recursively builds a binary tree, expanding forward and backward in time until a U-turn is detected or a pre-specified maximum tree depth is reached. One candidate state is then selected according to the standard NUTS transition rule. In our implementation, the step size is adapted during burn-in using a dual-averaging scheme \citep{hoffman2014}, targeting an average acceptance probability close to $0.8$. After the update, the anchor columns are reattached to the sampled free coordinates to reconstruct the full matrix $\mathbf{D}$.

\bmhead{Isotropic case}
In the isotropic case, deformation is suppressed by setting $\mathbf{D} = \mathbf{S}$, thus providing a baseline for model comparison.

\subsubsection*{Hybrid MCMC algorithm}

The proposed hybrid MCMC scheme alternates between sampling the missing responses $\uvec{\mathbf{y}}_{\mis}$ and updating the model parameters $\boldsymbol{\theta}$ through the sequence of block updates described earlier. These steps are repeated to generate a Markov chain with invariant distribution $[\uvec{\mathbf{Y}}_{\mis}, \boldsymbol{\theta} \mid \uvec{\mathbf{Y}}_{\obs} = \uvec{\mathbf{y}}_{\obs}]$.

Particular attention is required for the update of the state evolution covariance matrix $\mathbf{W}$. In this step, $\mathbf{W}$ is sampled from a marginal likelihood obtained by integrating out the latent state trajectory $\boldsymbol{\beta}_{0:T}$. The latent states are then sampled via the FFBS algorithm later in the same MCMC iteration, conditional on the updated value of $\mathbf{W}$. This update sequence is related to partially collapsed Gibbs strategies \citep{vandyk2008}, and is adopted to improve mixing by reducing posterior dependence between $\mathbf{W}$ and $\boldsymbol{\beta}_{0:T}$.

A detailed description of the full iterative procedure is given in Algorithm~\ref{alg:MCMC}.

%--- Section ---%
\section{Interpolation}
\label{sec:Interpolation}

This section addresses the prediction of responses at $N^{\ast}$ ungauged sites $\uvec{\mathbf{s}}_{N+1},\ldots,\uvec{\mathbf{s}}_{N+N^{\ast}}$ within the spatial domain $\mathcal{S}$. 
Let $\mathbf{S}^{\ast}$ denote the $2\times N^{\ast}$ matrix collecting these new locations, and let $\mathbf{D}^{\ast} = [\uvec{\mathbf{d}}_{N+1}, \ldots, \uvec{\mathbf{d}}_{N+N^{\ast}}]$ represent their corresponding coordinates in the deformed space $\mathcal{D}$, where $\uvec{\mathbf{d}}_{N+n}=d(\uvec{\mathbf{s}}_{N+n})$ for $n \in \{1, \ldots, N^{\ast}\}$. 
Conditional on $\mathbf{D}$ and $\boldsymbol{\sigma}_{d}^{2}$, the matrix-normal prior for $\mathbf{D}^{\ast}$ implies
\begin{equation}
\label{eq:Interpolation_D}
\bigl[\mathbf{D}^{\ast}\mid \mathbf{D}, \boldsymbol{\sigma}_{d}^{2}\bigr]
\sim \sfNor_{2\times N^{\ast}}\!\Big(
\mathbf{S}^{\ast}+(\mathbf{D}-\mathbf{S})\mathbf{R}_{d}^{-1}\mathbf{R}_{\gu}, \boldsymbol{\sigma}_{d}^{2}, \mathbf{R}_{d}^{\ast}-\mathbf{R}_{\gu}^{\top}\mathbf{R}_{d}^{-1}\mathbf{R}_{\gu}\Big),
\end{equation}
where $\mathbf{R}_{d}$, $\mathbf{R}_{d}^{\ast}$, and $\mathbf{R}_{\gu}$ are correlation matrices constructed using a Gaussian (squared-exponential) kernel with range parameter $\psi > 0$. 
Distribution~\eqref{eq:Interpolation_D} propagates the estimated spatial deformation to the ungauged sites, effectively weighting the mapping according to proximity in the geographic domain $\mathcal{S}$.

Given $\mathbf{D}^{\ast}$ and $\boldsymbol{\theta}$, the joint Gaussian model for $\{\mathbf{Y}_{t}, \mathbf{Y}_{t}^{\ast}\}$ leads to the usual conditional formulation:
\begin{equation}
\label{eq:Interpolation_Y}
[\mathbf{Y}_{t}^{\ast}\mid \mathbf{Y}_{t} = \mathbf{y}_{t},\boldsymbol{\theta},\mathbf{D}^{\ast}]
\sim \sfNor_{N^{\ast} \times q} \!\big(\mathbf{M}_{t,\cond}^{\ast},
 \mathbf{B}_{\cond}^{\ast}, \boldsymbol{\Sigma}\big),
\end{equation}
with
\begin{equation*}
\begin{split}
\mathcal{M}_{t,\cond}^{\ast} &= \mathbf{X}_{t}^{\ast}\boldsymbol{\beta}_{t}
+ \mathbf{B}_{\gu}^{\top} \mathbf{B}^{-1} \big(\mathbf{y}_{t}-\mathbf{X}_{t}\boldsymbol{\beta}_{t}\big), \\
\mathbf{B}_{\cond}^{\ast} &= \mathbf{B}^{\ast}-\mathbf{B}_{\gu}^{\top}\mathbf{B}^{-1}\mathbf{B}_{\gu},
\end{split}
\end{equation*}
where $\mathbf{B}$, $\mathbf{B}^{\ast}$, and $\mathbf{B}_{\gu}$ are spatial correlation matrices derived from deformed distances using the exponential kernel with range parameter $\phi > 0$. 
In practice, prediction at each ungauged location combines two components: 
a regression term $\mathbf{X}_{t}^{\ast}\boldsymbol{\beta}_{t}$ representing the large-scale trend, 
and a kriging-type correction $\mathbf{B}_{\gu}^{\top}\mathbf{B}^{-1}(\cdot)$ that borrows spatial information from the observed sites through the deformed space.

To embed the interpolation step within the hierarchical framework, the parameter set is augmented to $\boldsymbol{\theta}_{\interp}=\{\boldsymbol{\theta},\uvec{\mathbf{Y}}_{\mis},\mathbf{D}^{\ast}\}$, and we define $\mathbf{Y}_{\interp}=\{\mathbf{Y}_{t}^{\ast}\}_{t=1}^{T}$. 
The corresponding predictive density can then be written as
\begin{equation}
\label{eq:Interpolation}
f(\uvec{\mathbf{y}}_{\interp} \mid \uvec{\mathbf{y}}_{\obs})
=
\int\limits_{\boldsymbol{\Theta}_{\interp}}
\left[\prod_{t=1}^{T}
f(\uvec{\mathbf{y}}_{t}^{\ast}\mid \uvec{\mathbf{y}}_{t},\boldsymbol{\theta},\mathbf{D}^{\ast})\right]
f(\mathbf{D}^{\ast}\mid \mathbf{D}, \boldsymbol{\sigma}_{d}^{2})
f(\boldsymbol{\theta},\uvec{\mathbf{y}}_{\mis}\mid \uvec{\mathbf{y}}_{\obs})
\partial\boldsymbol{\theta}_{\interp}.
\end{equation}

It is worth noting that, as in Section~\ref{subsec:MCMC}, both vectorized and matrix representations are used in a complementary manner. While the predictive density in \eqref{eq:Interpolation} is naturally expressed in vectorized form, the conditional distributions in \eqref{eq:Interpolation_D} and \eqref{eq:Interpolation_Y} are evaluated using matrix-normal formulations, which lead to more efficient computations.

Since the integral in \eqref{eq:Interpolation} has no closed form, it is approximated via Monte Carlo simulation. For each posterior sample $\{\mathbf{y}^{(k)}, \boldsymbol{\theta}^{(k)}\}$, the deformation is propagated to $\mathbf{D}^{\ast(k)}$ using \eqref{eq:Interpolation_D}, and conditional samples of $\mathbf{Y}_{t}^{\ast(k)}$ are drawn using \eqref{eq:Interpolation_Y}. By accumulating these samples across iterations, the algorithm yields Monte Carlo approximations to the posterior predictive distributions of the responses at ungauged sites. This procedure is described in Algorithm~\ref{alg:Interpolation}.

%--- Section ---%
\section{Model comparison and checking}\label{sec:Model_Comparison-Checking}

We compare four Bayesian spatiotemporal models that differ in their treatment of spatial deformation and cross-variable dependence. These models are evaluated in a simulation study (Section~\ref{subsec:Sim2}) and in the real-data application (Section~\ref{sec:Application}). The models under comparison are defined as follows:
\begin{itemize}
    \item $\mathcal{M}_{1}$ (no deformation, independent responses): $\mathbf{D}=\mathbf{S}$ and $\boldsymbol{\Sigma}$ diagonal;
    \item $\mathcal{M}_{2}$ (no deformation, correlated responses): $\mathbf{D}=\mathbf{S}$ and $\boldsymbol{\Sigma}$ unrestricted;
    \item $\mathcal{M}_{3}$ (deformation, independent responses): $\mathbf{D}\neq\mathbf{S}$ and $\boldsymbol{\Sigma}$ diagonal;
    \item $\mathcal{M}_{4}$ (deformation, correlated responses): $\mathbf{D}\neq\mathbf{S}$ and $\boldsymbol{\Sigma}$ unrestricted, which corresponds to the proposed model.
\end{itemize}

To assess and compare the performance of these models, we consider goodness-of-fit, predictive accuracy, and model checking criteria. Goodness-of-fit evaluates how well a model explains the observed data, predictive accuracy assesses its ability to predict unobserved or interpolated responses, and model checking examines whether the model is compatible with key features of the observed data.

\bmhead{Goodness-of-fit}
The Deviance Information Criterion (DIC) is widely used for Bayesian model comparison \citep{spiegelhalter2002}. In the presence of missing data, the standard DIC can be ambiguous, as it depends on whether the focus is on the observed data or on the complete data (including latent variables). To address this issue, we adopt the formulation proposed by \citet{Celeux2006}, which evaluates the deviance based on the observed-data likelihood while integrating over the predictive distribution of the missing values. This criterion balances model fit and complexity by penalizing overparameterization. Among competing models fitted to the same dataset, the model with the smallest DIC is expected to yield the best predictive performance for a replicate dataset with the same structure \citep{tsiko2015}.

\bmhead{Predictive performance}
We evaluate predictive accuracy at ungauged locations. Let $\hat{Y}_{N+n,i,t}^{[\alpha]}$ denote the $\alpha$-percentile and $\bar{Y}_{N+n,i,t}$ the posterior mean of the interpolated samples $Y_{N+n,i,t}^{(1)}, \ldots, Y_{N+n,i,t}^{(K)}$. Following \citet[Sec.~18.1]{gelman2013}, define the inclusion indicator
\begin{equation*}
O_{n,i,t} =
\left\{
  \begin{array}{rl}
    1, & \textnormal{if }Y_{n,i,t}\textnormal{ is observed }(Y_{n,i,t}=Y_{n,i,t}^{\obs}), \\[2pt]
    0, & \textnormal{if }Y_{n,i,t}\textnormal{ is missing }(Y_{n,i,t}=Y_{n,i,t}^{\mis}). \\
  \end{array}
\right.
\end{equation*}

To quantify the accuracy and calibration of predictive intervals, we use the Interval Score (IS) proposed by \citet{Winkler1972}. The $(N+n,i,t)$th $\alpha$-score is defined as
\begin{equation*}
\mathrm{IS}_{N+n,i,t}^{[\alpha]} = \hat{Y}_{N+n,i,t}^{[1 - \alpha/2]} - \hat{Y}_{N+n,i,t}^{[\alpha/2]} +
\begin{dcases}
    \tfrac{2}{\alpha}(\hat{Y}_{N+n,i,t}^{[\alpha/2]} - Y_{N+n,i,t}), & \text{if } Y_{N+n,i,t} < \hat{Y}_{N+n,i,t}^{[\alpha/2]}, \\
    \tfrac{2}{\alpha}(Y_{N+n,i,t} - \hat{Y}_{N+n,i,t}^{[1 - \alpha/2]}), & \text{if } Y_{N+n,i,t} > \hat{Y}_{N+n,i,t}^{[1 - \alpha/2]}, \\
    0, & \text{otherwise},
\end{dcases}
\end{equation*}
and its average across time is given by
\begin{equation*}
\mathrm{IS}_{N+n,i}^{[\alpha]} = \dfrac{\sum\limits_{t=1}^{T} O_{N+n,i,t} \cdot \mathrm{IS}_{N+n,i,t}^{[\alpha]}}{\sum\limits_{t=1}^{T} O_{N+n,i,t}}.
\end{equation*}

We also consider the Predictive Mean Squared Error (PMSE), defined as
\begin{equation*}
\mathrm{PMSE} = \dfrac{\sum\limits_{n=1}^{N^{\ast}} \sum\limits_{i=1}^{q} \sum\limits_{t=1}^{T} O_{N+n,i,t}  (\bar{Y}_{N+n,i,t} - Y_{N+n,i,t})^{2}}{\sum\limits_{n=1}^{N^{\ast}} \sum\limits_{i=1}^{q} \sum\limits_{t=1}^{T} O_{N+n,i,t}},
\end{equation*}
which measures the average squared prediction error across all ungauged locations \citep{shen2019exploring, Morales2022}.

To assess the full predictive distribution, we employ the Continuous Ranked Probability Score (CRPS), a strictly proper scoring rule that evaluates both calibration and sharpness \citep{Gneiting2007}. The global CRPS is defined as
\begin{equation*}
\mathrm{CRPS} = \dfrac{\sum\limits_{n=1}^{N^{\ast}} \sum\limits_{i=1}^{q} \sum\limits_{t=1}^{T} O_{N+n,i,t} \cdot \mathrm{CRPS}_{N+n,i,t}}{\sum\limits_{n=1}^{N^{\ast}} \sum\limits_{i=1}^{q} \sum\limits_{t=1}^{T} O_{N+n,i,t}},
\end{equation*}
where the pointwise score is computed from the $K$ posterior samples as
\begin{equation*}
\mathrm{CRPS}_{N+n,i,t} = \frac{1}{K} \sum_{k=1}^{K} |Y_{N+n,i,t}^{(k)} - Y_{N+n,i,t}| - \frac{1}{2K^2} \sum_{k=1}^{K} \sum_{k'=1}^{K} |Y_{N+n,i,t}^{(k)} - Y_{N+n,i,t}^{(k')}|.
\end{equation*}

Lower values of IS, PMSE, and CRPS indicate better predictive performance.

\bmhead{Posterior predictive checking}
To assess model adequacy, we perform posterior predictive checking (PPC) using the Empirical Coverage Probability (ECP). For each posterior draw, we generate replicated observations $\mathbf{Y}_{t}^{(k)}$ from the matrix-normal distribution with parameters $\mathbf{X}_{t}\boldsymbol{\beta}_{t}^{(k)}$, $\mathbf{B}^{(k)}$, and $\boldsymbol{\Sigma}^{(k)}$, where $B_{n,n'}^{(k)} = \exp\{-\phi^{(k)} \|\uvec{\mathbf{d}}_{n}^{(k)} - \uvec{\mathbf{d}}_{n'}^{(k)}\|\}$ is an entry of $\mathbf{B}^{(k)}$. Let $Y_{n,i,t}^{(k)}$ denote the $(n,i)$th entry of $\mathbf{Y}_{t}^{(k)}$, and let $\hat{Y}_{n,i,t}^{[\alpha]}$ be the $\alpha$-percentile of the posterior predictive samples $Y_{n,i,t}^{(1)}, \ldots, Y_{n,i,t}^{(K)}$. The ECP for response variable $i$ is defined as
\begin{equation*}
\mathrm{ECP}_{i} = \dfrac{\sum\limits_{n=1}^{N} \sum\limits_{t=1}^{T} O_{n,i,t} \cdot \mathds{1}_{\left[ \hat{Y}_{n,i,t}^{[\alpha/2]}, \hat{Y}_{n,i,t}^{[1-\alpha/2]} \right]}(Y_{n,i,t})}{\sum\limits_{n=1}^{N} \sum\limits_{t=1}^{T} O_{n,i,t}},
\end{equation*}
which represents the empirical proportion of observed values falling within the nominal predictive intervals, where $\mathds{1}_{A}(\cdot)$ denotes the indicator function of a set $A$.

Values of $\mathrm{ECP}_{i}$ close to the nominal level $1-\alpha$ (e.g., 0.95 for 95\% predictive intervals) indicate well-calibrated predictive distributions. Values substantially below the nominal level suggest undercoverage, while values above it indicate overcoverage.

\bmhead{Residual analysis}
We further assess the quality of the fit using pointwise standardized residuals. For an observed value $Y_{n,i,t}$, the standardized residual is defined as
\begin{equation*}
Z_{n,i,t} = \dfrac{Y_{n,i,t} - \bar{Y}_{n,i,t}}{\hat{\sigma}_{n,i,t}},
\end{equation*}
where $\bar{Y}_{n,i,t}$ and $\hat{\sigma}_{n,i,t}$ denote the posterior predictive mean and standard deviation computed from the samples $Y_{n,i,t}^{(1)}, \ldots, Y_{n,i,t}^{(K)}$ obtained via PPC, respectively.

These residuals provide a diagnostic for detecting systematic bias, outliers, and local model misspecification.

\bmhead{MCMC convergence diagnostics}
To assess the efficiency of MCMC, we compute effective sample sizes (ESS) based on the unnormalized log-posterior distribution associated with \eqref{eq:Posterior}. This provides a global diagnostic of mixing performance, reflecting the joint behavior of all model parameters. Larger ESS values indicate more efficient exploration of the posterior distribution, while smaller values suggest higher autocorrelation in the Markov chains. As a practical guideline, ESS values above 100 are often considered sufficient for stable estimation of posterior summaries \citep[see, e.g.,][Sec.~11.5]{gelman2013}, although larger values are preferable. The ESS results are complemented by visual inspection of trace plots in selected challenging scenarios, providing additional evidence on the stability and mixing behavior of the MCMC chains.

\bmhead{Computational details}
All analyses were conducted on a personal computer with the following configuration: Intel Core i7-6500U CPU @ 2.50 GHz, 16 GB of RAM, and Windows 10 (64-bit). Reported execution times correspond to full MCMC runs under this configuration.

%--- Section ---%
\section{Simulation studies}\label{sec:Simulation}

We present two complementary simulation studies designed to evaluate different aspects of the proposed modeling framework and inference procedure.

The first study (Section~\ref{subsec:Sim1}) focuses on parameter recovery under the proposed model $\mathcal{M}_{4}$, assessing the ability of Algorithm~\ref{alg:MCMC} to accurately estimate model parameters in multivariate spatiotemporal settings with missing data. This study considers a three-dimensional response setting, evaluating performance under different numbers of spatial locations while keeping the temporal dimension and missingness level fixed.

The second study (Section~\ref{subsec:Sim2}) extends this evaluation to a comparative setting, examining both parameter recovery and predictive performance across competing models $\mathcal{M}_{1}$--$\mathcal{M}_{4}$. In addition to in-sample fit, this study emphasizes out-of-sample prediction at ungauged locations under varying temporal lengths and missing data proportions, focusing on a bivariate response configuration to facilitate comparison with existing approaches.

\subsection{Parameter recovery under the proposed model}\label{subsec:Sim1}

We conduct a simulation study to evaluate the ability of Algorithm~\ref{alg:MCMC} to accurately recover the parameters of the proposed model $\mathcal{M}_{4}$ in multivariate spatiotemporal settings with missing data. 
In particular, we assess the recovery of key model components under controlled settings, focusing on the impact of spatial resolution while keeping the temporal dimension and missingness level fixed.

\bmhead{Anisotropic data-generating mechanism}

\begin{figure}[htb!]
\centering
\begin{subfigure}{0.49\textwidth}
  \includegraphics[width=\linewidth]{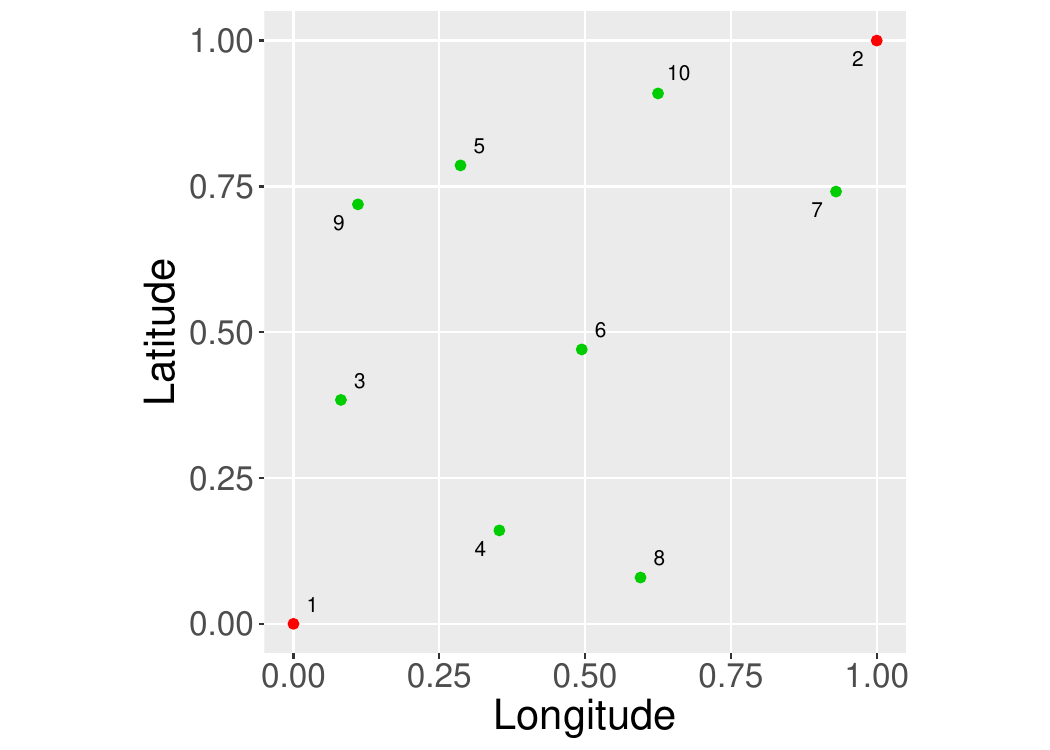}
  \caption{Spatial domain $\mathcal{S}$ with $N=10$ sites.}
  \label{fig:Sim1_S-10}
\end{subfigure}
\begin{subfigure}{0.49\textwidth}
  \includegraphics[width=\linewidth]{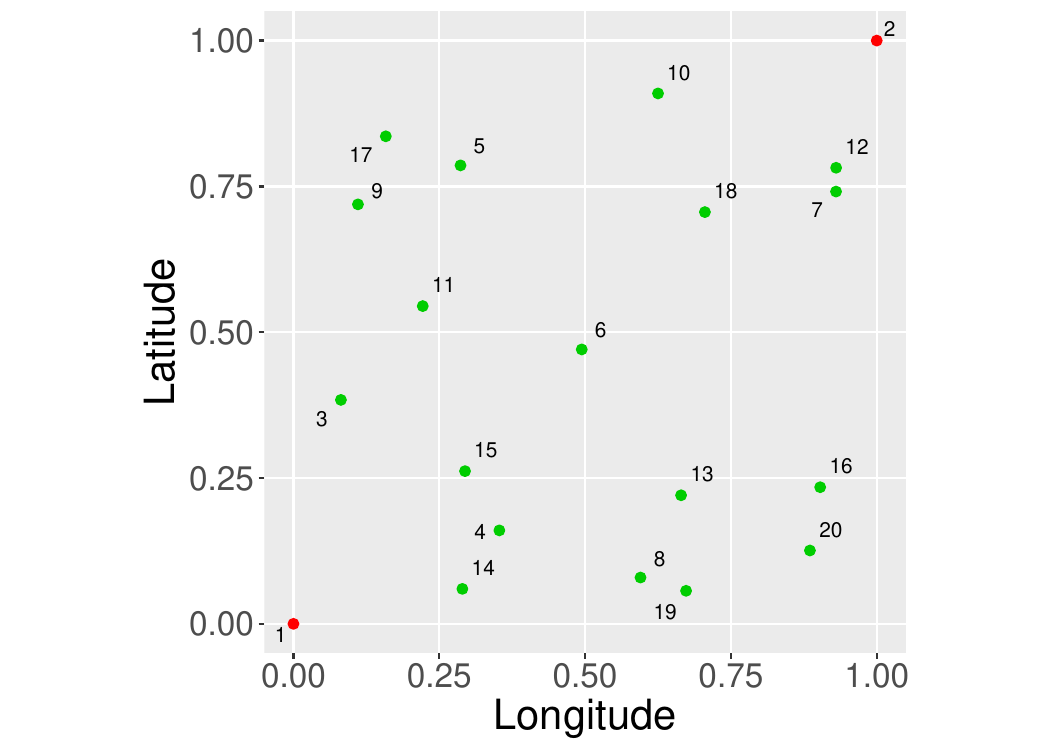}
  \caption{Spatial domain $\mathcal{S}$ with $N=20$ sites.}
  \label{fig:Sim1_S-20}
\end{subfigure}
\caption{Spatial domains in the simulation study of Section~\ref{subsec:Sim1}. Red circles denote the two anchor points ($\uvec{\mathbf{s}}_{1}$ and $\uvec{\mathbf{s}}_{2}$), and green circles represent the remaining gauged sites ($\uvec{\mathbf{s}}_{3}, \ldots, \uvec{\mathbf{s}}_{N}$).}
\label{fig:Sim1_S}
\end{figure}

Spatial locations are defined within the unit square $\mathcal{S}=[0,1]^2$. 
Two anchor sites are fixed at $\uvec{\mathbf{s}}_{1}=(0,0)$ and $\uvec{\mathbf{s}}_{2}=(1,1)$, while the remaining locations are sampled independently from a uniform distribution over $\mathcal{S}$. Spatial dependence is induced through a deformation-based covariance structure, where the latent coordinates $\mathbf{D}$ are generated according to the prior specification in \eqref{eq:D_Prior}. This construction yields a stochastic deformation of the spatial domain, allowing for flexible departures from isotropy. To assess the impact of spatial resolution, we consider two scenarios with $N \in \{10,20\}$ spatial locations. The corresponding spatial configurations are displayed in Figure~\ref{fig:Sim1_S}.

\bmhead{Design and evaluation targets}

We consider a three-dimensional response ($q=3$), $p=2$ regression coefficients per response and time, and a temporal length fixed at $T=600$. Missing data are introduced by randomly removing a fixed proportion $\gamma=0.10$ of observations at each time point and for each response. Synthetic datasets are generated from the matrix-variate dynamic model described in Section~\ref{subsec:ST_Modelling}. The latent regression coefficients evolve according to a dynamic linear model, and covariate matrices are generated independently at each time point. Conditional on the latent states and spatial covariance structure, the response matrices are drawn from \eqref{eq:Observation_Mat}. We evaluate the performance of Algorithm~\ref{alg:MCMC} in terms of its ability to recover the full set of model parameters $\boldsymbol{\theta}$, encompassing both static and dynamic components.

Complete details of the data-generating mechanism, including parameter values, prior specifications, and simulation settings, are provided in Appendix~\ref{app:SimProtocol_Sim1}.

To fit model~$\mathcal{M}_{4}$ through Algorithm~\ref{alg:MCMC}, we used the observed subvectors $\uvec{\mathbf{y}}_{1,\obs}, \ldots, \uvec{\mathbf{y}}_{600,\obs}$. The following hyperparameters were specified: $\lambda = 1$, $\tau_{j}^{2} = 625.0$ for $j \in \{1,\ldots, p\}$, $\mathbf{M}_{0} = \mathbf{0}_{p\times q}$, $\mathbf{C}_{0} = \mathbf{I}_{p}$, $a_{\sigma_{d,m}} = b_{\sigma_{d,m}} = 0.001$ for $m \in \{1, 2\}$, $a_{\boldsymbol{\Sigma}} = 1.001$ and $\mathbf{b}_{\boldsymbol{\Sigma}} = 0.001 \cdot \mathbf{I}_{q}$. These choices correspond to weakly informative prior distributions. We also specified $\psi = 1.0$, so as to differ from the value used in the data generation process ($\psi_{\simu} = 1.5$).

Algorithm~\ref{alg:MCMC} was executed for 80{,}000 iterations to sample from the joint posterior of the model parameters and missing responses under model~$\mathcal{M}_{4}$. The algorithm operates on an augmented state space, generating draws from $f(\uvec{\mathbf{y}}_{\mis}, \boldsymbol{\theta} \mid \uvec{\mathbf{y}}_{\obs})$, so that missing values are imputed jointly with parameter estimation. Convergence was achieved after approximately 50{,}000 iterations. To mitigate autocorrelation in the Markov chains, we retained every 15th iteration after burn-in, resulting in $K = 2000$ posterior samples. Execution times (in minutes) were 82.9 and 298.4 for $N = 10$ and $N = 20$, respectively. Effective sample sizes computed from the unnormalized log-posterior distribution were 1349.1 and 566.6, indicating adequate mixing in both scenarios, albeit with some degradation as dimensionality increases. For $N=10$, convergence diagnostics suggested a burn-in period of approximately 10{,}000 iterations; however, for comparability, the same number of total iterations was adopted in both cases.

Table~\ref{tab:Sim1_parameters} summarizes posterior means and 95\% highest posterior density (HPD) intervals for the model parameters. Overall, the spatial range parameter $\phi$ is accurately recovered in both scenarios, with posterior means close to the true value and HPD intervals that contain it. The diagonal elements of $\boldsymbol{\Sigma}$ are also well estimated, although a mild positive bias is observed, particularly for $\Sigma_{2,2}$ and $\Sigma_{3,3}$, and this effect is slightly more pronounced when $N=20$. The off-diagonal elements are reasonably well captured, with small deviations that remain within the corresponding HPD intervals.

The state evolution variances $W_{1,1}$ and $W_{2,2}$ are satisfactorily estimated in both scenarios. A slight underestimation is observed for $N=10$, whereas for $N=20$ the posterior means are closer to the true values. Taken together, these results indicate that the model is capable of recovering the main covariance components, although the estimation of certain parameters becomes more challenging as the number of spatial locations increases.

\begin{figure}[htb!]
    \centering
    \includegraphics[width=1.0\linewidth]{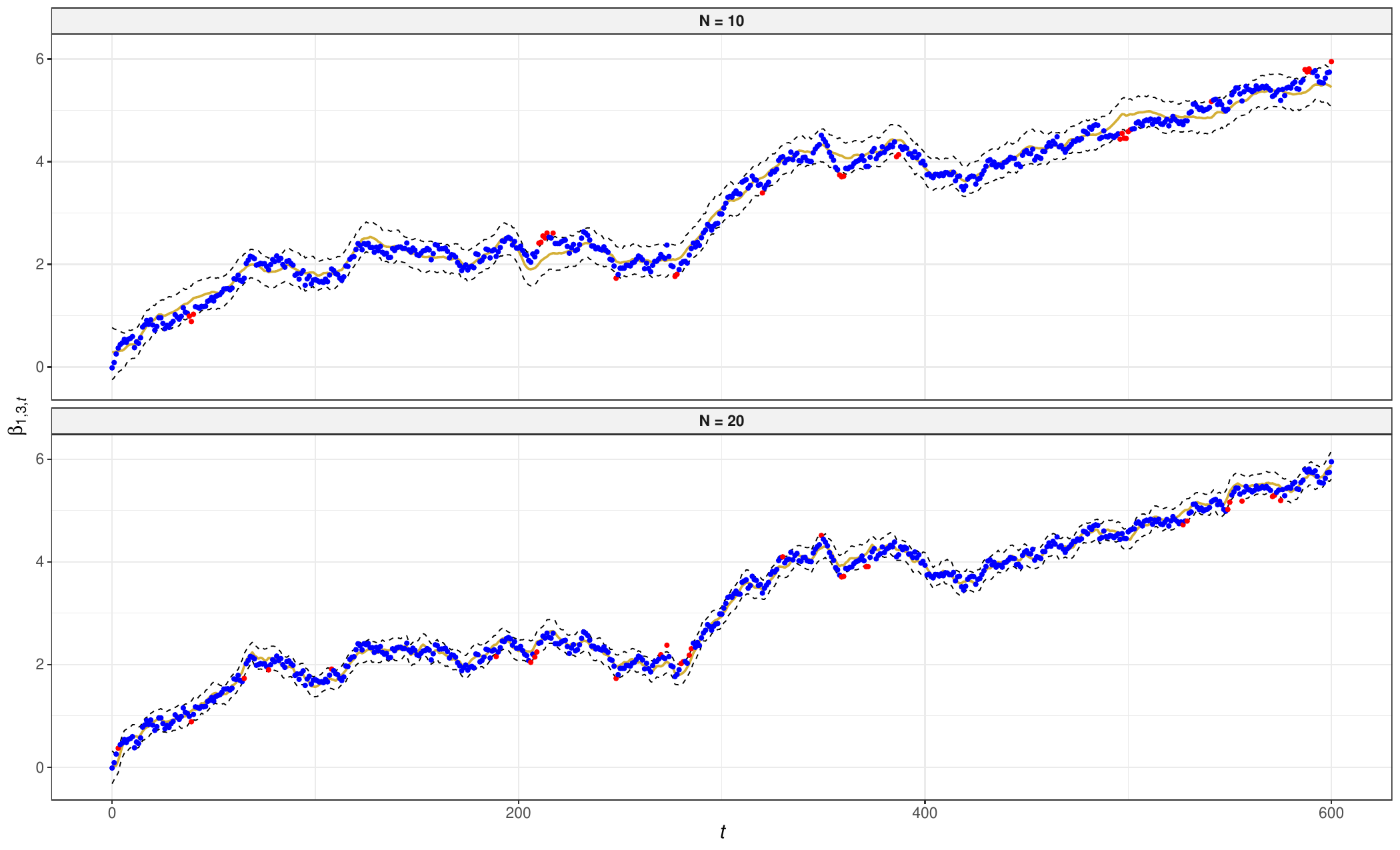}
    \caption{Posterior trajectories of $\beta_{1,3,t}$ for $t \in \{0, 1, \ldots,600\}$ under $\mathcal{M}_4$, for $N=10$ (top panel) and $N=20$ (bottom panel). Black dashed lines denote 95\% highest posterior density intervals, the solid golden line represents the posterior mean, and points indicate the true values (blue if within the interval and red otherwise). Results from the simulation study in Section~\ref{subsec:Sim1}.}
\label{fig:Sim1_Beta}
\end{figure}

The recovery of latent temporal states is illustrated in Figure~\ref{fig:Sim1_Beta}, which shows the posterior trajectories of $\beta_{1,3,t}$. For $N = 10$, the posterior mean follows the true trajectory reasonably well. Although the corresponding credible intervals are somewhat broader, they still capture the spread of the observed point cloud appropriately, resulting in a larger number of time points for which the true values lie within the intervals. In contrast, for $N=20$, the posterior trajectories exhibit improved agreement with the true values, with narrower credible intervals and fewer discrepancies. This suggests that increasing the number of spatial locations provides additional information for estimating the latent states, resulting in improved temporal reconstruction.

\begin{table}[htbp!]
\centering
\caption{Posterior mean, 95\% highest posterior density (HPD) intervals, and true values for the model parameters under $\mathcal{M}_4$, by number of spatial locations \mbox{($N \in \{10,20\}$)}. Results from the simulation study in Section~\ref{subsec:Sim1}.}
\label{tab:Sim1_parameters}
\setlength{\tabcolsep}{8pt}
\begin{tabular}{crlcc}
\toprule
\textbf{Parameter} & \textbf{True value} & \textbf{Metric} & \textbf{$N = 10$} & \textbf{$N = 20$} \\ \midrule

\multirow{2}{*}{$\phi$} 
& \multirow{2}{*}{0.5} & Mean & 0.4730 & 0.4700 \\
&  & HPD & 0.4372--0.5113 & 0.4334--0.5030 \\ \midrule

\multirow{2}{*}{$\Sigma_{1,1}$} 
& \multirow{2}{*}{1} & Mean & 1.0170 & 1.0380 \\
&  & HPD & 0.9371--1.1011 & 0.9668--1.1188 \\ \midrule

\multirow{2}{*}{$\Sigma_{1,2}$} 
& \multirow{2}{*}{0.1} & Mean & 0.0970 & 0.1110 \\
&  & HPD & 0.0640--0.1298 & 0.0878--0.1330 \\ \midrule

\multirow{2}{*}{$\Sigma_{1,3}$} 
& \multirow{2}{*}{0.05} & Mean & 0.0560 & 0.0440 \\
&  & HPD & 0.0258--0.0854 & 0.0232--0.0654 \\ \midrule

\multirow{2}{*}{$\Sigma_{2,2}$} 
& \multirow{2}{*}{1} & Mean & 1.0680 & 1.0800 \\
&  & HPD & 0.9786--1.1592 & 1.0020--1.1583 \\ \midrule

\multirow{2}{*}{$\Sigma_{2,3}$} 
& \multirow{2}{*}{0.15} & Mean & 0.1430 & 0.1650 \\
&  & HPD & 0.1101--0.1758 & 0.1396--0.1882 \\ \midrule

\multirow{2}{*}{$\Sigma_{3,3}$} 
& \multirow{2}{*}{1} & Mean & 1.0630 & 1.0540 \\
&  & HPD & 0.9800--1.1505 & 0.9705--1.1286 \\ \midrule

\multirow{2}{*}{$W_{1,1}$} 
& \multirow{2}{*}{0.008} & Mean & 0.0060 & 0.0070 \\
&  & HPD & 0.0034--0.0094 & 0.0040--0.0103 \\ \midrule

\multirow{2}{*}{$W_{2,2}$} 
& \multirow{2}{*}{0.008} & Mean & 0.0070 & 0.0080 \\
&  & HPD & 0.0052--0.0095 & 0.0061--0.0096 \\ \bottomrule

\end{tabular}
\end{table}

\citet{Schmidt2003} remark that $\boldsymbol{\sigma}_{d}^{2}$ is not identifiable in the sense of \citet{dawid1979conditional}. As shown in \eqref{eq:sigma2d_m-FCD}, the scale parameter of the full conditional distribution for $\sigma_{d_{m,m}}^{2}$ ($m \in \{1, 2\}$) depends on $\mathbf{R}_{d}$. Consequently, the specification of the hyperparameter $\psi > 0$ directly influences the estimation of the entries of $\boldsymbol{\sigma}_{d}^{2}$. Thus, the issue of non-identifiability regarding $\boldsymbol{\sigma}_{d}^{2}$ is particularly evident in scenarios where the value of $\psi$ used for estimation differs from the true value $\psi_{\simu}$ used in the data generation process.

To assess whether Algorithm~\ref{alg:MCMC} accurately recovers the original covariance structure, we define a metric to compare the true matrix $\mathbf{R}_{d}(\psi_{\simu}) \otimes \boldsymbol{\sigma}_{d}^{2}$ with the posterior samples $\{\mathbf{R}_{d}(\psi) \otimes \boldsymbol{\sigma}_{d}^{2(k)}\}_{k=1}^{K}$. For the sake of both simplicity and numerical stability, this recovery is evaluated based on the average of the log-eigenvalues of the total covariance structure. Specifically, we evaluate the proximity of the posterior distribution $\{\xi^{(k)}\}_{k=1}^{K}$ to the true value $\xi$, defined as follows:
\begin{equation}
\label{eq:log_eigen}
\xi = \frac{1}{2N} \ln \det \left[ \mathbf{R}_{d}(\psi_{\simu}) \otimes \boldsymbol{\sigma}_{d}^{2} \right],
\quad
\xi^{(k)} = \frac{1}{2N} \ln \det \left[ \mathbf{R}_{d}(\psi) \otimes \boldsymbol{\sigma}_{d}^{2(k)} \right].
\end{equation}

To evaluate how accurately the spatial deformation is recovered, we compute, for each anisotropic model $\mathcal{M}_{h}$ ($h\in\{3,4\}$), the squared Frobenius norm between the true deformation matrix $\mathbf{D}$ and its estimate at the $k$th MCMC iteration, $\mathbf{D}_{\mathcal{M}_{h}}^{(k)}$:
\begin{equation}
\label{eq:Frobenius_Norm}
\|\mathbf{D}-\mathbf{D}_{\mathcal{M}_{h}}^{(k)}\|_{\mathrm{F}}^{2}
=\tr\!\left[(\mathbf{D}-\mathbf{D}_{\mathcal{M}_{h}}^{(k)})(\mathbf{D}-\mathbf{D}_{\mathcal{M}_{h}}^{(k)})^{\top}\right].
\end{equation}

Table~\ref{tab:Sim1_samples} summarizes posterior means and 95\% highest posterior density (HPD) intervals for both recovery measures, namely $\{\xi^{(k)}\}_{k=1}^{K}$ and $\{\|\mathbf{D}-\mathbf{D}_{\mathcal{M}_{4}}^{(k)}\|_{\mathrm{F}}^{2}\}_{k=1}^{K}$, under the two spatial configurations.

For the average of the log-eigenvalues of the total covariance structure, the results indicate satisfactory recovery when $N=10$, as the posterior mean is close to the true value $\xi$ and the corresponding HPD interval contains it. For $N=20$, both the posterior mean and the HPD interval are slightly shifted downward relative to the true value, indicating a mild underestimation, although the overall behavior remains broadly consistent with the expected pattern.

For the squared Frobenius norm, the results indicate that the reconstruction error remains small in both scenarios, but is noticeably larger when $N=20$. This suggests that, although the deformation is reasonably well recovered, uncertainty increases with the number of spatial locations.

\begin{figure}[htb!]
\centering
\begin{subfigure}{0.49\textwidth}
  \includegraphics[width=\linewidth]{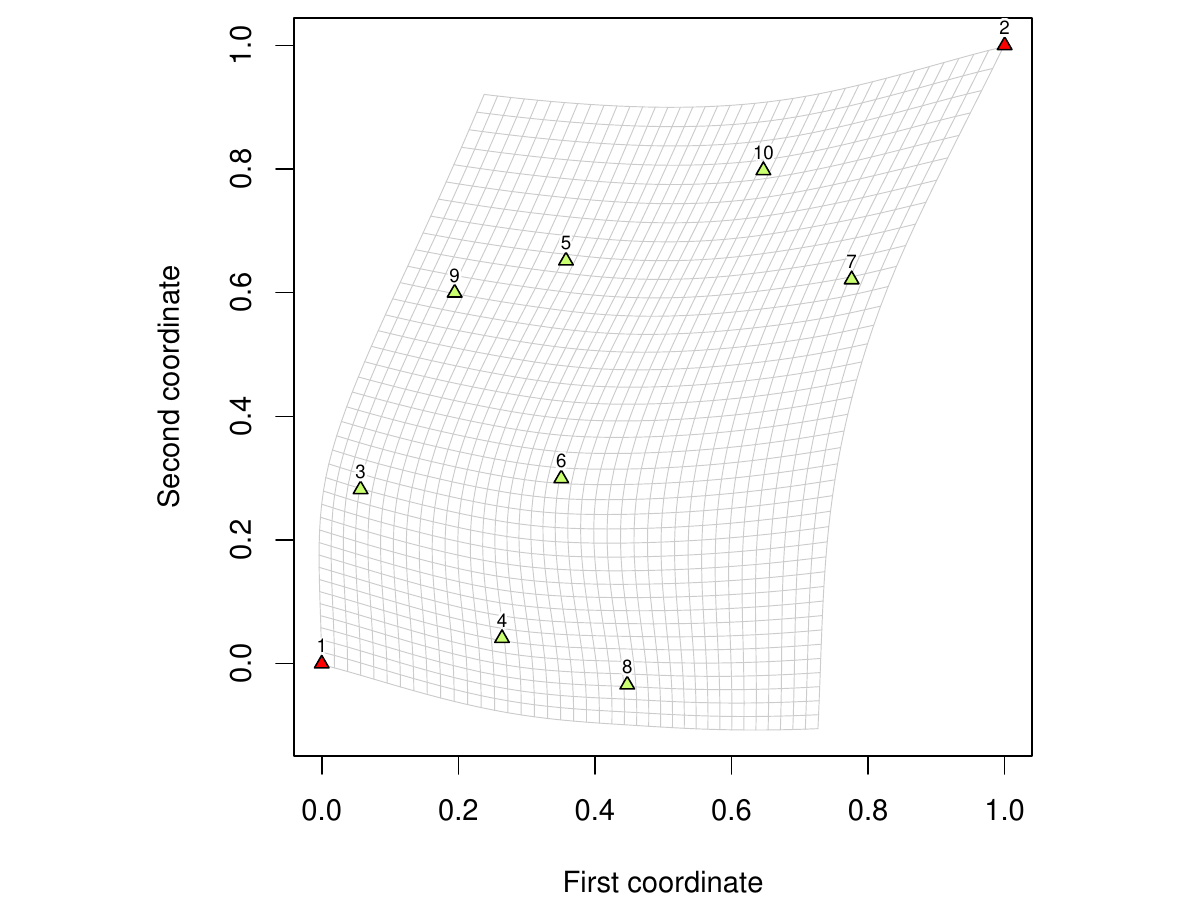}
  \caption{True spatial deformation ($N=10$).}
  \label{fig:Sim1_D_True-10}
\end{subfigure}
\begin{subfigure}{0.49\textwidth}
  \includegraphics[width=\linewidth]{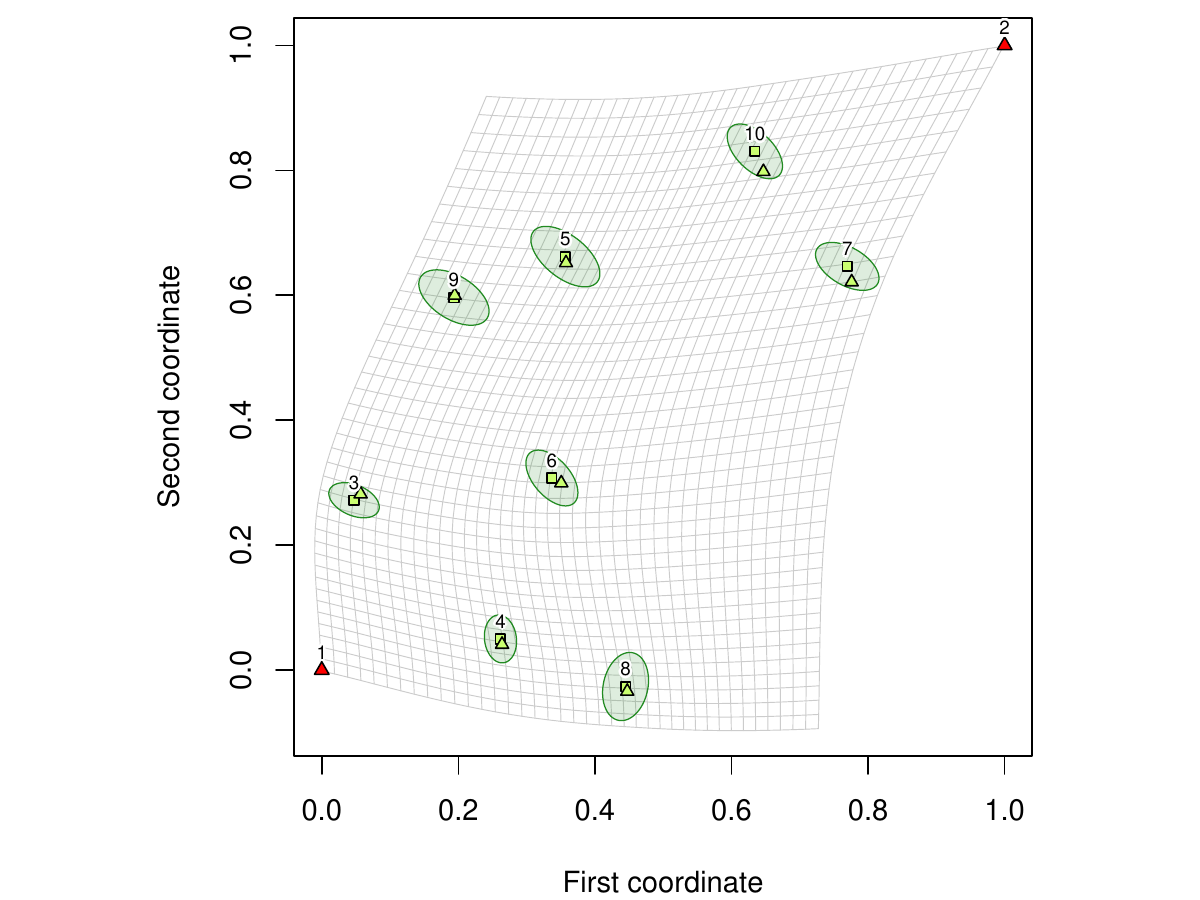}
  \caption{Estimated spatial deformation ($N=10$).}
  \label{fig:Sim1_D_Est-10}
\end{subfigure}
\begin{subfigure}{0.49\textwidth}
  \includegraphics[width=\linewidth]{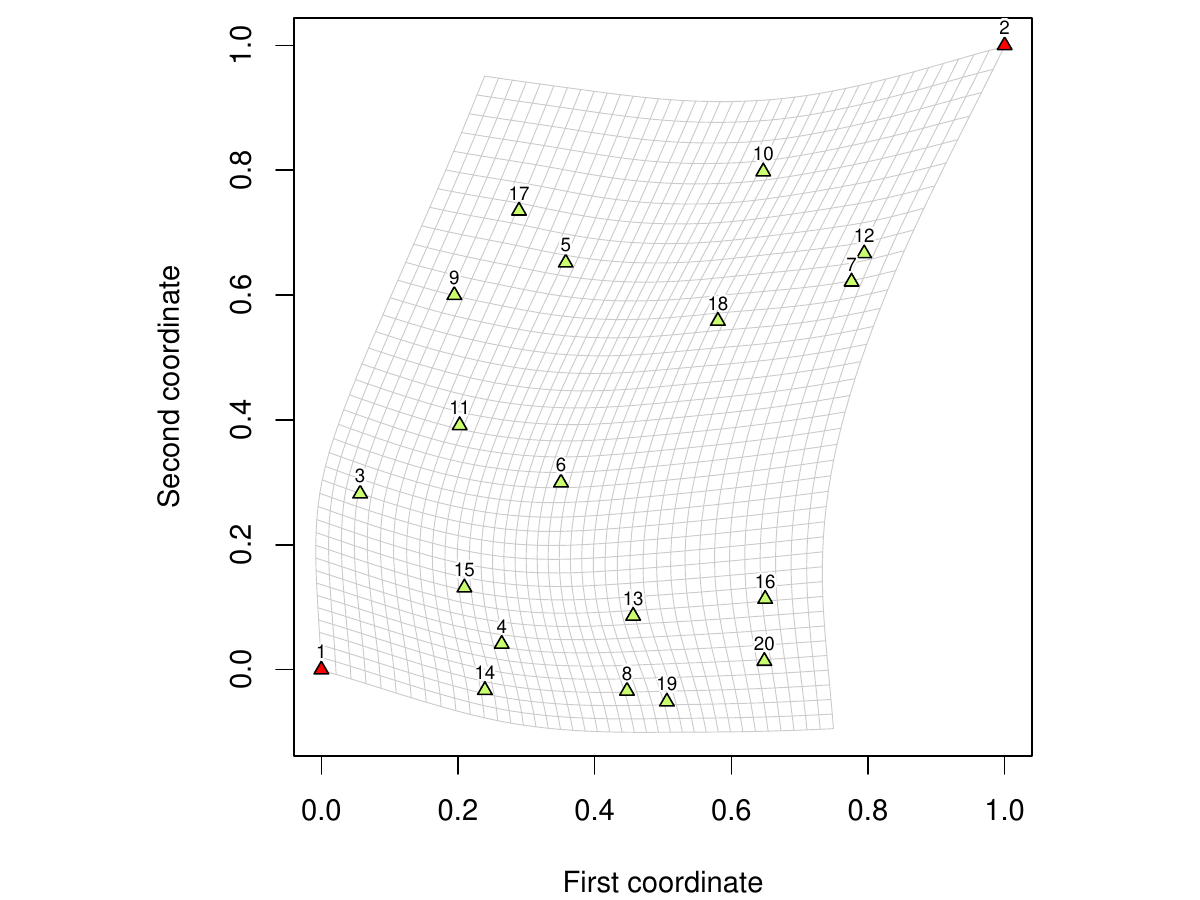}
  \caption{True spatial deformation ($N=20$).}
  \label{fig:Sim1_D_True-20}
\end{subfigure}
\begin{subfigure}{0.49\textwidth}
  \includegraphics[width=\linewidth]{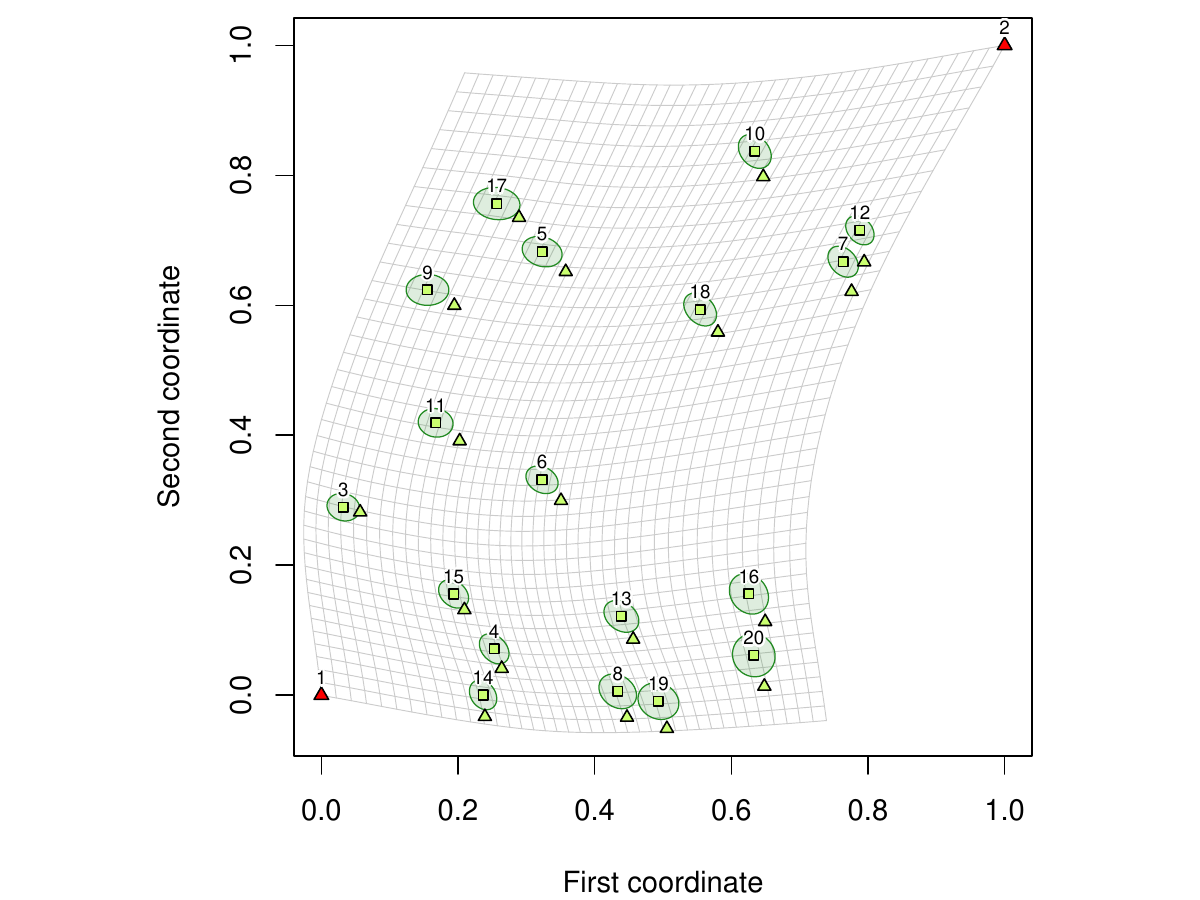}
  \caption{Estimated spatial deformation ($N=20$).}
  \label{fig:Sim1_D_Est-20}
\end{subfigure}
\caption{True and estimated spatial deformations in $\mathcal{D}$-space for $N=10$ and $N=20$. Estimated deformations are obtained under $\mathcal{M}_4$. Triangles denote true deformations (red for anchor points and green for gauged sites), while squares denote posterior means of estimated deformations. Ellipses denote 95\% credible regions. Results from the simulation study in Section~\ref{subsec:Sim1}.}
\label{fig:Sim1_D}
\end{figure}

These findings are consistent with the visual evidence provided in Figure~\ref{fig:Sim1_D}. For $N=10$, the estimated deformation closely matches the true configuration, with posterior means nearly overlapping the true points and relatively small credible regions. For $N=20$, the overall geometric structure of the deformation is still well captured; however, the estimated locations exhibit a slight displacement relative to the true positions. This indicates that the model is able to recover the main features of the deformation, although accurate alignment of individual locations becomes more challenging as the number of spatial sites increases, particularly under a fixed temporal length ($T = 600$).

\begin{table}[htbp!]
\centering
\caption{Posterior summaries (mean and 95\% highest posterior density (HPD) intervals) and true values for two recovery measures under $\mathcal{M}_4$, by number of spatial locations \mbox{($N \in \{10,20\}$)}. The measures correspond to the average of the log-eigenvalues of the total covariance structure and the squared Frobenius norm, defined in \eqref{eq:log_eigen} and \eqref{eq:Frobenius_Norm}, respectively. Results from the simulation study in Section~\ref{subsec:Sim1}.}
\label{tab:Sim1_samples}
\setlength{\tabcolsep}{10pt}
\begin{tabular}{clcc}
\toprule
\textbf{Posterior sample} & \textbf{Quantity} & \textbf{$N = 10$} & \textbf{$N = 20$} \\ \midrule

\multirow{3}{*}{$\{\xi^{(k)}\}_{k=1}^{K}$}
& True value & $-6.1400$ & $-9.1290$ \\
& Mean & $-6.0842$ & $-9.8997$ \\
& HPD & $(-6.8325, -5.3176)$ & $(-10.5664, -9.1839)$ \\ \midrule

\multirow{3}{*}{$\{\|\mathbf{D}-\mathbf{D}_{\mathcal{M}_{4}}^{(k)}\|_{\mathrm{F}}^{2}\}_{k=1}^{K}$}
& True value & \multicolumn{2}{c}{Less is better} \\
& Mean & $0.0073$ & $0.0357$ \\
& HPD & $(0.0013, 0.0179)$ & $(0.0142, 0.0596)$ \\ \bottomrule

\end{tabular}
\end{table}

In summary, the results indicate that the model performs well in recovering the main covariance parameters and the spatial deformation structure for both values of $N$. Increasing the number of spatial locations improves the estimation of latent temporal states, as evidenced by the results for $\beta_{1,3,t}$, but introduces additional challenges for the estimation of certain covariance components and deformation-related quantities. These findings suggest that larger values of $T$ may be required to achieve more stable inference when both $N$ and $q$ increase. Despite these challenges, the MCMC algorithm exhibits satisfactory mixing behavior and remains computationally feasible, although at a substantially increased cost.

\subsection{Comparative predictive performance across competing models}\label{subsec:Sim2}

We conduct a simulation study to evaluate the practical benefits of the proposed model relative to simpler alternatives. 
In particular, we examine whether incorporating spatial deformation and cross-variable correlation improves parameter estimation and predictive performance. 
To this end, we generate synthetic anisotropic spatiotemporal data under controlled settings, varying the time series length and the fraction of missing observations. 
The results are then compared across four model specifications that differ in their treatment of anisotropy and multivariate dependence. 
The four models under comparison ($\mathcal{M}_{h}$, $h \in \{1, 2, 3, 4\}$) are defined in Section~\ref{sec:Model_Comparison-Checking}.

\bmhead{Anisotropic data-generating mechanism}
To simulate geometric anisotropy, we follow \citet[Sec.~4]{maity2012testing} and specify the exponential spatial kernel
\begin{equation}
\label{Sim-Structure}
\exp\left\{-\phi \sqrt{(\uvec{\mathbf{s}}_{n}-\uvec{\mathbf{s}}_{n'})^{\top}\mathbf{A}(\uvec{\mathbf{s}}_{n}-\uvec{\mathbf{s}}_{n'})}\right\},
\end{equation}
where $\mathbf{A} \in \SPD(2)$ controls the orientation and magnitude of anisotropy. 
By writing $\mathbf{A}=\boldsymbol{\Lambda}^{\top}\boldsymbol{\Lambda}$, the structure in \eqref{Sim-Structure} is equivalent to an isotropic kernel in the linearly transformed coordinates $\uvec{\mathbf{d}}=\boldsymbol{\Lambda}\uvec{\mathbf{s}}$:
\begin{eqnarray*}
\exp\left\{-\phi \sqrt{(\uvec{\mathbf{s}}_{n}-\uvec{\mathbf{s}}_{n'})^{\top}\mathbf{A}(\uvec{\mathbf{s}}_{n}-\uvec{\mathbf{s}}_{n'})}\right\} &=& \exp\{-\phi \|\boldsymbol{\Lambda}\uvec{\mathbf{s}}_{n}-\boldsymbol{\Lambda}\uvec{\mathbf{s}}_{n'}\|\} \\
    &=& \exp\{-\phi \|d(\uvec{\mathbf{s}}_{n})-d(\uvec{\mathbf{s}}_{n'})\|\}.
\end{eqnarray*}
Hence, $d(\uvec{\mathbf{s}})=\boldsymbol{\Lambda}\uvec{\mathbf{s}}$ defines a deterministic deformation that induces anisotropy. 
In contrast to the stochastic deformation induced by the prior specification in Section~\ref{subsec:Sim1}, the present construction is fully deterministic and does not rely on \eqref{eq:D_Prior}. 
As a result, the data-generating mechanism does not inherently favor models that assume a prior-driven deformation, providing a neutral setting for assessing the impact of spatial deformation on predictive performance. The factorization $\mathbf{A}=\boldsymbol{\Lambda}^{\top}\boldsymbol{\Lambda}$ is not unique; here, $\boldsymbol{\Lambda}$ denotes one convenient choice that yields the desired deformation.
Following \citet[Sec.~4]{maity2012testing}, we adopt $\mathbf{A}=9 \cdot \mathbf{I}_{2}-4 \cdot \mathbf{1}_{2\times2}$. One such decomposition is given by
\begin{equation}
\label{eq:Lambda}
\boldsymbol{\Lambda}=
\begin{bmatrix}
1 & 0\\
0 & 3
\end{bmatrix}
\left[
  \begin{array}{rr}
\cos(\pi/4) & \sin(\pi/4)\\
-\sin(\pi/4) & \cos(\pi/4)
  \end{array}
\right],
\end{equation}
which stretches one spatial axis and rotates the field by $45^\circ$.

\bmhead{Design and evaluation targets}

\begin{figure}[htb!]
  \centering
  {\includegraphics[width=0.7\linewidth]{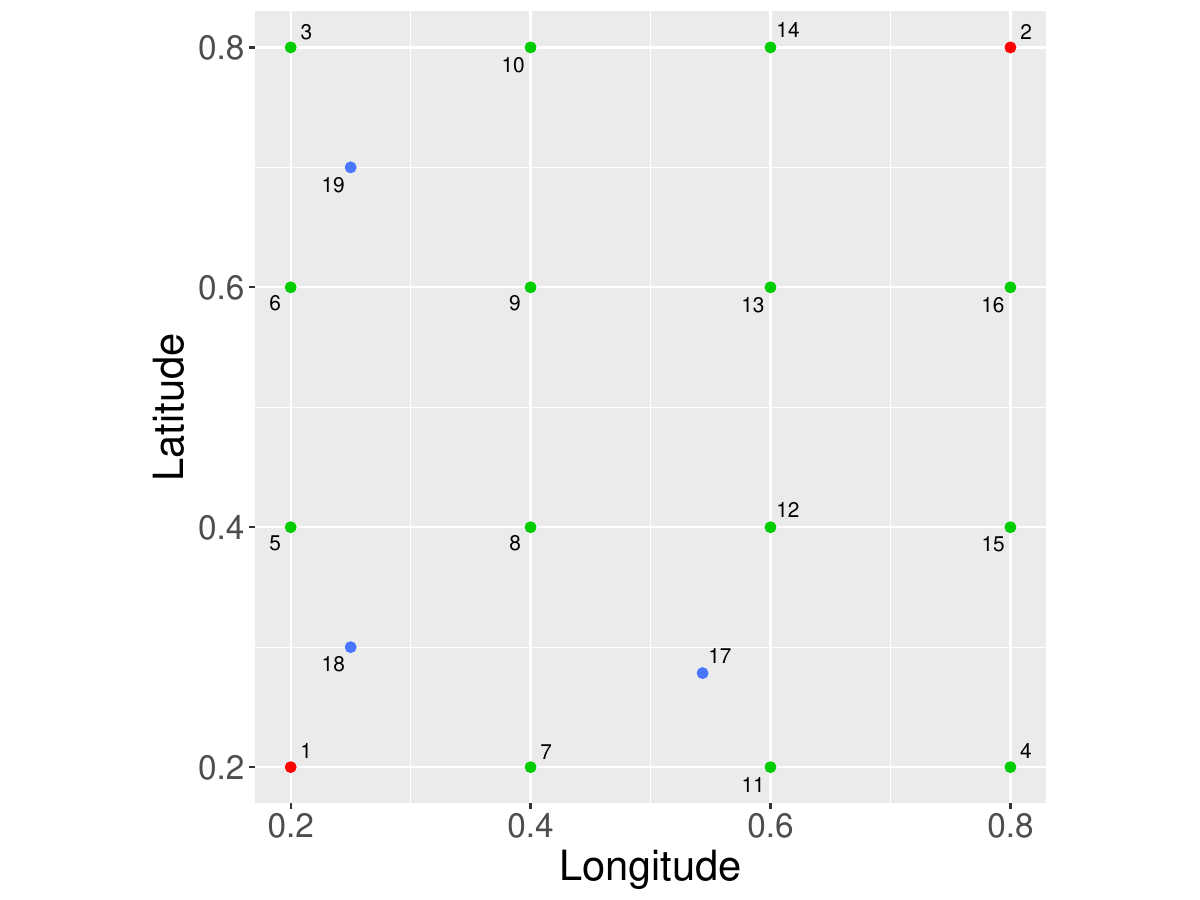}}
  \caption{
Spatial domain $\mathcal{S}$ for the simulation study in Section~\ref{subsec:Sim2}.
Red circles denote the two anchor points ($\uvec{\mathbf{s}}_{1}$ and $\uvec{\mathbf{s}}_{2}$), 
green circles denote the remaining gauged sites ($\uvec{\mathbf{s}}_{3},\ldots,\uvec{\mathbf{s}}_{16}$), 
and blue circles denote the ungauged locations used for interpolation 
($\uvec{\mathbf{s}}_{17}$, $\uvec{\mathbf{s}}_{18}$, and $\uvec{\mathbf{s}}_{19}$).
}
  \label{fig:Sim2_S}
\end{figure}

We consider $q=2$ response variables, time lengths $T\in\{100,500\}$, and $p=2$ regression coefficients per response and time. 
Synthetic datasets are generated from the matrix-variate dynamic model described in Section~\ref{subsec:ST_Modelling}, with latent states evolving according to a dynamic linear model and observations drawn from \eqref{eq:Observation_Mat}. 
Realizations of the response process are generated at $N=16$ equally spaced sites within the unit square, and jointly at $N^{\ast}=3$ additional unequally spaced sites for interpolation assessment. 
We compare models $\mathcal{M}_{1}$–$\mathcal{M}_{4}$ in terms of (i) accuracy of parameter recovery and (ii) predictive performance at ungauged locations. 
Figure~\ref{fig:Sim2_S} illustrates the spatial domain, highlighting the two fixed anchor sites ($\uvec{\mathbf{s}}_{1}$ and $\uvec{\mathbf{s}}_{2}$), the remaining gauged locations ($\uvec{\mathbf{s}}_{3},\ldots,\uvec{\mathbf{s}}_{16}$), and the ungauged sites $\mathbf{S}^{\ast}=[\uvec{\mathbf{s}}_{17}\ \uvec{\mathbf{s}}_{18}\ \uvec{\mathbf{s}}_{19}]$ used for evaluating interpolation.

Data generation and MCMC configuration details, including the construction of the deformation matrices, parameter settings, covariate generation, and missing-value imposition, are fully described in Appendix~\ref{app:SimProtocol_Sim2}. 
In brief, we simulate latent states and observations according to the model equations, with varying time lengths \mbox{($T\in\{100,500\}$)} and missing-value fractions \mbox{($\gamma\in\{0.0625, 0.25\}$)}. 
These fractions arise from the discrete structure of the data: since each response at time $t$ consists of $N=16$ observations, we remove either one value ($1/16=0.0625$) or four values ($4/16=0.25$) per response and time point. 
Each scenario produces complete and partially observed datasets, which are analyzed under models $\mathcal{M}_{1}$–$\mathcal{M}_{4}$ using the same prior and MCMC specifications.

To implement Algorithm~\ref{alg:MCMC}, we fit models $\mathcal{M}_{1}$--$\mathcal{M}_{4}$ using the observed subvectors $\uvec{\mathbf{y}}_{1,\obs}, \ldots, \uvec{\mathbf{y}}_{T,\obs}$. The following hyperparameters were specified: $\lambda = 1$, $\tau_{j}^{2} = 625.0$ for $j \in \{1,\ldots, p\}$, $\mathbf{M}_{0} = \mathbf{0}_{p\times q}$, $\mathbf{C}_{0} = \mathbf{I}_{p}$, and $a_{\sigma_{d,m}} = b_{\sigma_{d,m}} = 0.001$ for $m \in \{1, 2\}$. For models with diagonal cross-variable covariance ($h \in \{1,3\}$), we set $a_{\Sigma_{i,i}} = b_{\Sigma_{i,i}} = 0.001$ for $i \in \{1,\ldots,q\}$. For models with full covariance structure ($h \in \{2,4\}$), we specified $a_{\boldsymbol{\Sigma}} = 1.001$ and $\mathbf{b}_{\boldsymbol{\Sigma}} = 0.001 \cdot \mathbf{I}_{q}$. These choices correspond to weakly informative prior distributions. 

For the spatial deformation parameter $\psi > 0$, which controls the decay of the Gaussian correlation function in $\mathbf{R}_{d}$, we conducted an empirical sensitivity analysis by considering the values $0.1$, $0.5$, $1.0$, $2.5$, $5.0$, $10.0$, $50.0$, $100.0$, and $500.0$ under the simulation setting corresponding to model~$\mathcal{M}_{4}$ with $(T, \gamma) = (500, 0.0625)$. The value $\psi = 2.5$ was selected based on predictive performance as measured by the CRPS. This value was then used for both models with spatial deformation ($\mathcal{M}_{3}$ and $\mathcal{M}_{4}$) and across all considered simulation scenarios to ensure a consistent comparison across specifications, in line with the primary goal of accurate spatial interpolation.

Algorithm~\ref{alg:MCMC} was executed for 30{,}000 iterations to sample from the joint posterior of the model parameters and missing responses under model~$\mathcal{M}_{4}$. The algorithm operates on an augmented state space, generating draws from $f(\uvec{\mathbf{y}}_{\mis}, \boldsymbol{\theta} \mid \uvec{\mathbf{y}}_{\obs})$, so that missing values are imputed jointly with parameter estimation. Convergence was achieved after approximately 10{,}000 iterations. To mitigate autocorrelation in the Markov chains, we retained every 10th iteration after burn-in, resulting in $K = 2000$ posterior samples. Using posterior samples of the model parameters and completed data, Algorithm~\ref{alg:Interpolation} was applied to approximate the predictive distribution $f(\uvec{\mathbf{y}}_{\interp} \mid \uvec{\mathbf{y}}_{\obs})$, which is then used to interpolate the responses at the ungauged locations. The same configuration was employed for the competing models $\mathcal{M}_{1}$, $\mathcal{M}_{2}$, and $\mathcal{M}_{3}$. 

Tables~\ref{tab:Sim2_phi}--\ref{tab:Sim2_W} summarize, for each model and all $(T,\gamma)$ scenarios, the posterior mean and the 95\% highest posterior density (HPD) interval for $\phi$, the entries of $\boldsymbol{\Sigma}$, and the diagonal components of $\mathbf{W}$. Table~\ref{tab:Sim2_sigma2d} reports analogous summaries for the diagonal elements of $\boldsymbol{\sigma}_{d}^{2}$ under the anisotropic models. Several patterns emerge clearly.

A first important result concerns the spatial range parameter $\phi$. The isotropic models $\mathcal{M}_{1}$ and $\mathcal{M}_{2}$ systematically overestimate this parameter in all scenarios. Their posterior means remain far above the true value $\phi=0.4$, and for $T=100$ and also for $T=500$ the corresponding HPD intervals do not cover the truth. This behavior is consistent with model misspecification: when the true anisotropy is ignored, the isotropic models compensate by favoring a much longer-range correlation structure over the original geographic domain. In contrast, the anisotropic models $\mathcal{M}_{3}$ and $\mathcal{M}_{4}$ recover $\phi$ substantially better. For $T=100$, both models already produce posterior means close to the true value, although with moderate uncertainty. For $T=500$, recovery becomes very accurate, with posterior means essentially equal to $0.4$ and narrow HPD intervals containing the truth in both missing-data scenarios. Thus, once the deformation is incorporated, the model no longer needs to inflate the spatial range parameter in order to reproduce the observed dependence pattern.

The posterior summaries of $\boldsymbol{\Sigma}$ also reveal an interpretable structure. As imposed by construction, $\Sigma_{1,2}$ is fixed at zero under $\mathcal{M}_{1}$ and $\mathcal{M}_{3}$, whereas under $\mathcal{M}_{2}$ and $\mathcal{M}_{4}$ it is estimated freely. In these two models, the posterior mean of $\Sigma_{1,2}$ remains close to the true value $0.85$ across all scenarios, and the HPD intervals always include the truth. Likewise, the marginal variances $\Sigma_{1,1}$ and $\Sigma_{2,2}$ are satisfactorily recovered by all models, with posterior concentration increasing as $T$ grows from $100$ to $500$. The main effect of increasing the sample size is a marked reduction in posterior uncertainty, whereas the effect of larger missingness is comparatively mild. Overall, the contemporaneous covariance structure is well identified in the present simulation design, even when anisotropy is misspecified.

The diagonal entries of $\mathbf{W}$ provide information on the temporal innovation variability of the regression coefficients. Here the impact of spatial misspecification is more visible, especially for $W_{1,1}$. When $T=100$, the isotropic models tend to overestimate this quantity substantially: posterior means range from $0.019$ to $0.046$, well above the true value $0.005$, and the associated HPD intervals are quite wide. The anisotropic models also overestimate $W_{1,1}$ in the smaller-sample scenarios, but to a much lesser extent, particularly under $\mathcal{M}_{3}$. This suggests that, when anisotropy is ignored, part of the unresolved spatial structure is spuriously absorbed as excess temporal innovation in the state evolution. As the sample size increases to $T=500$, all models improve considerably, and the posterior means move much closer to the truth. Even in this more favorable setting, however, the anisotropic formulations remain at least as accurate as the isotropic ones.

The recovery of $W_{2,2}$ is generally easier. Posterior means are already close to $0.005$ even for $T=100$, and by $T=500$ all four models provide narrow HPD intervals centered near the true value. This difference between $W_{1,1}$ and $W_{2,2}$ indicates that not all state-innovation components are equally sensitive to spatial misspecification. In the present setup, the first diagonal element of $\mathbf{W}$ is more affected by the omission of anisotropy, whereas the second remains comparatively stable.

The interpretation of $\boldsymbol{\sigma}_{d}^{2}$ is necessarily different. Because the data were generated from a deterministic deformation, there is no true stochastic deformation variance to be recovered. For this reason, Table~\ref{tab:Sim2_sigma2d} should not be read as a parameter-recovery exercise analogous to Tables~\ref{tab:Sim2_phi}--\ref{tab:Sim2_W}. Instead, these posterior summaries are mainly descriptive of how the anisotropic models allocate uncertainty to the latent deformation coordinates.

Even with this caveat, Table~\ref{tab:Sim2_sigma2d} conveys a consistent qualitative message. In all scenarios, $\sigma_{d_{2,2}}^{2}$ is much larger than $\sigma_{d_{1,1}}^{2}$ under both $\mathcal{M}_{3}$ and $\mathcal{M}_{4}$. This directional imbalance is compatible with the way the true deformation was constructed, since the transformation stretches one axis more strongly than the other. At the same time, the HPD intervals for both parameters are wide, and they do not contract monotonically with increasing $T$ or decreasing $\gamma$. This lack of systematic concentration is not surprising, given that these variance components do not correspond to identifiable truth parameters in the present deterministic setup. Their role is better understood as controlling the prior variability of the latent deformation rather than representing directly observed features of the data-generating mechanism.

\begin{figure}[htb!]
    \centering
    \includegraphics[width=1.0\linewidth]{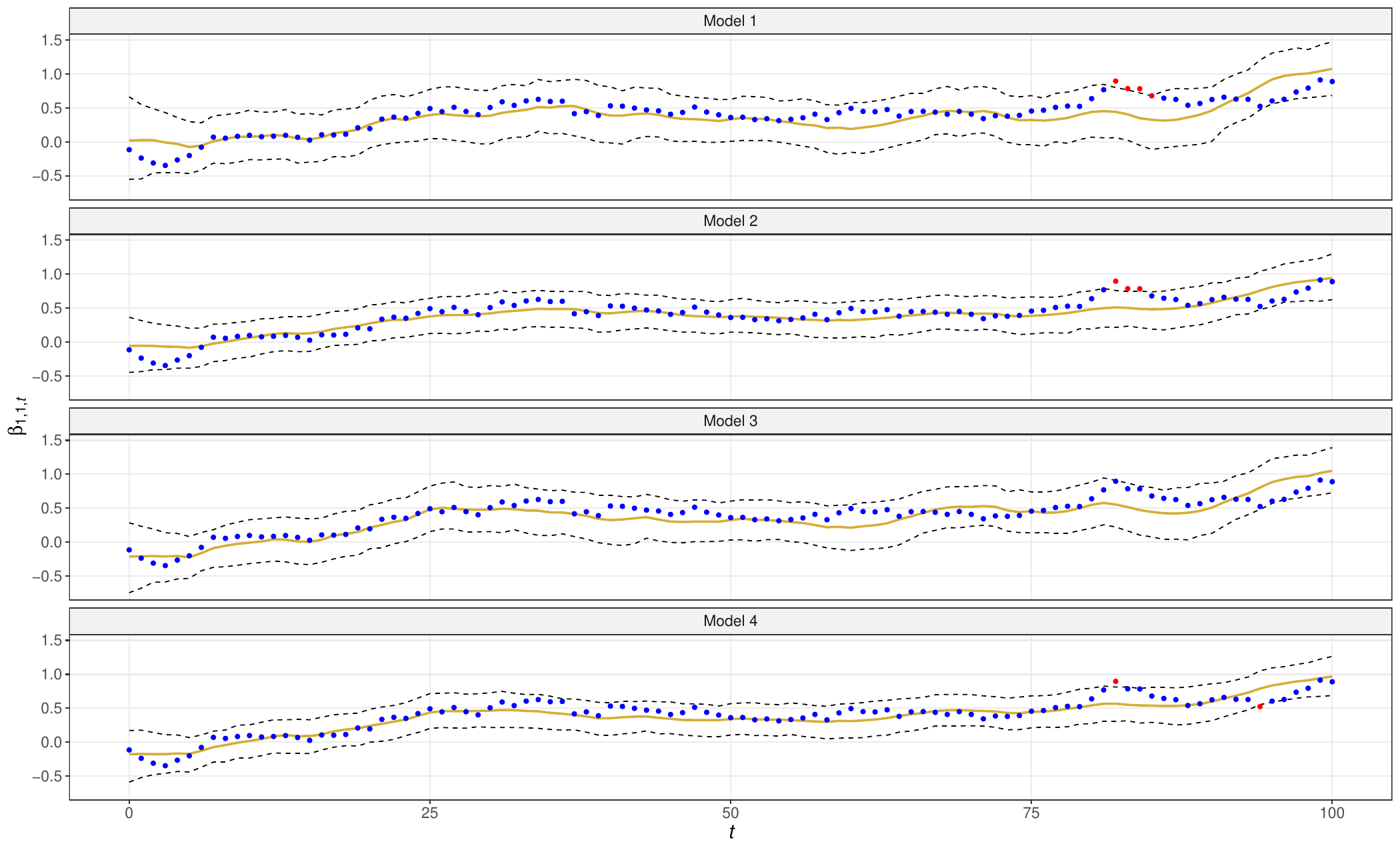}
    \caption{Posterior trajectories of $\beta_{1,1,t}$ for $t \in \{0, 1, \ldots,100\}$ across models ($\mathcal{M}_1$--$\mathcal{M}_4$), with $T = 100$ and $\gamma = 0.25$. Black dashed lines denote 95\% highest posterior density intervals, the solid golden line represents the posterior mean, and points indicate the true values (blue if within the interval and red otherwise). Results from the simulation study in Section~\ref{subsec:Sim2}.}
\label{fig:Sim2_Beta}
\end{figure}

Estimates of the state trajectories are accurate and quite similar across models. Figure~\ref{fig:Sim2_Beta} illustrates this for $\beta_{1,1,t}$ when $T=100$ and $\gamma=0.25$. All four models recover the broad temporal evolution of the coefficient, and the posterior mean trajectories follow the true values closely over time. The isotropic models tend to produce slightly wider HPD bands, whereas the anisotropic models yield somewhat more concentrated inference, especially over the middle portion of the series. The overall differences are modest, which is itself informative: even when spatial structure is misspecified, the dynamic regression component remains reasonably well estimated in this simulation design. Thus, the main benefits of the anisotropic formulations are more clearly expressed in spatial recovery and prediction than in the estimation of $\boldsymbol{\beta}_{0:T}$.

\begin{table}[htbp!]
\centering
\caption{Posterior mean and 95\% highest posterior density (HPD) intervals for the spatial decay parameter $\phi$, by model ($\mathcal{M}_1$--$\mathcal{M}_4$), sample size ($T \in \{100, 500\}$), and missing-value proportion ($\gamma \in \{0.0625, 0.25\}$). The true value is $\phi = 0.4$. Results from the simulation study in Section~\ref{subsec:Sim2}.}
\label{tab:Sim2_phi}
\setlength{\tabcolsep}{4pt}
\begin{tabular}{cclcccc}
\toprule
$T$ & $\gamma$ (\%) & \textbf{Metric} & $\mathcal{M}_1$ & $\mathcal{M}_2$ & 
$\mathcal{M}_3$ & $\mathcal{M}_4$ \\
\midrule
\multirow{4}{*}{100} & \multirow{2}{*}{6.25} & Mean & 1.224 & 1.122 & 
0.494 & 0.460 \\
 &  & HPD & 0.9508-1.4983 & 0.8812-1.3668 & 0.3828-0.5981 & 0.3554-0.5549 \\
\cmidrule{2-7}
 & \multirow{2}{*}{25.0} & Mean & 1.096 & 1.056 & 0.451 & 0.435 \\
 &  & HPD & 0.8566-1.3091 & 0.8592-1.2909 & 0.3558-0.5581 & 0.3403-0.5486 \\
\midrule
\multirow{4}{*}{500} & \multirow{2}{*}{6.25} & Mean & 0.947 & 0.989 & 
0.405 & 0.408 \\
 &  & HPD & 0.8638-1.0302 & 0.9121-1.0766 & 0.3661-0.4457 & 0.3679-0.4513 \\
\cmidrule{2-7}
 & \multirow{2}{*}{25.0} & Mean & 0.904 & 0.928 & 0.407 & 0.411 \\
 &  & HPD & 0.8162-0.9825 & 0.8406-1.0061 & 0.3618-0.4490 & 0.3682-0.4575 \\
\bottomrule
\end{tabular}
\end{table}

\begin{table}[htbp!]
\centering
\caption{Posterior mean and 95\% highest posterior density (HPD) intervals for the entries of $\boldsymbol{\Sigma}$, by model ($\mathcal{M}_1$--$\mathcal{M}_4$), sample size ($T \in \{100, 500\}$), and missing-value proportion ($\gamma \in \{0.0625, 0.25\}$). The true values are $\Sigma_{1,1} = 1.0$, $\Sigma_{1,2} = 0.85$, and $\Sigma_{2,2} = 1.0$. Results from the simulation study in Section~\ref{subsec:Sim2}.}
\label{tab:Sim2_Sigma}
\setlength{\tabcolsep}{4pt}
\begin{tabular}{ccclcccc}
\toprule
\textbf{Param.} & $T$ & $\gamma$ (\%) & \textbf{Metric} & $\mathcal{M}_1$ & $\mathcal{M}_2$ & 
$\mathcal{M}_3$ & $\mathcal{M}_4$ \\
\midrule
\multirow{8}{*}{$\Sigma_{1,1}$} & \multirow{4}{*}{100} & \multirow{2}{*}{6.25} & 
Mean & 0.9080 & 0.9820 & 0.9360 & 1.0080 \\
 &  &  & HPD & 0.7153-1.1126 & 0.7997-1.2075 & 0.7603-1.1323 & 0.8260-1.2069 \\
\cmidrule{3-8}
 &  & \multirow{2}{*}{25.0} & Mean & 0.9700 & 1.0030 & 0.9780 & 1.0330 \\
 &  &  & HPD & 0.7680-1.2032 & 0.8102-1.2320 & 0.7715-1.1977 & 0.8347-1.2619 \\
\cmidrule{2-8}
 & \multirow{4}{*}{500} & \multirow{2}{*}{6.25} & Mean & 1.0450 & 1.0060 & 
1.0690 & 1.0310 \\
 &  &  & HPD & 0.9594-1.1482 & 0.9233-1.0866 & 0.9665-1.1693 & 0.9359-1.1204 \\
\cmidrule{3-8}
 &  & \multirow{2}{*}{25.0} & Mean & 1.0680 & 1.0490 & 1.0880 & 1.0590 \\
 &  &  & HPD & 0.9656-1.1875 & 0.9535-1.1489 & 0.9724-1.1970 & 0.9559-1.1638 \\
\midrule
\multirow{8}{*}{$\Sigma_{1,2}$} & \multirow{4}{*}{100} & \multirow{2}{*}{6.25} & 
Mean & 0.0000 & 0.8270 & 0.0000 & 0.8500 \\
 &  &  & HPD & 0.0000-0.0000 & 0.6578-1.0053 & 0.0000-0.0000 & 0.6870-1.0108 \\
\cmidrule{3-8}
 &  & \multirow{2}{*}{25.0} & Mean & 0.0000 & 0.8440 & 0.0000 & 0.8700 \\
 &  &  & HPD & 0.0000-0.0000 & 0.6649-1.0175 & 0.0000-0.0000 & 0.6962-1.0542 \\
\cmidrule{2-8}
 & \multirow{4}{*}{500} & \multirow{2}{*}{6.25} & Mean & 0.0000 & 0.8410 & 
0.0000 & 0.8650 \\
 &  &  & HPD & 0.0000-0.0000 & 0.7724-0.9115 & 0.0000-0.0000 & 0.7934-0.9490 \\
\cmidrule{3-8}
 &  & \multirow{2}{*}{25.0} & Mean & 0.0000 & 0.8880 & 0.0000 & 0.8940 \\
 &  &  & HPD & 0.0000-0.0000 & 0.8066-0.9706 & 0.0000-0.0000 & 0.8060-0.9784 \\
\midrule
\multirow{8}{*}{$\Sigma_{2,2}$} & \multirow{4}{*}{100} & \multirow{2}{*}{6.25} & 
Mean & 0.8740 & 0.9530 & 0.9050 & 0.9820 \\
 &  &  & HPD & 0.6734-1.0539 & 0.7648-1.1588 & 0.7372-1.0899 & 0.8061-1.1760 \\
\cmidrule{3-8}
 &  & \multirow{2}{*}{25.0} & Mean & 0.9250 & 0.9760 & 0.9210 & 1.0020 \\
 &  &  & HPD & 0.7453-1.1106 & 0.7774-1.1726 & 0.7537-1.1043 & 0.7994-1.1981 \\
\cmidrule{2-8}
 & \multirow{4}{*}{500} & \multirow{2}{*}{6.25} & Mean & 1.0050 & 0.9750 & 
1.0460 & 1.0110 \\
 &  &  & HPD & 0.9147-1.0956 & 0.8927-1.0526 & 0.9552-1.1515 & 0.9281-1.1034 \\
\cmidrule{3-8}
 &  & \multirow{2}{*}{25.0} & Mean & 1.0480 & 1.0350 & 1.0880 & 1.0540 \\
 &  &  & HPD & 0.9569-1.1405 & 0.9467-1.1255 & 0.9928-1.1927 & 0.9584-1.1529 \\
\bottomrule
\end{tabular}
\end{table}

\begin{table}[htbp!]
\centering
\caption{Posterior mean and 95\% highest posterior density (HPD) intervals for the diagonal elements of $\mathbf{W}$, by model ($\mathcal{M}_1$--$\mathcal{M}_4$), sample size ($T \in \{100, 500\}$), and missing-value proportion ($\gamma \in \{0.0625, 0.25\}$). The true values are $W_{1,1} = W_{2,2} = 0.005$. Results from the simulation study in Section~\ref{subsec:Sim2}.}
\label{tab:Sim2_W}
\setlength{\tabcolsep}{4pt}
\begin{tabular}{ccclcccc}
\toprule
\textbf{Param.} & $T$ & $\gamma$ (\%) & \textbf{Metric} & $\mathcal{M}_1$ & $\mathcal{M}_2$ & 
$\mathcal{M}_3$ & $\mathcal{M}_4$ \\
\midrule
\multirow{8}{*}{$W_{1,1}$} & \multirow{4}{*}{100} & \multirow{2}{*}{6.25} & 
Mean & 0.0380 & 0.0460 & 0.0130 & 0.0210 \\
 &  &  & HPD & 0.0003-0.1198 & 0.0034-0.1200 & 0.0000-0.0364 & 0.0030-0.0500 \\
\cmidrule{3-8}
 &  & \multirow{2}{*}{25.0} & Mean & 0.0190 & 0.0310 & 0.0110 & 0.0200 \\
 &  &  & HPD & 0.0001-0.0573 & 0.0027-0.0805 & 0.0001-0.0322 & 0.0031-0.0471 \\
\cmidrule{2-8}
 & \multirow{4}{*}{500} & \multirow{2}{*}{6.25} & Mean & 0.0040 & 0.0080 & 
0.0030 & 0.0070 \\
 &  &  & HPD & 0.0011-0.0076 & 0.0038-0.0143 & 0.0010-0.0061 & 0.0032-0.0119 \\
\cmidrule{3-8}
 &  & \multirow{2}{*}{25.0} & Mean & 0.0030 & 0.0070 & 0.0030 & 0.0060 \\
 &  &  & HPD & 0.0009-0.0067 & 0.0033-0.0124 & 0.0010-0.0057 & 0.0028-0.0104 \\
\midrule
\multirow{8}{*}{$W_{2,2}$} & \multirow{4}{*}{100} & \multirow{2}{*}{6.25} & 
Mean & 0.0090 & 0.0050 & 0.0070 & 0.0040 \\
 &  &  & HPD & 0.0021-0.0198 & 0.0008-0.0097 & 0.0021-0.0135 & 0.0012-0.0089 \\
\cmidrule{3-8}
 &  & \multirow{2}{*}{25.0} & Mean & 0.0130 & 0.0050 & 0.0100 & 0.0050 \\
 &  &  & HPD & 0.0022-0.0275 & 0.0008-0.0111 & 0.0022-0.0203 & 0.0010-0.0101 \\
\cmidrule{2-8}
 & \multirow{4}{*}{500} & \multirow{2}{*}{6.25} & Mean & 0.0040 & 0.0030 & 
0.0040 & 0.0040 \\
 &  &  & HPD & 0.0023-0.0063 & 0.0020-0.0051 & 0.0025-0.0056 & 0.0022-0.0051 \\
\cmidrule{3-8}
 &  & \multirow{2}{*}{25.0} & Mean & 0.0040 & 0.0040 & 0.0030 & 0.0030 \\
 &  &  & HPD & 0.0017-0.0054 & 0.0018-0.0056 & 0.0020-0.0052 & 0.0019-0.0049 \\
\bottomrule
\end{tabular}
\end{table}

\begin{table}[htbp!]
\centering
\caption{Posterior mean and 95\% highest posterior density (HPD) intervals for the diagonal elements of $\boldsymbol{\sigma}_{d}^{2}$, by model ($\mathcal{M}_3$ and $\mathcal{M}_4$), sample size ($T \in \{100, 500\}$), and missing-value proportion ($\gamma \in \{0.0625, 0.25\}$). Results from the simulation study in Section~\ref{subsec:Sim2}.}
\label{tab:Sim2_sigma2d}
\setlength{\tabcolsep}{4pt}
\begin{tabular}{ccclcc}
\toprule
\textbf{Parameter} & $T$ & $\gamma$ (\%) & \textbf{Metric} & $\mathcal{M}_3$ & $\mathcal{M}_4$ \\
\midrule
\multirow{8}{*}{$\sigma_{d_{1,1}}^{2}$} & \multirow{4}{*}{100} & 
\multirow{2}{*}{6.25} & Mean & 0.8550 & 0.4560 \\
 &  &  & HPD & 0.0870-2.2453 & 0.0392-1.2007 \\
\cmidrule{3-6}
 &  & \multirow{2}{*}{25.0} & Mean & 0.5700 & 0.3740 \\
 &  &  & HPD & 0.0459-1.6343 & 0.0293-1.0767 \\
\cmidrule{2-6}
 & \multirow{4}{*}{500} & \multirow{2}{*}{6.25} & Mean & 1.3010 & 1.8560 \\
 &  &  & HPD & 0.3699-2.6954 & 0.5058-3.8800 \\
\cmidrule{3-6}
 &  & \multirow{2}{*}{25.0} & Mean & 1.0250 & 1.6080 \\
 &  &  & HPD & 0.1724-2.2947 & 0.3480-3.5548 \\
\midrule
\multirow{8}{*}{$\sigma_{d_{2,2}}^{2}$} & \multirow{4}{*}{100} & 
\multirow{2}{*}{6.25} & Mean & 10.2430 & 13.3530 \\
 &  &  & HPD & 2.2907-22.4841 & 2.8519-28.6220 \\
\cmidrule{3-6}
 &  & \multirow{2}{*}{25.0} & Mean & 11.2200 & 11.5940 \\
 &  &  & HPD & 2.7013-24.0803 & 2.6377-24.7267 \\
\cmidrule{2-6}
 & \multirow{4}{*}{500} & \multirow{2}{*}{6.25} & Mean & 43.8960 & 36.5440 \\
 &  &  & HPD & 6.0577-96.5181 & 4.7917-82.9137 \\
\cmidrule{3-6}
 &  & \multirow{2}{*}{25.0} & Mean & 18.3120 & 22.7770 \\
 &  &  & HPD & 3.0807-42.8860 & 4.0568-54.0609 \\
\bottomrule
\end{tabular}
\end{table}

Table~\ref{tab:Sim2_sq-frobenius} reports posterior summaries of the squared Frobenius norm defined in \eqref{eq:Frobenius_Norm}. As expected, the estimated deformations approach the true one as the sample size increases and the proportion of missing data decreases. For $T=100$, both anisotropic models still exhibit substantial uncertainty, and the difference between $\mathcal{M}_{3}$ and $\mathcal{M}_{4}$ is small. In fact, under the more difficult case $(T,\gamma)=(100,0.25)$, the posterior means are nearly identical, indicating that when information is limited the inclusion of cross-response covariance has only a minor effect on deformation recovery. The picture changes more clearly at $T=500$. In both missing-data scenarios, the squared Frobenius distance becomes much smaller, and model~$\mathcal{M}_{4}$ attains the best results, especially when $\gamma=0.0625$, for which the posterior mean drops to $0.081$. Thus, the latent deformation becomes estimable only once the temporal replication is sufficiently informative, and under those richer scenarios the full model enjoys a modest but clear advantage.

\begin{figure}[htb!]
\centering
\begin{subfigure}{0.50\textwidth}
  \includegraphics[width=\linewidth]{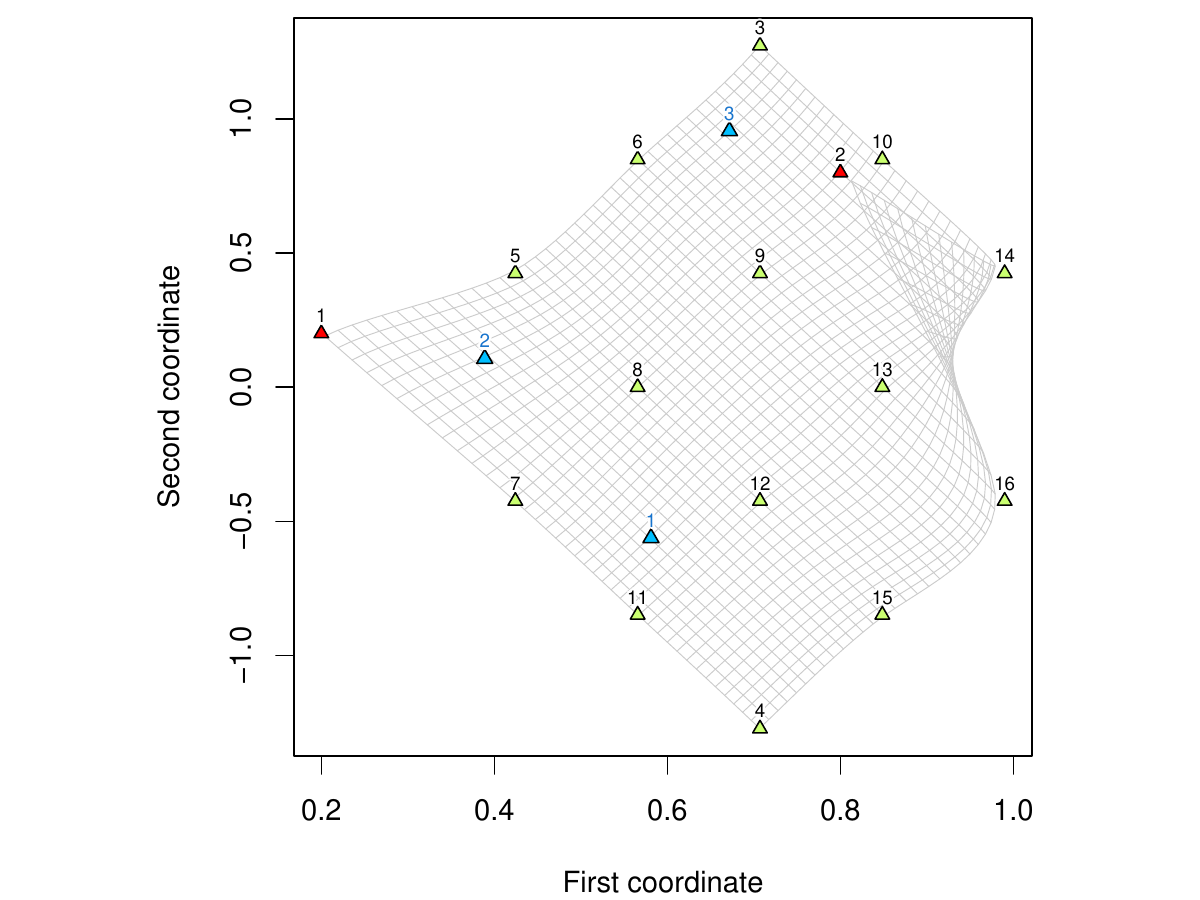}
  \caption{True spatial deformation.}\label{fig:Sim2_D-True}
\end{subfigure}
\begin{subfigure}{0.50\textwidth}
  \includegraphics[width=\linewidth]{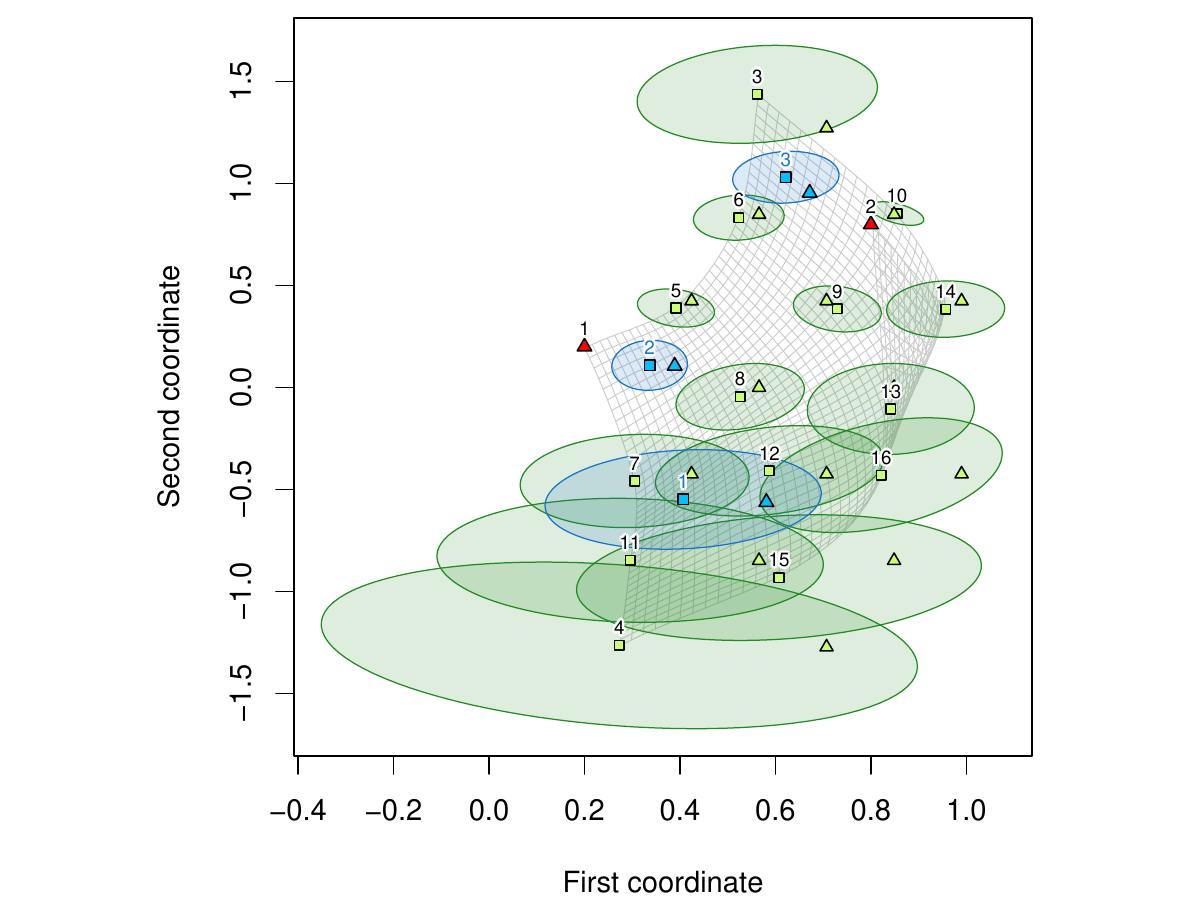}
  \caption{Estimated spatial deformation under model $\mathcal{M}_3$, with $T = 100$ and $\gamma = 0.25$.}\label{fig:Sim2_D-Worst}
\end{subfigure}
\begin{subfigure}{0.50\textwidth}
  \includegraphics[width=\linewidth]{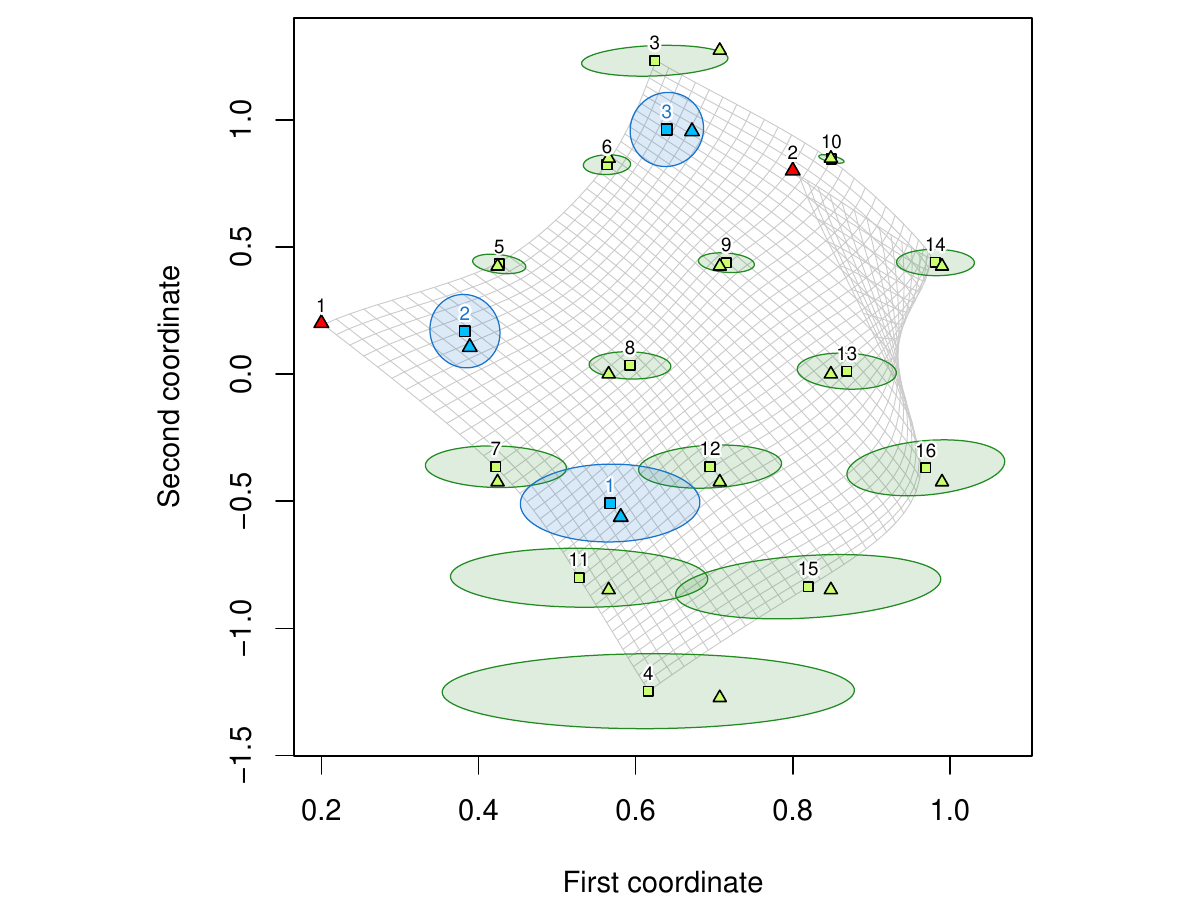}
  \caption{Estimated spatial deformation under model $\mathcal{M}_4$, with $T = 500$ and $\gamma = 0.0625$.}\label{fig:Sim2_D-Best}
\end{subfigure}
\caption{True, estimated, and interpolated spatial deformations in $\mathcal{D}$-space under two scenarios. Triangles denote true deformations (red for anchor points, green for gauged sites, and blue for ungauged sites), while green squares denote posterior means of estimated deformations at gauged sites and blue squares denote posterior means of interpolated deformations at ungauged sites. Ellipses denote 95\% credible regions. Results from the simulation study in Section~\ref{subsec:Sim2}.}
\label{fig:Sim2_D}
\end{figure}

The visual evidence in Figure~\ref{fig:Sim2_D} is fully consistent with these numerical summaries. Panel~\ref{fig:Sim2_D-True} shows the true deformation used in data generation, with the regular spatial grid mapped into a rotated and directionally stretched latent domain. Panel~\ref{fig:Sim2_D-Worst}, corresponding to the low-information scenario $T=100$ and $\gamma=0.25$ under $\mathcal{M}_{3}$, displays visibly larger credible ellipses and several estimated locations that remain noticeably displaced from the true ones. The uncertainty is particularly pronounced for some non-anchor gauged sites, and the interpolated locations of the ungauged sites are also more diffuse. By contrast, panel~\ref{fig:Sim2_D-Best}, obtained under $\mathcal{M}_{4}$ with $T=500$ and $\gamma=0.0625$, shows posterior means that lie much closer to the true deformed positions, together with substantially tighter credible ellipses. In this most favorable case, both the estimated gauged locations and the interpolated ungauged locations respect quite well the geometry of the true deformed domain. Hence, the deformation figures reinforce the central message from Table~\ref{tab:Sim2_sq-frobenius}: recovering the latent geometry is difficult under weak information, but becomes increasingly reliable as the sample size grows.

\begin{table}[htbp!]
\centering
\caption{Posterior mean and 95\% highest posterior density (HPD) intervals for the squared Frobenius norm between the true and estimated deformation, for the anisotropic models ($\mathcal{M}_3$ and $\mathcal{M}_4$), sample sizes ($T \in \{100, 500\}$) and missing-value proportions ($\gamma \in \{0.0625, 0.25\}$). Results from the simulation study in Section~\ref{subsec:Sim2}.}
\label{tab:Sim2_sq-frobenius}
\small
\setlength{\tabcolsep}{10pt}
\begin{tabular}{cclcc}
\toprule
$T$ & $\gamma$ (\%) & \textbf{Metric} & $\mathcal{M}_3$ & $\mathcal{M}_4$ \\
\midrule
\multirow{4}{*}{100} & \multirow{2}{*}{6.25} & Mean & 0.614 & 0.449 \\
 & & HPD & 0.0639-1.4114 & 0.0388-1.3597 \\
 \cmidrule(lr){2-5}
 & \multirow{2}{*}{25.0} & Mean & 0.756 & 0.809 \\
 & & HPD & 0.0806-1.8552 & 0.0748-1.9667 \\
\midrule
\multirow{4}{*}{500} & \multirow{2}{*}{6.25} & Mean & 0.158 & 0.081 \\
 & & HPD & 0.0391-0.3149 & 0.0129-0.1841 \\
 \cmidrule(lr){2-5}
 & \multirow{2}{*}{25.0} & Mean & 0.186 & 0.141 \\
 & & HPD & 0.0436-0.3387 & 0.0317-0.2824 \\
\bottomrule
\end{tabular}
\end{table}

Table~\ref{tab:Sim2_model-comparison} reports the global model comparison criteria. According to the DIC, the proposed model $\mathcal{M}_{4}$ provides the best overall fit in every scenario, followed systematically by $\mathcal{M}_{2}$, then $\mathcal{M}_{3}$, and finally $\mathcal{M}_{1}$. This ranking indicates that modeling cross-response dependence yields a substantial gain in in-sample fit, and that the full model combining anisotropy and contemporaneous covariance is the most adequate description of the simulated data.

For predictive performance, the picture is closely related but slightly more nuanced. The anisotropic models clearly outperform the isotropic competitors in terms of both PMSE and CRPS across all scenarios, which shows that accounting for spatial deformation is the dominant ingredient for accurate interpolation. Between $\mathcal{M}_{3}$ and $\mathcal{M}_{4}$, the differences are generally small. Model~$\mathcal{M}_{3}$ attains the smallest PMSE and CRPS in the scenario $(T,\gamma)=(100,0.0625)$, but only by a very narrow margin. In the remaining three scenarios, $\mathcal{M}_{4}$ yields the best values. This pattern suggests that the main predictive gains come from representing anisotropy correctly, while explicit cross-response dependence provides an additional refinement that becomes more beneficial as the problem becomes harder, either because missingness is larger or because the interaction between spatial and multivariate structure becomes more consequential.

\begin{table}[htbp!]
\centering
\caption{Model comparison metrics (DIC, PMSE, and CRPS) across models ($\mathcal{M}_1$--$\mathcal{M}_4$), sample sizes ($T \in \{100, 500\}$), and missing-value proportions ($\gamma \in \{0.0625, 0.25\}$). Results from the simulation study in Section~\ref{subsec:Sim2}. The smallest values are highlighted in bold.}
\label{tab:Sim2_model-comparison}
\setlength{\tabcolsep}{8pt}
\begin{tabular}{cclrrrr}
\toprule
$T$ & $\gamma$ (\%) & \textbf{Metric} & $\mathcal{M}_1$ & $\mathcal{M}_2$ & 
$\mathcal{M}_3$ & $\mathcal{M}_4$ \\
\midrule
\multirow{8}{*}{100} & \multirow{4}{*}{6.25} & DIC & 4749.74 & 2748.27 & 
3915.09 & \textbf{1905.35} \\
 & & PMSE & 0.0930 & 0.0930 & \textbf{0.0761} & 0.0771 \\
 & & CRPS & 0.1734 & 0.1727 & \textbf{0.1564} & 0.1570 \\
 \cmidrule(lr){2-7}
 & \multirow{4}{*}{25.0} & DIC & 3708.90 & 2042.24 & 3031.27 & \textbf{1326.48} \\
 & & PMSE & 0.1095 & 0.1020 & 0.0919 & \textbf{0.0839} \\
 & & CRPS & 0.1870 & 0.1803 & 0.1701 & \textbf{0.1625} \\
\midrule
\multirow{8}{*}{500} & \multirow{4}{*}{6.25} & DIC & 22594.62 & 12785.34 & 
18359.24 & \textbf{8612.87} \\
 & & PMSE & 0.0944 & 0.0918 & 0.0801 & \textbf{0.0779} \\
 & & CRPS & 0.1734 & 0.1713 & 0.1600 & \textbf{0.1578} \\
 \cmidrule(lr){2-7}
 & \multirow{4}{*}{25.0} & DIC & 17845.13 & 9596.20 & 14382.46 & \textbf{6250.98} \\
 & & PMSE & 0.1091 & 0.0984 & 0.0913 & \textbf{0.0826} \\
 & & CRPS & 0.1854 & 0.1771 & 0.1701 & \textbf{0.1623} \\
\bottomrule
\end{tabular}
\end{table}

Table~\ref{tab:Sim2_IS} provides a more detailed view through the interval score, broken down by ungauged site and response. The proposed model $\mathcal{M}_{4}$ achieves the lowest IS in the majority of cases and becomes particularly dominant when $T=500$, especially under $\gamma=0.25$, where it is best for all sites and both responses. For $T=100$, the comparison is more mixed. There are isolated cases in which $\mathcal{M}_{3}$ or even one of the simpler models attains a marginally smaller score. This is expected in finite samples, since IS combines interval width and calibration at a local level and may therefore exhibit small scenario-specific fluctuations. Still, the general pattern remains clear: once enough information is available, the full anisotropic and multivariate model tends to produce the most accurate and best calibrated predictive intervals.

\begin{figure}[!htb]
    \centering
    \includegraphics[width=1.0\linewidth]{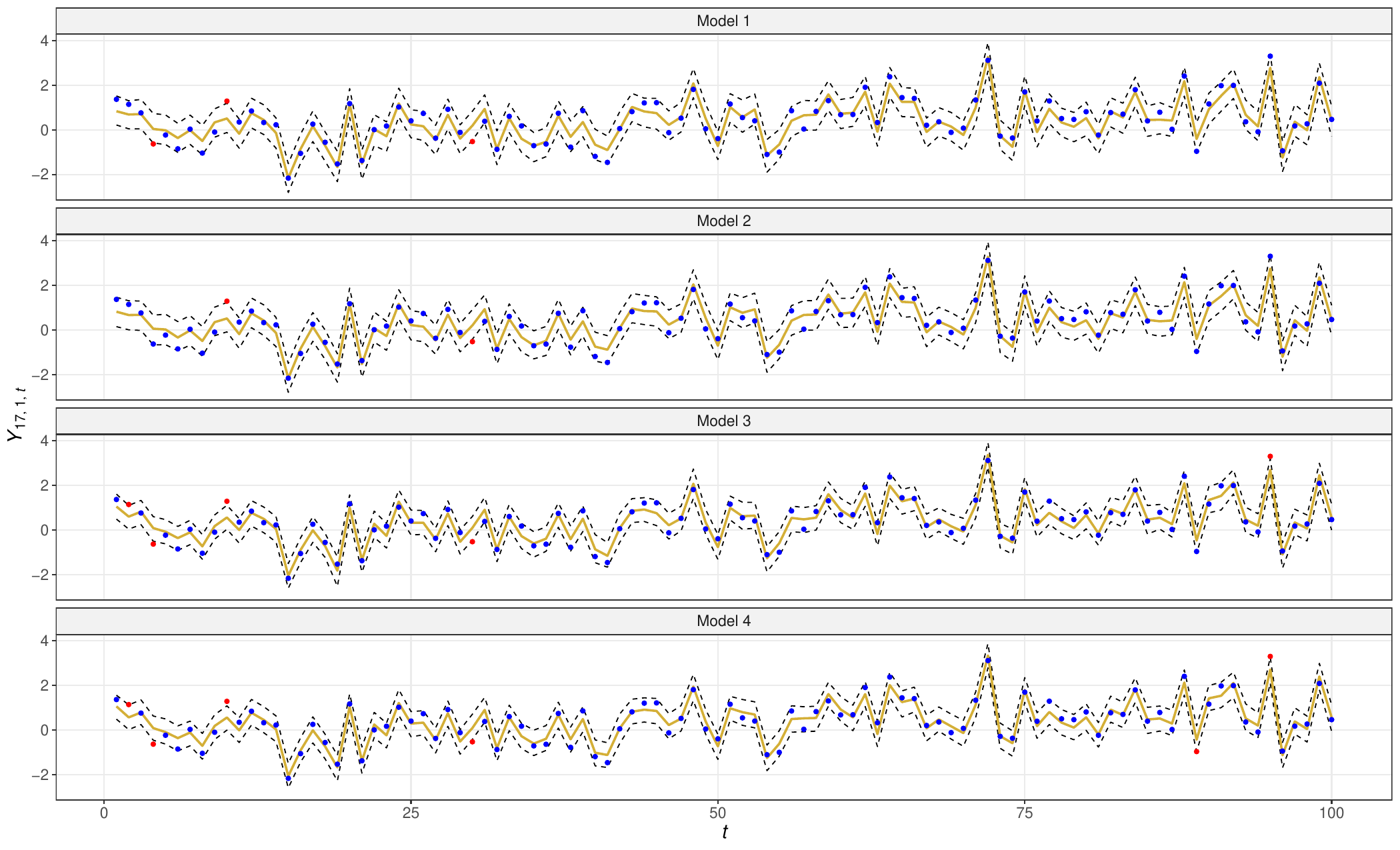}
    \caption{Posterior summaries of $Y_{17,1,t}$ for $t \in \{1,\ldots,100\}$ across models ($\mathcal{M}_1$--$\mathcal{M}_4$), with $T = 100$ and $\gamma = 0.0625$. Black dashed lines denote the 2.5th and 97.5th posterior quantiles, the solid golden line represents the posterior mean, and points indicate the true values (blue if within the interval and red otherwise). Results from the simulation study in Section~\ref{subsec:Sim2}.}
    \label{fig:Sim2_interpolation}
\end{figure}

Figure~\ref{fig:Sim2_interpolation} reinforces this interpretation graphically for site $\uvec{\mathbf{s}}_{17}$ and Response~1. All four models recover the broad temporal pattern of the held-out series, but the isotropic models typically produce wider 95\% posterior intervals. The anisotropic models, particularly $\mathcal{M}_{3}$ and $\mathcal{M}_{4}$, deliver more concentrated predictive bands while keeping the posterior mean close to the true trajectory. The difference between these two models is again modest, which matches the numerical results in Table~\ref{tab:Sim2_model-comparison}: the main gain comes from allowing deformation, whereas the additional cross-response covariance in $\mathcal{M}_{4}$ mainly refines the predictive distribution.

\begin{table}[htbp!]
\centering
\caption{Mean interval score (IS) across models ($\mathcal{M}_1$--$\mathcal{M}_4$), ungauged sites ($\uvec{\mathbf{s}}_{17}$, $\uvec{\mathbf{s}}_{18}$, and $\uvec{\mathbf{s}}_{19}$), response variables ($i \in \{1,2\}$), sample sizes ($T \in \{100, 500\}$), and missing-value proportions ($\gamma \in \{0.0625, 0.25\}$). Results from the simulation study in Section~\ref{subsec:Sim2}. The smallest values are highlighted in bold.}
\label{tab:Sim2_IS}
\begin{tabular}{ccclrrrr}
\toprule
$T$ & $\gamma$ (\%) & \textbf{Site} & $i$ & $\mathcal{M}_{1}$ & 
$\mathcal{M}_{2}$ & $\mathcal{M}_{3}$ & $\mathcal{M}_{4}$ \\
\midrule

% --- BLOCK T = 100 ---
\multirow{17}{*}{100} & \multirow{8}{*}{6.25} & \multirow{2}{*}{$\uvec{\mathbf{s}}_{17}$} 
    & 1 & 0.0352 & 0.0347 & 0.0334 & \textbf{0.0327} \\
    & & & 2 & 0.0349 & 0.0345 & \textbf{0.0291} & 0.0308 \\
    \cmidrule(lr){3-8}
    & & \multirow{2}{*}{$\uvec{\mathbf{s}}_{18}$} 
    & 1 & 0.0380 & 0.0406 & \textbf{0.0358} & 0.0378 \\
    & & & 2 & 0.0371 & 0.0373 & 0.0360 & \textbf{0.0352} \\
    \cmidrule(lr){3-8}
    & & \multirow{2}{*}{$\uvec{\mathbf{s}}_{19}$} 
    & 1 & 0.0349 & 0.0341 & 0.0337 & \textbf{0.0321} \\
    & & & 2 & 0.0332 & 0.0330 & 0.0333 & \textbf{0.0321} \\
\cmidrule{2-8}
    & \multirow{8}{*}{25.0} & \multirow{2}{*}{$\uvec{\mathbf{s}}_{17}$} 
    & 1 & 0.0364 & 0.0352 & 0.0367 & \textbf{0.0349} \\
    & & & 2 & 0.0405 & 0.0370 & \textbf{0.0304} & 0.0322 \\
    \cmidrule(lr){3-8}
    & & \multirow{2}{*}{$\uvec{\mathbf{s}}_{18}$} 
    & 1 & 0.0385 & \textbf{0.0356} & 0.0376 & \textbf{0.0356} \\
    & & & 2 & \textbf{0.0403} & 0.0404 & 0.0421 & 0.0405 \\
    \cmidrule(lr){3-8}
    & & \multirow{2}{*}{$\uvec{\mathbf{s}}_{19}$} 
    & 1 & 0.0375 & 0.0341 & 0.0361 & \textbf{0.0330} \\
    & & & 2 & 0.0365 & \textbf{0.0343} & 0.0376 & 0.0344 \\
\midrule

% --- BLOCK T = 500 ---
\multirow{17}{*}{500} & \multirow{8}{*}{6.25} & \multirow{2}{*}{$\uvec{\mathbf{s}}_{17}$} 
    & 1 & 0.0347 & 0.0340 & 0.0321 & \textbf{0.0319} \\
    & & & 2 & 0.0344 & 0.0340 & 0.0308 & \textbf{0.0307} \\
    \cmidrule(lr){3-8}
    & & \multirow{2}{*}{$\uvec{\mathbf{s}}_{18}$} 
    & 1 & 0.0362 & 0.0357 & \textbf{0.0328} & 0.0330 \\
    & & & 2 & 0.0377 & 0.0372 & 0.0353 & \textbf{0.0349} \\
    \cmidrule(lr){3-8}
    & & \multirow{2}{*}{$\uvec{\mathbf{s}}_{19}$} 
    & 1 & 0.0352 & 0.0346 & 0.0325 & \textbf{0.0324} \\
    & & & 2 & 0.0371 & 0.0358 & 0.0339 & \textbf{0.0333} \\
\cmidrule{2-8}
    & \multirow{8}{*}{25.0} & \multirow{2}{*}{$\uvec{\mathbf{s}}_{17}$} 
    & 1 & 0.0369 & 0.0350 & 0.0333 & \textbf{0.0326} \\
    & & & 2 & 0.0377 & 0.0358 & 0.0333 & \textbf{0.0318} \\
    \cmidrule(lr){3-8}
    & & \multirow{2}{*}{$\uvec{\mathbf{s}}_{18}$} 
    & 1 & 0.0407 & 0.0375 & 0.0370 & \textbf{0.0349} \\
    & & & 2 & 0.0412 & 0.0387 & 0.0372 & \textbf{0.0357} \\
    \cmidrule(lr){3-8}
    & & \multirow{2}{*}{$\uvec{\mathbf{s}}_{19}$} 
    & 1 & 0.0384 & 0.0369 & 0.0349 & \textbf{0.0332} \\
    & & & 2 & 0.0399 & 0.0370 & 0.0352 & \textbf{0.0339} \\
\bottomrule
\end{tabular}
\end{table}

Table~\ref{tab:Sim2_model-checking} complements the comparison with posterior predictive checking, residual summaries, execution times, and effective sample sizes based on the unnormalized log-posterior. The empirical coverage probabilities are reassuring across all models and scenarios. For both responses, the values remain close to the nominal $0.95$ level, typically ranging from about $0.934$ to $0.953$. This indicates that all four formulations produce reasonably calibrated predictive intervals in the simulation setting. At the same time, the anisotropic and/or correlated models often yield slightly better coverages, especially for Response~2 and in the more difficult scenarios, although these differences are not dramatic. Hence, the main distinctions among the models are not driven by severe failures of calibration, but rather by differences in fit quality, predictive sharpness, and spatial recovery.

The residual summaries tell a similar story. Across all scenarios, the residual means and medians are positive, which suggests a mild tendency toward underprediction on average. However, these location summaries are generally smaller under the more flexible models, especially under $\mathcal{M}_{4}$ for $T=100$ and under the anisotropic models for $T=500$. The upper quartiles also tend to be slightly lower for $\mathcal{M}_{4}$, while the minimum and maximum residuals remain of similar magnitude across formulations. Thus, although the residual differences are not large, they point in the same direction as the model-comparison criteria: the richer models provide a somewhat more balanced predictive fit.

From a computational perspective, the anisotropic models require only a moderate additional cost. For $T=100$, the execution times increase from about six minutes under the isotropic models to approximately nine or ten minutes under $\mathcal{M}_{3}$ and $\mathcal{M}_{4}$. For $T=500$, all models remain in the range of roughly 25 to 32 minutes, so the added flexibility does not induce a prohibitive computational burden. The ESS values are also informative. In most scenarios they remain satisfactory, but they decrease noticeably in the most challenging case $(T,\gamma)=(500,0.25)$, particularly for $\mathcal{M}_{2}$ and $\mathcal{M}_{4}$. This is consistent with the greater posterior complexity of the correlated models under heavier missingness. Even so, the chains still provide stable inference for the summaries reported here.

\begin{figure}[htb!]
\centering
\begin{subfigure}{0.49\textwidth}
  \includegraphics[width=\linewidth]{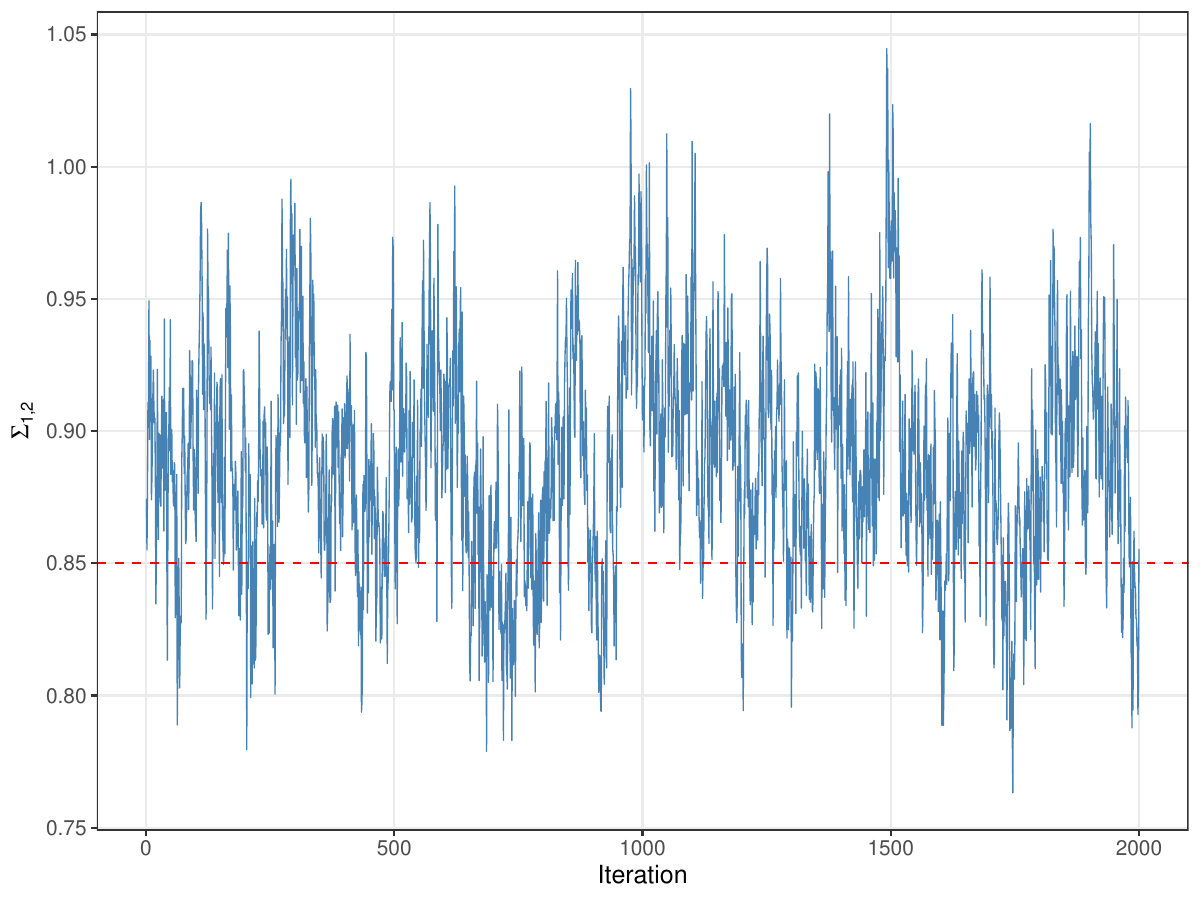}
  \caption{Trace plot of $\Sigma_{1,2}$ under $\mathcal{M}_{2}$.}\label{fig:Sim2_Sigma12_M2}
\end{subfigure}
\begin{subfigure}{0.49\textwidth}
  \includegraphics[width=\linewidth]{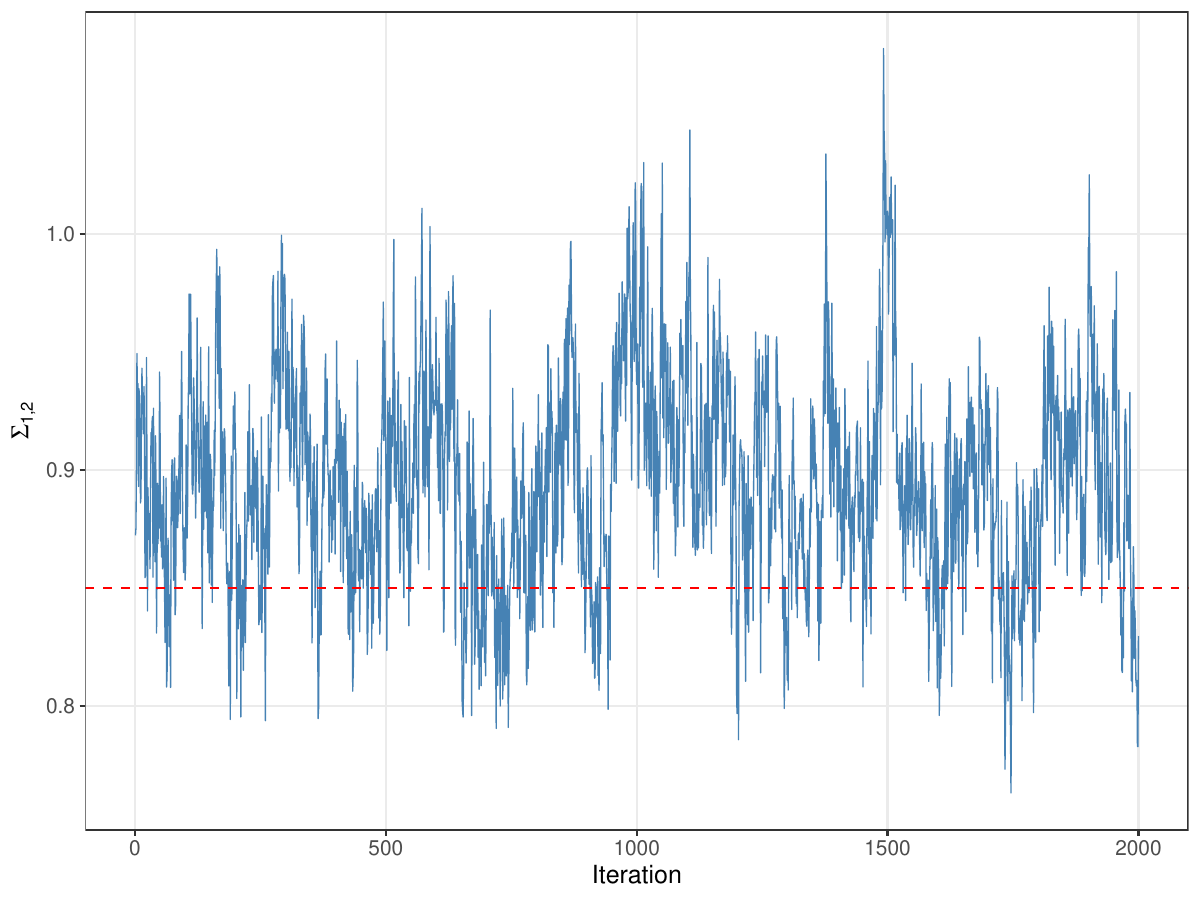}
  \caption{Trace plot of $\Sigma_{1,2}$ under $\mathcal{M}_{4}$.}\label{fig:Sim2_Sigma12_M4}
\end{subfigure}
\begin{subfigure}{0.49\textwidth}
  \includegraphics[width=\linewidth]{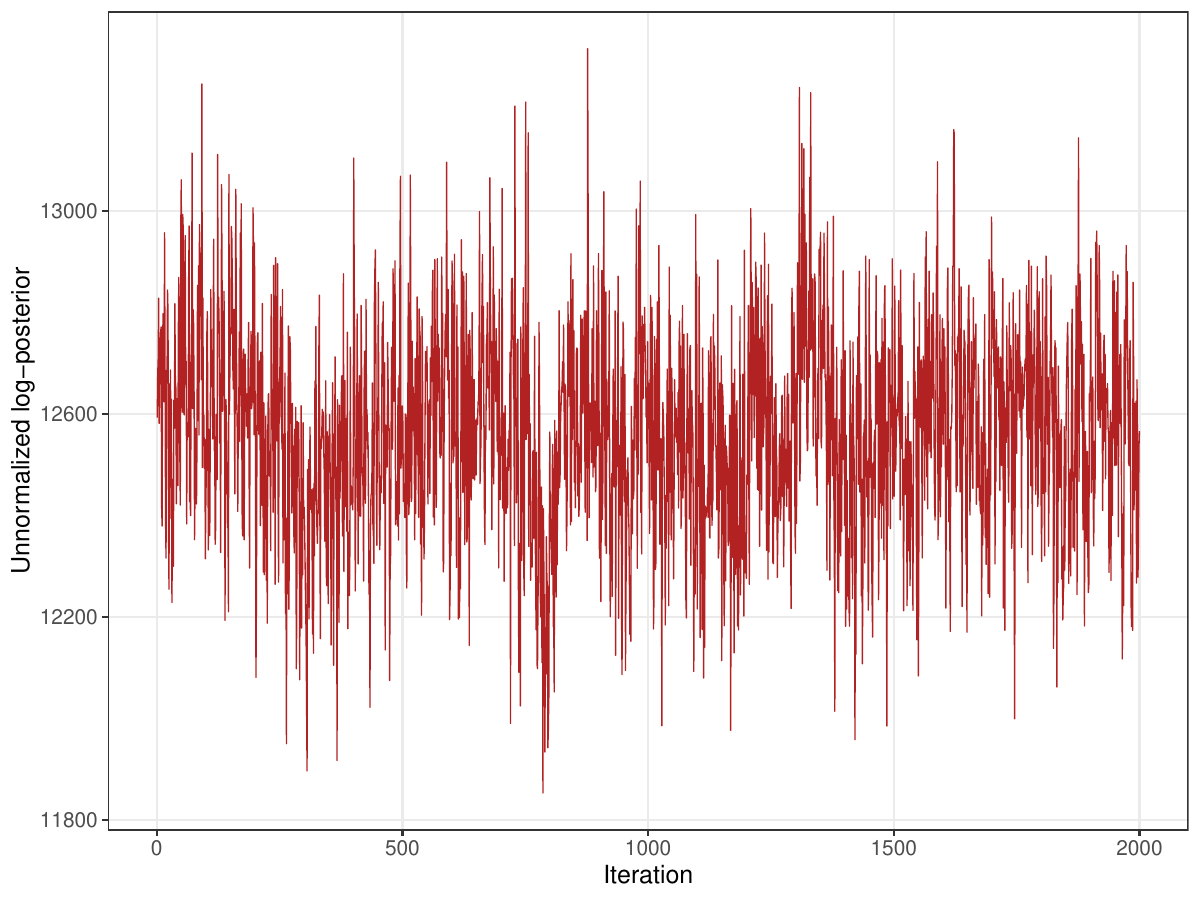}
  \caption{Trace plot of the unnormalized log-posterior under $\mathcal{M}_{2}$.}\label{fig:Sim2_LP_M2}
\end{subfigure}
\begin{subfigure}{0.49\textwidth}
  \includegraphics[width=\linewidth]{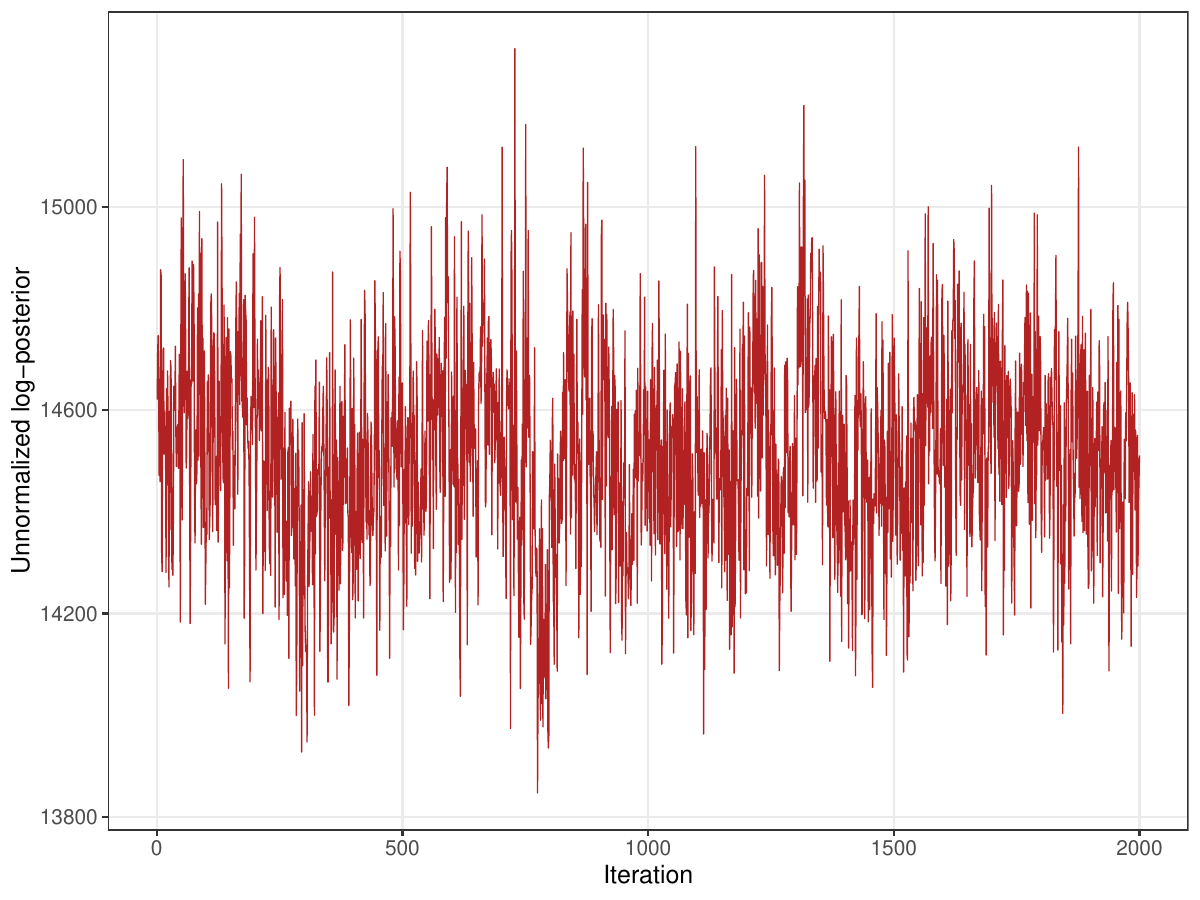}
  \caption{Trace plot of the unnormalized log-posterior under $\mathcal{M}_{4}$.}\label{fig:Sim2_LP_M4}
\end{subfigure}
\caption{MCMC trace plots for the cross-covariance $\Sigma_{1,2}$ and the unnormalized log-posterior distribution under models $\mathcal{M}_{2}$ and $\mathcal{M}_{4}$ in a challenging scenario ($T = 500$ and $\gamma = 0.25$). The horizontal red dashed lines in Figures~\ref{fig:Sim2_Sigma12_M2} and~\ref{fig:Sim2_Sigma12_M4} indicate the true value $\Sigma_{1,2} = 0.85$. Results from the simulation study in Section~\ref{subsec:Sim2}.}
\label{fig:Sim2_TracePlots}
\end{figure}

Figure~\ref{fig:Sim2_TracePlots} illustrates this point in the challenging scenario $(T,\gamma)=(500,0.25)$. The trace plots of $\Sigma_{1,2}$ under $\mathcal{M}_{2}$ and $\mathcal{M}_{4}$ fluctuate around the true value $0.85$ without visible drift, which supports the adequacy of the posterior summaries in Table~\ref{tab:Sim2_Sigma}. Likewise, the trace plots of the unnormalized log-posterior show no evident long-term trend or abrupt regime switching, although the chain under $\mathcal{M}_{4}$ explores its support somewhat more slowly, in line with its lower ESS. Taken together, Table~\ref{tab:Sim2_model-checking} and Figure~\ref{fig:Sim2_TracePlots} indicate that the proposed MCMC scheme remains practically effective even in the most demanding simulated setting, albeit with some loss of mixing efficiency for the most complex model.

\begin{table}[htbp!]
\centering
\caption{Posterior predictive checking via empirical coverage probability (ECP), residual analysis, execution time, and effective sample size (ESS) based on the unnormalized log-posterior distribution, across models ($\mathcal{M}_1$--$\mathcal{M}_4$), sample sizes ($T \in \{100, 500\}$), and missing-value proportions ($\gamma \in \{0.0625, 0.25\}$). Results from the simulation study in Section~\ref{subsec:Sim2}.}
\label{tab:Sim2_model-checking}
\footnotesize
\begin{tabular}{cc lrrrr}
\toprule
$T$ & $\gamma$ (\%) & \textbf{Metric} & $\mathcal{M}_{1}$ & 
$\mathcal{M}_{2}$ & $\mathcal{M}_{3}$ & $\mathcal{M}_{4}$ \\
\midrule

% --- BLOCK T = 100 ---
\multirow{29}{*}{100} & \multirow{14}{*}{6.25} & \textit{ECP} & & & & \\
    &        & Response 1   & 0.9488 & 0.9519 & 0.9406 & 0.9488 \\
    &        & Response 2   & 0.9369 & 0.9469 & 0.9381 & 0.9456 \\
\cmidrule(lr){3-7}
    &        & \textit{Residual statistics} & & & & \\
    &        & Mean         & 0.2666 & 0.2615 & 0.2583 & 0.2500 \\
    &        & Median       & 0.2229 & 0.2176 & 0.2071 & 0.2009 \\
    &        & 1st quartile & -0.3831 & -0.3581 & -0.3990 & -0.3750 \\
    &        & 3rd quartile & 0.8547 & 0.8233 & 0.8664 & 0.8131 \\
    &        & Minimum      & -3.1578 & -3.0429 & -3.2817 & -3.1703 \\
    &        & Maximum      & 3.9931 & 3.7731 & 3.9450 & 3.8474 \\
\cmidrule(lr){3-7}
    &        & \textit{Bayesian computation} & & & & \\
    &        & Execution time (min.)& 6.2    & 6.2    & 8.7    & 9.0    \\
    &        & ESS (log-post.) & 697.4 & 782.9 & 951.3 & 880.3  \\
\cmidrule{2-7}
    & \multirow{14}{*}{25.0} & \textit{ECP} & & & & \\
    &        & Response 1   & 0.9494 & 0.9531 & 0.9450 & 0.9519 \\
    &        & Response 2   & 0.9388 & 0.9450 & 0.9356 & 0.9450 \\
\cmidrule(lr){3-7}
    &        & \textit{Residual statistics} & & & & \\
    &        & Mean         & 0.2618 & 0.2529 & 0.2585 & 0.2534 \\
    &        & Median       & 0.2248 & 0.2151 & 0.2113 & 0.2083 \\
    &        & 1st quartile & -0.3894 & -0.3656 & -0.3986 & -0.3652 \\
    &        & 3rd quartile & 0.8403 & 0.8052 & 0.8554 & 0.8018 \\
    &        & Minimum      & -3.2468 & -3.1102 & -3.2693 & -3.0526 \\
    &        & Maximum      & 3.8257 & 3.7846 & 4.0231 & 3.8086 \\
\cmidrule(lr){3-7}
    &        & \textit{Bayesian computation} & & & & \\
    &        & Execution time (min.)& 6.1    & 6.5    & 8.7    & 10.4   \\
    &        & ESS (log-post.) & 983.7 & 469.6 & 950.7 & 407.6  \\
\midrule

% --- BLOCK T = 500 ---
\multirow{29}{*}{500} & \multirow{14}{*}{6.25} & \textit{ECP} & & & & \\
    &        & Response 1   & 0.9408 & 0.9408 & 0.9439 & 0.9426 \\
    &        & Response 2   & 0.9346 & 0.9338 & 0.9381 & 0.9371 \\
\cmidrule(lr){3-7}
    &        & \textit{Residual statistics} & & & & \\
    &        & Mean         & 0.1607 & 0.1632 & 0.1548 & 0.1561 \\
    &        & Median       & 0.1511 & 0.1544 & 0.1416 & 0.1472 \\
    &        & 1st quartile & -0.5091 & -0.5139 & -0.5084 & -0.5115 \\
    &        & 3rd quartile & 0.8371 & 0.8434 & 0.8212 & 0.8231 \\
    &        & Minimum      & -3.8293 & -3.8501 & -3.8624 & -3.7065 \\
    &        & Maximum      & 3.8114 & 3.9256 & 3.9650 & 3.8513 \\
\cmidrule(lr){3-7}
    &        & \textit{Bayesian computation} & & & & \\
    &        & Execution time (min.)& 27.4   & 26.2   & 32.2   & 29.1   \\
    &        & ESS (log-post.) & 1266.3 & 1011.8 & 1031.2 & 758.7  \\
\cmidrule{2-7}
    & \multirow{14}{*}{25.0} & \textit{ECP} & & & & \\
    &        & Response 1   & 0.9441 & 0.9450 & 0.9455 & 0.9454 \\
    &        & Response 2   & 0.9368 & 0.9405 & 0.9420 & 0.9425 \\
\cmidrule(lr){3-7}
    &        & \textit{Residual statistics} & & & & \\
    &        & Mean         & 0.1635 & 0.1613 & 0.1585 & 0.1584 \\
    &        & Median       & 0.1534 & 0.1523 & 0.1477 & 0.1454 \\
    &        & 1st quartile & -0.4983 & -0.4968 & -0.4997 & -0.4963 \\
    &        & 3rd quartile & 0.8295 & 0.8264 & 0.8152 & 0.8170 \\
    &        & Minimum      & -3.8500 & -3.8606 & -3.7450 & -3.7416 \\
    &        & Maximum      & 3.6725 & 3.7615 & 3.7416 & 3.8863 \\
\cmidrule(lr){3-7}
    &        & \textit{Bayesian computation} & & & & \\
    &        & Execution time (min.)& 26.8   & 24.5   & 27.5   & 28.6   \\
    &        & ESS (log-post.) & 682.7 & 267.8 & 631.6 & 205.0  \\
\bottomrule
\end{tabular}
\end{table}

Overall, this simulation study provides consistent evidence in favor of the anisotropic formulations. When the data are generated from a geometrically anisotropic mechanism, the deformed models $\mathcal{M}_{3}$ and $\mathcal{M}_{4}$ recover the spatial range parameter much more accurately than the isotropic alternatives, estimate the latent deformation increasingly well as $T$ grows, and achieve clearly superior interpolation performance. Among them, $\mathcal{M}_{4}$ delivers the best overall combination of goodness-of-fit and predictive accuracy, although the gains over $\mathcal{M}_{3}$ are sometimes modest, particularly in the smaller-sample settings. These results indicate that correctly representing anisotropy is the main determinant of performance in this simulation design, while explicit cross-response dependence provides an additional but secondary improvement.

%--- Section ---%
\section{Application}\label{sec:Application}

To illustrate the performance of the proposed model in a real-world context, we apply it to fine particulate matter (PM) data from the Central Valley region of California, USA. \citet[Sec.~4]{Hasheminassab2014} analyzed long-term PM\textsubscript{2.5} records from multiple monitoring stations in the San Joaquin Valley (i.e., the southern portion of the Central Valley) and identified persistent spatial gradients in particulate concentrations. Their findings indicated that secondary aerosol formation and regional transport processes dominate the PM\textsubscript{2.5} burden, resulting in markedly higher concentrations in the southern and central subregions compared with the northern area. Such spatial heterogeneity makes this region a suitable case study for the proposed anisotropic modeling framework.

We focus on PM\textsubscript{10} (Response~1) and PM\textsubscript{2.5} (Response~2), which are expected to be positively associated because they are influenced by related emission, transport, and atmospheric processes. Hourly data were obtained from $21$ monitoring stations that measured both pollutants between January~1 and December~31,~2024. Daily means were then computed, yielding $T = 366$ observations. The dataset was retrieved from the \citet{EPA2025}.

\begin{figure}[htb!]
  \centering
  {\includegraphics[scale=0.50]{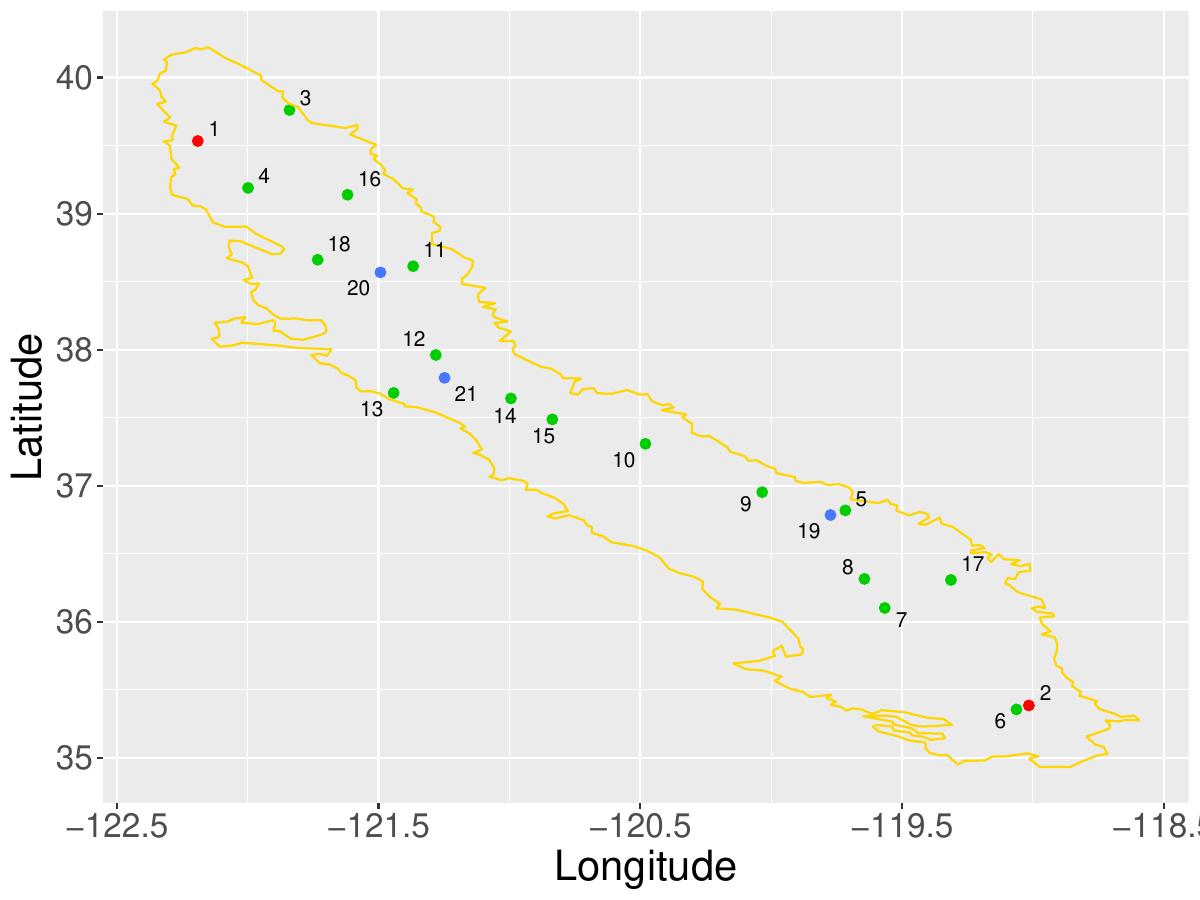}}
  \caption{Spatial domain $\mathcal{S}$ corresponding to the Central Valley region of California, USA. Red circles denote the two anchor sites ($\uvec{\mathbf{s}}_{1}$ and $\uvec{\mathbf{s}}_{2}$), green circles denote the non-anchor gauged sites ($\uvec{\mathbf{s}}_{3}, \ldots, \uvec{\mathbf{s}}_{18}$), and blue circles denote the ungauged sites used for interpolation ($\uvec{\mathbf{s}}_{19}$, $\uvec{\mathbf{s}}_{20}$, and $\uvec{\mathbf{s}}_{21}$). Results from the application in Section~\ref{sec:Application}.}
  \label{fig:Application_S}
\end{figure}

Of the $21 \times 366 = 7686$ potential daily observations for each variable, $10.71\%$ were missing in Response~1 and $1.61\%$ in Response~2. We selected $N = 18$ stations for model fitting and reserved $N^{\ast} = 3$ additional stations for evaluating interpolation performance. Figure~\ref{fig:Application_S} depicts the spatial configuration of the monitoring sites. The observed sites span the Central Valley over a broad north--south extent, while the three ungauged sites are distributed across the domain rather than concentrated in a single subregion. This design provides a nontrivial interpolation setting, since predictions must be produced at locations surrounded by distinct local spatial neighborhoods. In addition, the elongated geographic configuration of the network reinforces the practical relevance of allowing anisotropic spatial dependence.

The analysis was performed with two explanatory covariates in addition to the intercept, so that the matrices $\mathbf{X}_{t}$ and $\mathbf{X}_{t}^{\ast}$ were built with $p = 3$, including altitude and daily mean temperature. These covariates were incorporated to account for important physical drivers of particulate concentrations. Altitude captures broad topographic variation that may affect dispersion and accumulation processes, whereas temperature is related to meteorological conditions and secondary aerosol formation. We consider $\mathbf{G}_{t} = \mathbf{I}_{p}$, which yields a dynamic specification in which the regression coefficients evolve over time according to a random-walk structure.

To implement Algorithm~\ref{alg:MCMC}, we use the observed subvectors $\uvec{\mathbf{y}}_{1,\obs}, \ldots, \uvec{\mathbf{y}}_{366,\obs}$. The prior specification and MCMC configuration follow those adopted in the second simulation study (Section~\ref{subsec:Sim2}), and are therefore not repeated here.

The only tuning parameter requiring dataset-specific calibration is the deformation parameter $\psi$. An empirical sensitivity analysis over the grid $\{0.1, 0.5, 1.0, 2.5, 5.0, 10.0, 50.0, 100.0, 500.0\}$ was conducted, and the value $\psi = 5.0$ was selected based on predictive performance as measured by the CRPS.

Posterior inference was carried out using 30{,}000 MCMC iterations, with a burn-in period of 10{,}000 iterations and thinning of 10, yielding $K = 2000$ posterior samples. This configuration was applied uniformly across models $\mathcal{M}_{1}$--$\mathcal{M}_{4}$. The resulting posterior draws of the model parameters and completed data were then used via Algorithm~\ref{alg:Interpolation} to approximate \eqref{eq:Interpolation} and obtain predictions at ungauged locations.

Table~\ref{tab:Application_parameters} summarizes posterior estimates of $\phi$, the elements of $\boldsymbol{\Sigma}$, the diagonal entries of $\mathbf{W}$, and the deformation variance parameters in $\boldsymbol{\sigma}_{d}^{2}$. Reported metrics include the posterior mean and the 95\% highest posterior density (HPD) interval. A first important pattern concerns the spatial range parameter $\phi$. The isotropic models $\mathcal{M}_{1}$ and $\mathcal{M}_{2}$ yield substantially larger posterior means for $\phi$ than the anisotropic models $\mathcal{M}_{3}$ and $\mathcal{M}_{4}$. This suggests that, when deformation is not available, the model compensates by favoring a longer-range isotropic correlation structure over the original geographic space. Once spatial deformation is introduced, the range parameter can remain much smaller because part of the large-scale spatial organization is absorbed by the latent geometry.

As expected, $\Sigma_{1,2}$ is fixed at zero under $\mathcal{M}_{1}$ and $\mathcal{M}_{3}$, whereas under $\mathcal{M}_{2}$ and $\mathcal{M}_{4}$ its posterior mean is positive, indicating positive contemporaneous covariance between PM\textsubscript{10} and PM\textsubscript{2.5}. Moreover, this covariance is substantially larger under $\mathcal{M}_{4}$ than under $\mathcal{M}_{2}$, suggesting that once anisotropy is properly accounted for, the contemporaneous dependence structure between the two responses becomes more clearly identifiable. The marginal variances $\Sigma_{1,1}$ and $\Sigma_{2,2}$ are also larger under the anisotropic models, reflecting a redistribution of variability after the spatial structure is more flexibly modeled.

\begin{figure}[!htb]
    \centering
    \includegraphics[width=\linewidth]{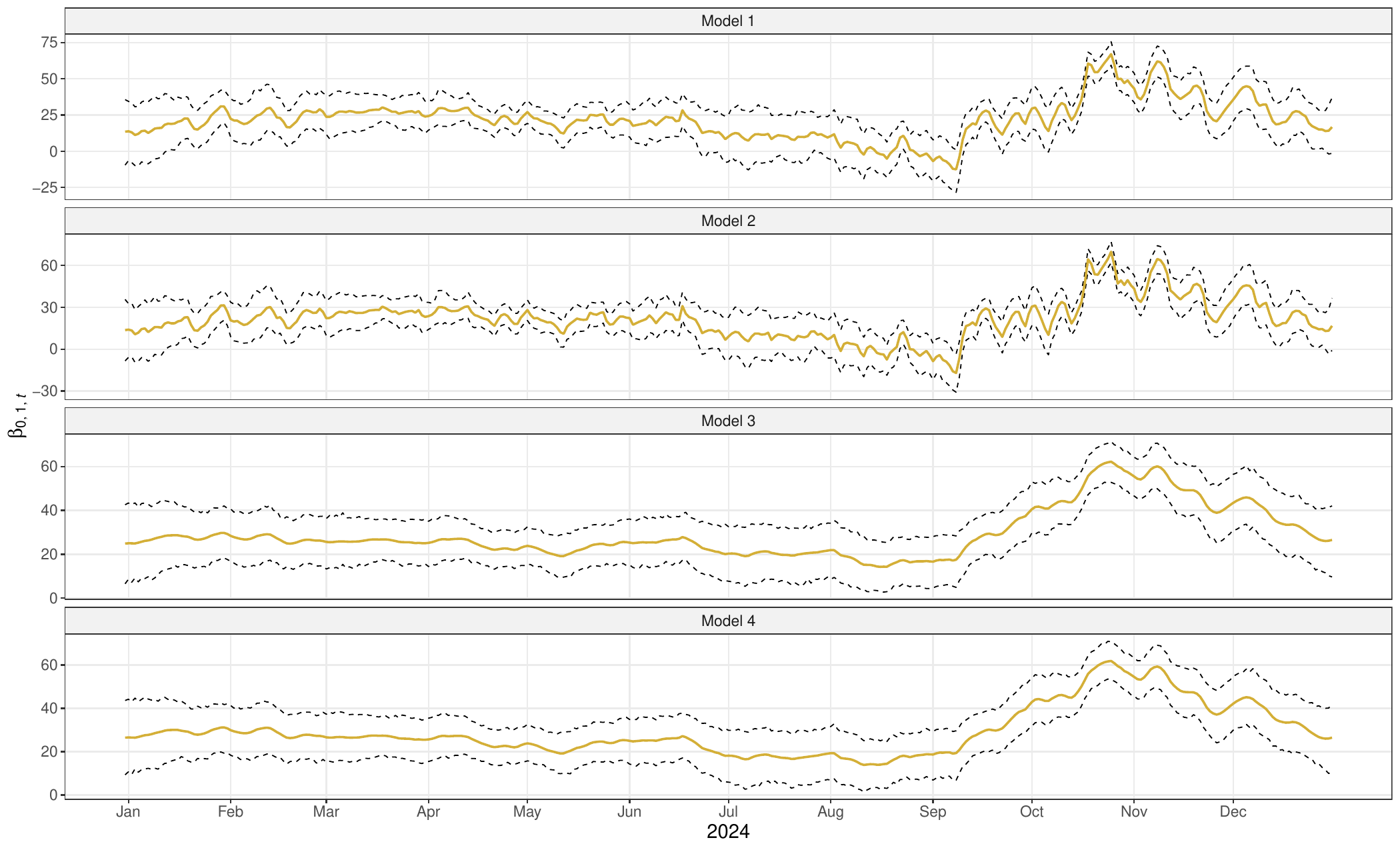}
    \caption{Posterior distribution of $\beta_{0,1,t}$ for $t \in \{0,1,\ldots,366\}$ across models ($\mathcal{M}_1$--$\mathcal{M}_4$). Black dashed lines denote 95\% highest posterior density intervals and the solid golden line represents the posterior mean. Results from the application in Section~\ref{sec:Application}.}
    \label{fig:Application_beta}
\end{figure}

The diagonal entries of $\mathbf{W}$ provide information on the temporal variability of the regression coefficients. For all three components, the anisotropic models tend to produce smaller posterior means than the isotropic competitors, especially for $W_{1,1}$ and $W_{3,3}$. This indicates that, once the spatial dependence is better represented, less temporal innovation is required to explain the observed series. In other words, part of what would otherwise be absorbed by coefficient fluctuations in an isotropic model is instead captured by the deformation. This interpretation is supported by Figure~\ref{fig:Application_beta}, where the posterior trajectories under $\mathcal{M}_{3}$ and $\mathcal{M}_{4}$ appear smoother and more stable over time.

The posterior summaries of $\boldsymbol{\sigma}_{d}^{2}$ show that $\sigma^2_{d_{2,2}}$ is markedly larger than $\sigma^2_{d_{1,1}}$ in both anisotropic models. This indicates greater deformation variability along the second coordinate of the latent domain, suggesting that anisotropy is not uniform across directions. Such directional imbalance is compatible with the geographic and environmental structure of the Central Valley, where pollution transport and accumulation mechanisms may act more strongly along one dominant axis.

\begin{table}[htbp!]
\centering
\caption{Posterior mean and 95\% highest posterior density (HPD) intervals for model parameters across models ($\mathcal{M}_1$--$\mathcal{M}_4$). Results from the application in Section~\ref{sec:Application}.}
\label{tab:Application_parameters}
\setlength{\tabcolsep}{4pt}
\begin{tabular}{ccccccc}
\toprule
\textbf{Parameter} & \textbf{Metric} & $\mathcal{M}_1$ & $\mathcal{M}_2$ & 
$\mathcal{M}_3$ & $\mathcal{M}_4$ \\
\midrule
\multirow{2}{*}{$\phi$} & Mean & 1.5500 & 2.0590 & 0.2330 & 0.2350 \\
 & HPD & 1.3450-1.7470 & 1.7930-2.3260 & 0.2000-0.2670 & 0.2050-0.2650 \\
\midrule
\multirow{2}{*}{$\Sigma_{1,1}$} & Mean & 191.0 & 165.6 & 312.5 & 300.4 \\
 & HPD & 175.28-210.26 & 154.23-178.32 & 274.09-360.80 & 268.18-340.51 \\
\midrule
\multirow{2}{*}{$\Sigma_{1,2}$} & Mean & 0.0 & 25.7 & 0.0 & 56.6 \\
 & HPD & 0.00-0.00 & 23.81-27.84 & 0.00-0.00 & 50.06-64.68 \\
\midrule
\multirow{2}{*}{$\Sigma_{2,2}$} & Mean & 22.4 & 19.4 & 41.9 & 41.7 \\
 & HPD & 20.59-24.70 & 17.99-21.05 & 36.64-48.49 & 36.85-47.63 \\
\midrule
\multirow{2}{*}{$W_{1,1}$} & Mean & 0.1400 & 0.1880 & 0.0250 & 0.0250 \\
 & HPD & 0.0960-0.1890 & 0.1380-0.2420 & 0.0090-0.0470 & 0.0110-0.0450 \\
\midrule
\multirow{2}{*}{$W_{2,2}$} & Mean & 0.0020 & 0.0030 & 0.0010 & 0.0010 \\
 & HPD & 0.0010-0.0040 & 0.0010-0.0060 & 0.0010-0.0030 & 0.0000-0.0020 \\
\midrule
\multirow{2}{*}{$W_{3,3}$} & Mean & 0.0360 & 0.0340 & 0.0130 & 0.0110 \\
 & HPD & 0.0180-0.0620 & 0.0170-0.0600 & 0.0060-0.0250 & 0.0050-0.0210 \\
\midrule
\multirow{2}{*}{$\sigma^2_{d_{1,1}}$} & Mean & -- & -- & 0.4800 & 0.5540 \\
 & HPD & -- & -- & 0.2350-0.9260 & 0.2750-1.1020 \\
\midrule
\multirow{2}{*}{$\sigma^2_{d_{2,2}}$} & Mean & -- & -- & 1.9570 & 2.4970 \\
 & HPD & -- & -- & 0.9740-3.7800 & 1.2410-4.7330 \\
\bottomrule
\end{tabular}
\end{table}

Figure~\ref{fig:Application_beta} displays the posterior distribution of $\beta_{0,1,t}$ for $t \in \{0,1,\ldots,366\}$ across the four models. Several features are worth noting. First, all models identify a similar broad temporal pattern, with moderate levels at the beginning of the year, a decline around midyear, and a pronounced increase during the last quarter. This common structure suggests that the underlying temporal signal is robust across specifications. However, the uncertainty bands are visibly wider under the isotropic models, especially under $\mathcal{M}_{1}$ and $\mathcal{M}_{2}$, and their posterior mean trajectories are more irregular. In contrast, the anisotropic models yield smoother posterior means and narrower HPD intervals, particularly over long stretches of the year. The trajectories under $\mathcal{M}_{3}$ and $\mathcal{M}_{4}$ are also very similar, which is consistent with the relatively close posterior summaries obtained for several parameters in Table~\ref{tab:Application_parameters}. Overall, the figure suggests that allowing deformation stabilizes the dynamic regression component and reduces posterior uncertainty about its temporal evolution.

\begin{figure}[htb!]
\centering
\begin{subfigure}{1.00\textwidth}
  \includegraphics[width=\linewidth]{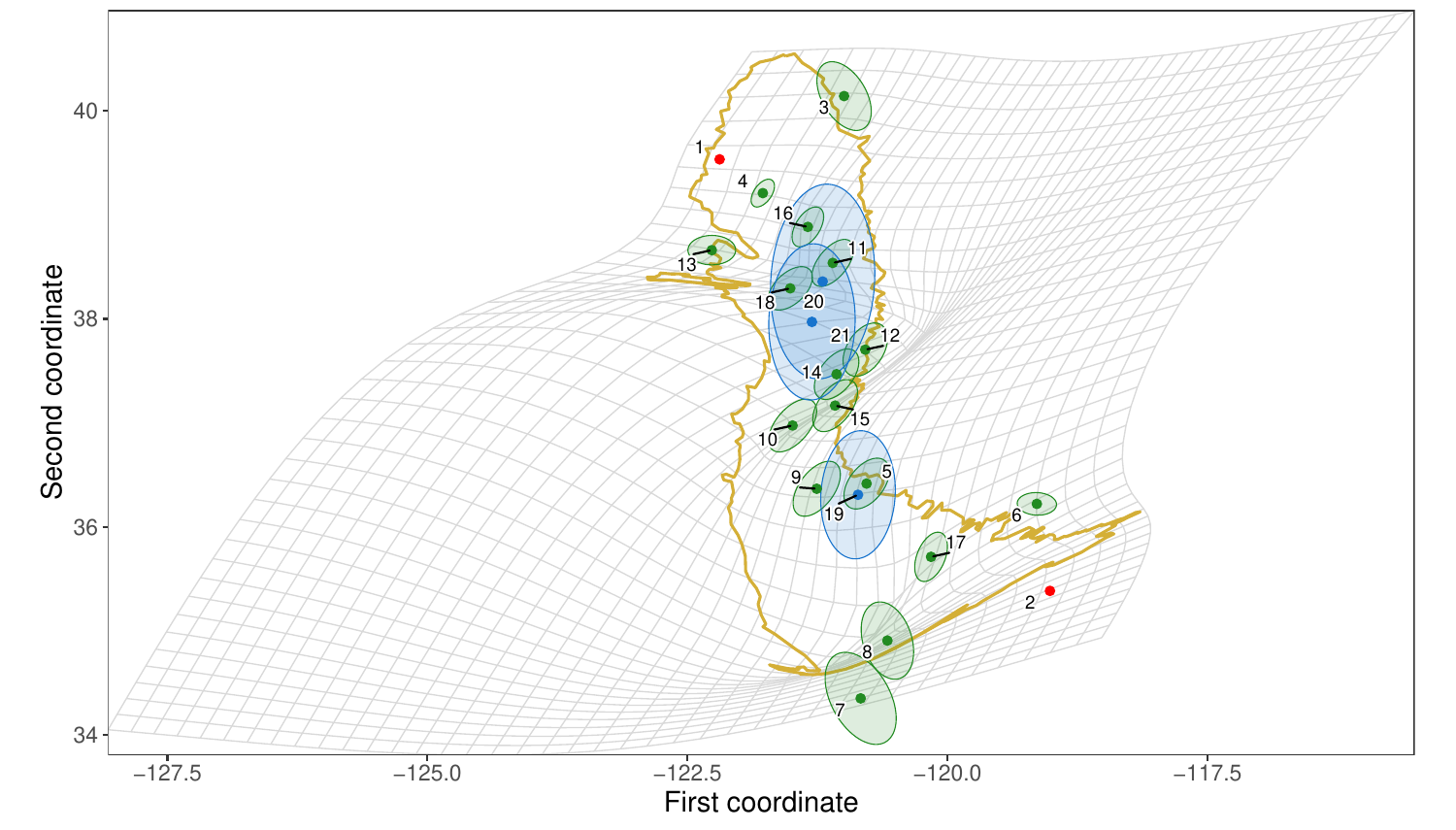}
  \caption{Estimated locations in $\mathcal{D}$-space under model $\mathcal{M}_3$.}\label{fig:Application_D-M3}
\end{subfigure}
\begin{subfigure}{1.00\textwidth}
  \includegraphics[width=\linewidth]{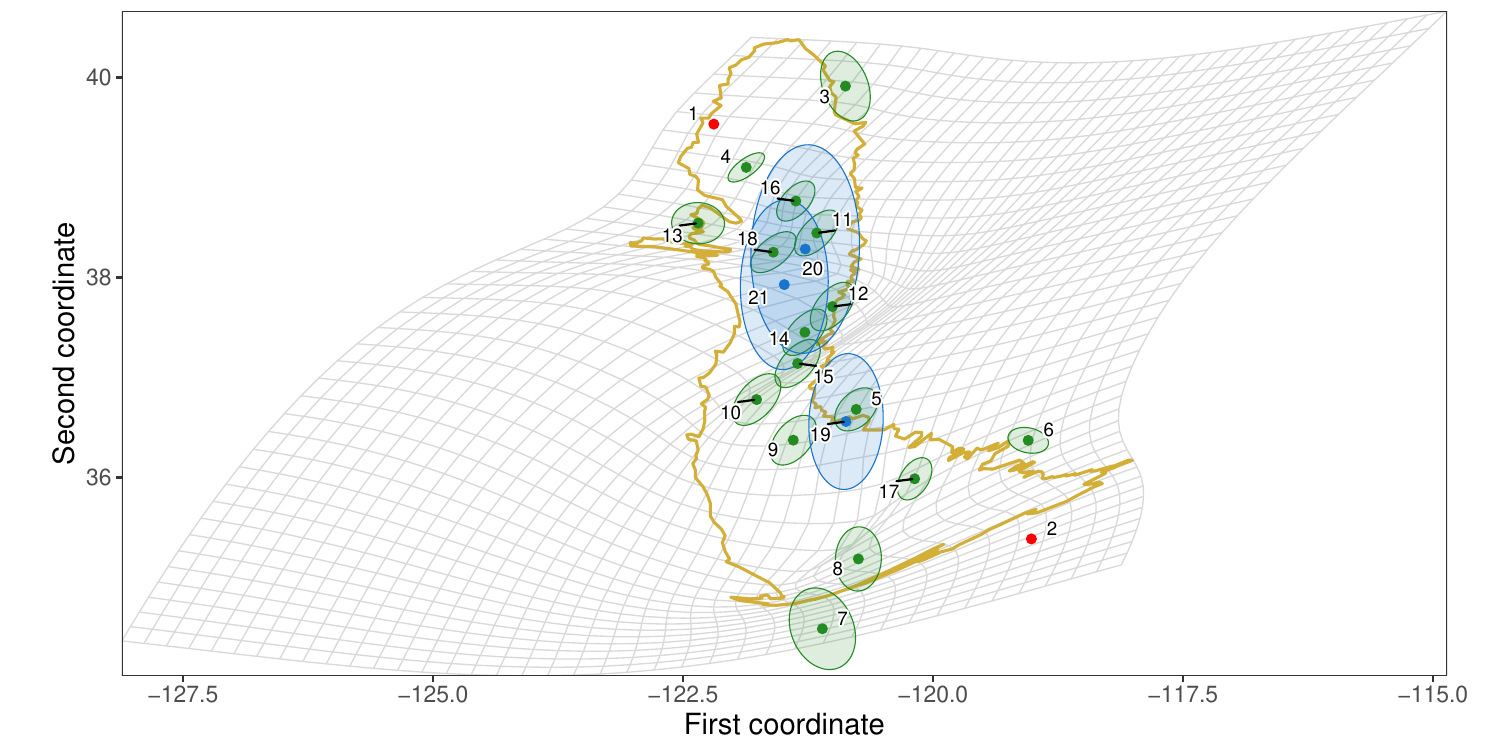}
  \caption{Estimated locations in $\mathcal{D}$-space under model $\mathcal{M}_4$.}\label{fig:Application_D-M4}
\end{subfigure}
\caption{Estimated and interpolated locations in $\mathcal{D}$-space under anisotropic models $\mathcal{M}_3$ and $\mathcal{M}_4$. Red circles denote the anchor points, green circles denote posterior means of estimated deformations at gauged sites, and blue circles denote posterior means of interpolated deformations at ungauged sites. Ellipses denote 95\% credible regions. Results from the application in Section~\ref{sec:Application}.}
\label{fig:Application_D}
\end{figure}

Figure~\ref{fig:Application_D} presents the posterior mean estimated deformations for gauged sites and the interpolated deformations for ungauged sites under the anisotropic models $\mathcal{M}_{3}$ and $\mathcal{M}_{4}$. The two panels are strikingly similar, indicating that the deformation pattern is robust to whether cross-response dependence is explicitly included. In both cases, the latent domain departs meaningfully from the original geographic configuration, which provides direct graphical evidence against the adequacy of a purely isotropic spatial representation. The 95\% credible ellipses are reasonably concentrated for most locations, suggesting that the deformation is estimated with useful precision. The three ungauged sites are mapped to positions that are coherent with the surrounding gauged network, which supports the use of the interpolated latent coordinates in the predictive step. Taken together, the figure reinforces two conclusions: anisotropy is relevant in this application, and the deformation learned by the model is stable across the two anisotropic specifications.

Table~\ref{tab:Application_model-comparison} reports the model comparison criteria. According to the DIC, model~$\mathcal{M}_{4}$ provides the best fit to the data, followed by $\mathcal{M}_{3}$, with both anisotropic models clearly outperforming the isotropic alternatives. The same ranking is observed for PMSE and CRPS, for which $\mathcal{M}_{4}$ again attains the smallest values. Thus, the improvement is not restricted to a single criterion: the proposed model offers the best overall combination of model fit and predictive performance.

The interval score (IS) provides additional detail by separating performance across sites and responses. For Response~1, model~$\mathcal{M}_{4}$ attains the best score at all three ungauged sites, indicating more accurate and better calibrated interpolation for PM\textsubscript{10}. For Response~2, the picture is slightly more mixed. Model~$\mathcal{M}_{4}$ still performs best at sites $\uvec{\mathbf{s}}_{19}$ and $\uvec{\mathbf{s}}_{21}$, whereas $\mathcal{M}_{3}$ is marginally better at $\uvec{\mathbf{s}}_{20}$. This suggests that the main predictive gains come from accommodating anisotropy, while the additional cross-response covariance in $\mathcal{M}_{4}$ yields a further but more modest refinement. This interpretation is consistent with the strong similarity already seen between $\mathcal{M}_{3}$ and $\mathcal{M}_{4}$ in the latent deformation patterns.

\begin{table}[htbp!]
\centering
\caption{Model comparison metrics (DIC, PMSE, and CRPS) and interval scores (IS) for ungauged sites ($\uvec{\mathbf{s}}_{19}$, $\uvec{\mathbf{s}}_{20}$, and $\uvec{\mathbf{s}}_{21}$) and response variables ($i \in \{1,2\}$) across models ($\mathcal{M}_1$--$\mathcal{M}_4$). Results from the application in Section~\ref{sec:Application}. The smallest values are highlighted in bold.}
\label{tab:Application_model-comparison}
\setlength{\tabcolsep}{8pt}
\begin{tabular}{lccrrrr}
\toprule
\textbf{Metric} & \textbf{Site} & $i$ & $\mathcal{M}_1$ & $\mathcal{M}_2$ & 
$\mathcal{M}_3$ & $\mathcal{M}_4$ \\
\midrule
DIC & -- & -- & 80370.0 & 78962.1 & 75388.6 & \textbf{73715.0} \\
PMSE & -- & -- & 27.7764 & 27.9194 & 25.4949 & \textbf{23.5658} \\
CRPS & -- & -- & 2.5631 & 2.5558 & 2.3389 & \textbf{2.2448} \\
\midrule
\multirow{6}{*}{IS} & \multirow{2}{*}{$\uvec{\mathbf{s}}_{19}$} & 1 & 0.85000 & 0.81100 & 0.83500 & \textbf{0.77200} \\
 & & 2 & 0.31800 & 0.31200 & 0.26300 & \textbf{0.25700} \\
 \cmidrule(lr){2-7}
 & \multirow{2}{*}{$\uvec{\mathbf{s}}_{20}$} & 1 & 1.12000 & 1.08000 & 0.72210 & \textbf{0.66600} \\
 & & 2 & 0.27300 & 0.27900 & \textbf{0.24000} & 0.24800 \\
 \cmidrule(lr){2-7}
 & \multirow{2}{*}{$\uvec{\mathbf{s}}_{21}$} & 1 & 1.30000 & 1.32000 & 1.16000 & \textbf{1.12000} \\
 & & 2 & 0.29200 & 0.29500 & 0.24000 & \textbf{0.23700} \\
\bottomrule
\end{tabular}
\end{table}

\begin{figure}[!htb]
    \centering
    \includegraphics[width=\linewidth]{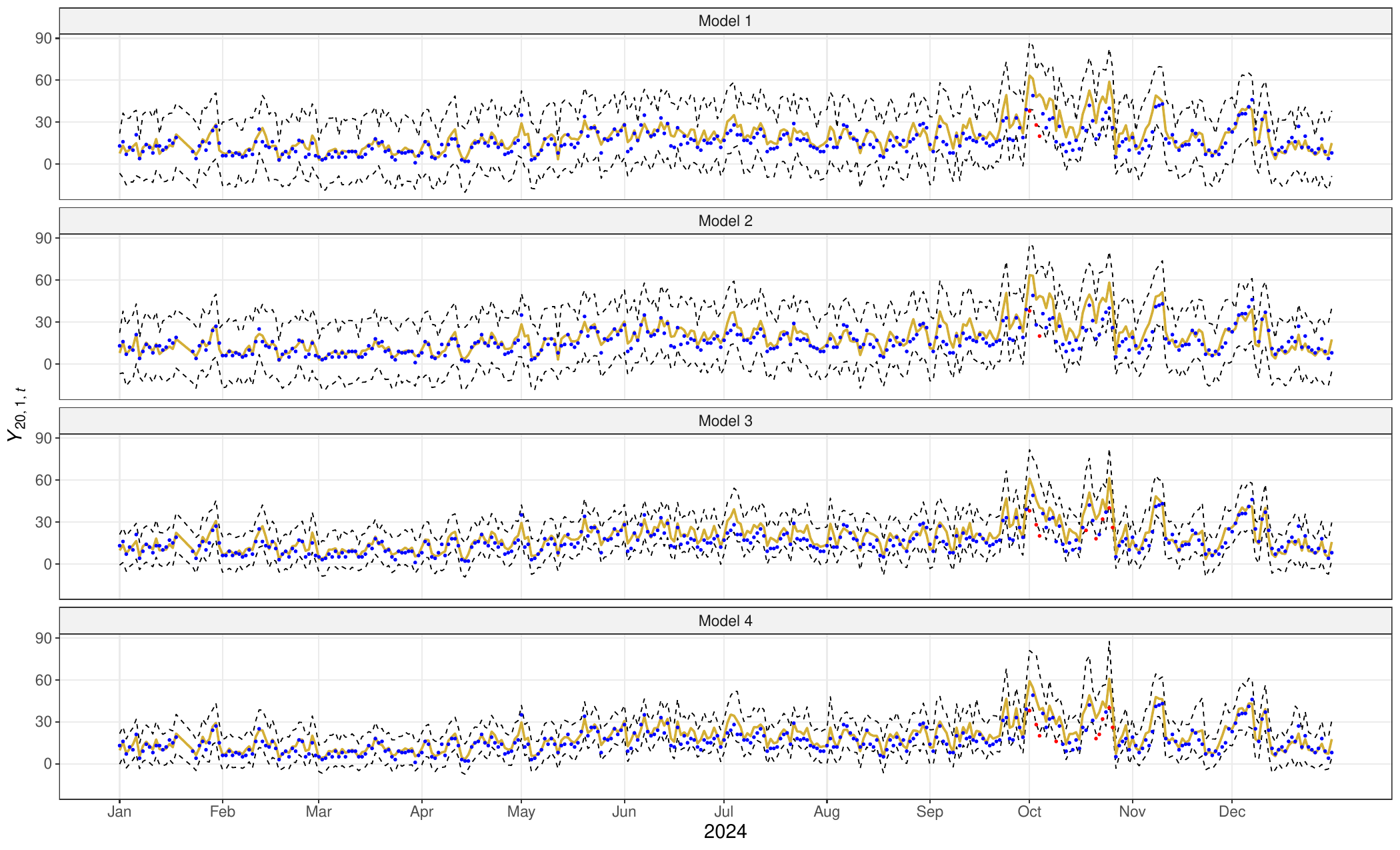}
    \caption{Posterior summaries of $Y_{20,1,t}$ for $t \in \{1,\ldots,366\}$ across models ($\mathcal{M}_1$--$\mathcal{M}_4$). Black dashed lines denote the 2.5th and 97.5th posterior quantiles, the solid golden line represents the posterior mean, and points indicate the observed values (blue if within the interval and red otherwise). Results from the application in Section~\ref{sec:Application}.}
    \label{fig:Application_interpolation}
\end{figure}

Figure~\ref{fig:Application_interpolation} illustrates the posterior summaries of $Y_{20,1,t}$ for site $\uvec{\mathbf{s}}_{20}$ and Response~1. All four models recover the general temporal pattern of the observed series, including the increase in variability toward the final months of the year. However, the differences in predictive uncertainty are substantial. The isotropic models, particularly $\mathcal{M}_{1}$ and $\mathcal{M}_{2}$, produce visibly wider 95\% posterior intervals throughout most of the year, reflecting greater uncertainty in interpolation. The anisotropic models, by contrast, yield more concentrated intervals while maintaining posterior means that remain close to the observed trajectory. Among them, $\mathcal{M}_{4}$ provides the best overall balance between tracking the local fluctuations and controlling predictive uncertainty. It is especially competitive during the more challenging period around October and November, when concentrations become more volatile and all models face greater difficulty. Therefore, the graphical evidence is fully compatible with the numerical superiority of $\mathcal{M}_{4}$ in PMSE, CRPS, and the interval score for Response~1.

Table~\ref{tab:Application_model-checking} presents posterior predictive checking, residual diagnostics, execution times, and effective sample sizes based on the unnormalized log-posterior distribution. The empirical coverage probability (ECP) values show that the anisotropic models provide substantially improved predictive calibration for Response~1, with coverages around $0.88$, whereas the isotropic models remain notably lower, especially $\mathcal{M}_{2}$. For Response~2, all models perform reasonably well, but again the best values are attained by the anisotropic specifications, particularly $\mathcal{M}_{4}$. Although some ECP values slightly exceed the nominal level, the overall pattern suggests that the deformed models produce predictive intervals that are more reliable than those from the isotropic alternatives.

The residual summaries lead to the same conclusion. Under $\mathcal{M}_{3}$ and $\mathcal{M}_{4}$, the residual mean and median are much closer to zero than under $\mathcal{M}_{1}$ and $\mathcal{M}_{2}$, indicating reduced systematic bias. The upper quartile and maximum residuals are also smaller for the anisotropic models, showing that the largest positive predictive discrepancies are attenuated once deformation is incorporated. In practical terms, this means that the isotropic models tend to miss some higher observed concentrations more severely, whereas the anisotropic models are better calibrated across the distribution.

From a computational perspective, the gain in predictive performance is obtained at only a modest additional cost. The execution times for $\mathcal{M}_{3}$ and $\mathcal{M}_{4}$ are only about three minutes higher than those of $\mathcal{M}_{1}$ and $\mathcal{M}_{2}$. At the same time, the ESS values remain satisfactory for all models and are actually higher for the anisotropic specifications than for the isotropic ones, with $\mathcal{M}_{3}$ attaining the largest ESS and $\mathcal{M}_{4}$ also performing well. Hence, the more flexible models are not only more accurate, but also computationally viable in this real-data setting.

\begin{table}[htbp!]
\centering
\caption{Posterior predictive checking via empirical coverage probability (ECP), residual analysis, execution time, and effective sample size (ESS) based on the unnormalized log-posterior distribution, across models ($\mathcal{M}_1$--$\mathcal{M}_4$). Results from the application in Section~\ref{sec:Application}.}
\label{tab:Application_model-checking}
\begin{tabular}{lrrrr}
\toprule
\textbf{Metric} & $\mathcal{M}_{1}$ & $\mathcal{M}_{2}$ & $\mathcal{M}_{3}$ & $\mathcal{M}_{4}$ \\
\midrule
\textit{ECP} & & & & \\
Response 1 & 0.8009 & 0.7763 & 0.8875 & 0.8830 \\
Response 2 & 0.9243 & 0.9172 & 0.9536 & 0.9585 \\
\midrule
\textit{Residual statistics} & & & & \\
Mean     & 0.4676  & 0.5125  & 0.1795  & 0.1614 \\
Median   & 0.2157  & 0.2356  & 0.0296  & 0.0287 \\
1st quartile & -0.4605 & -0.4708 & -0.5238 & -0.5919 \\
3rd quartile & 1.0779  & 1.1601  & 0.6601  & 0.6925 \\
Minimum  & -3.3345 & -3.8196 & -2.9533 & -3.0526 \\
Maximum  & 14.7325 & 14.9823 & 11.1625 & 11.5209 \\
\midrule
\textit{Bayesian computation} & & & & \\
Execution time (min.) & 29.0 & 29.3 & 32.2 & 32.0 \\
ESS (log-post.)    & 397.0 & 333.0 & 584.0 & 506.4 \\
\bottomrule
\end{tabular}
\end{table}

Overall, the application provides consistent evidence in favor of the anisotropic formulations. The deformed models $\mathcal{M}_{3}$ and $\mathcal{M}_{4}$ improve goodness-of-fit, predictive accuracy, interval calibration, and residual behavior relative to the isotropic alternatives. Among them, $\mathcal{M}_{4}$ delivers the best overall performance, combining the benefits of spatial deformation with explicit cross-response dependence. These results indicate that, for the Central Valley PM data, accounting for anisotropy is essential, while modeling the contemporaneous association between PM\textsubscript{10} and PM\textsubscript{2.5} yields an additional predictive gain.

%--- Section ---%
\section{Conclusions}\label{sec:Conclusions}

In this paper, we developed a Bayesian spatiotemporal model to jointly analyze two or more response variables measured at fixed spatial locations and discrete, equally spaced time points. The main contribution lies in relaxing the assumption of spatial isotropy by introducing a spatial deformation component within a matrix-variate dynamic modeling framework. The model also accommodates incomplete response matrices under the assumption that missing values are Missing Completely at Random (MCAR).

Our findings, based on two complementary simulation studies and a real-data application, provide consistent evidence on the role of spatial deformation in multivariate spatiotemporal modeling. The first simulation study is primarily designed to assess parameter recovery and structural identifiability. In this setting, the proposed framework is able to recover the main features of the deformation and covariance structure, supporting the validity of the modeling assumptions and inference procedure. The second simulation study focuses on predictive performance under more challenging conditions, including smaller sample sizes and higher proportions of missing data.

The results from the second simulation study indicate that incorporating spatial deformation substantially improves interpolation accuracy in settings with matrix-variate responses. At the same time, we observed that small sample sizes and large proportions of missing data may adversely affect both parameter estimation and predictive performance. In particular, under more challenging configurations (e.g., $T=100$ with $\gamma=0.25$), the most flexible specification with spatial deformation ($\mathcal{M}_4$) may lose predictive accuracy relative to simpler models, reflecting the increased difficulty of estimating richer dependence structures with limited information. These findings highlight the importance of balancing model flexibility with data availability.

In the empirical application discussed in Section~\ref{sec:Application}, the proposed model achieved the best overall interpolation performance, particularly for the first response variable. A comparison between the anisotropic models $\mathcal{M}_3$ and $\mathcal{M}_4$ shows that their predictive performance is very similar in both the second simulation study and the real-data application (see Tables~\ref{tab:Sim2_model-comparison}, \ref{tab:Sim2_IS}, and~\ref{tab:Application_model-comparison}). The main improvements arise from the introduction of spatial deformation, whereas the additional flexibility provided by modeling cross-variable dependence yields comparatively smaller gains. This pattern is also reflected in posterior predictive summaries (Figures~\ref{fig:Sim2_interpolation} and~\ref{fig:Application_interpolation}), where both models produce nearly indistinguishable predictive trajectories and uncertainty bands.

Posterior predictive checks based on empirical coverage probabilities and residual summaries (see Tables~\ref{tab:Sim2_model-checking} and~\ref{tab:Application_model-checking}) indicate that the proposed models provide a generally adequate fit to the data across the scenarios considered. Coverage levels are close to the nominal 95\% in several cases, particularly for the second response variable under the anisotropic specifications in the real-data application. Residual summaries show means close to zero and dispersion patterns broadly consistent with Gaussian assumptions, although some deviations are observed in the tails.

These results consistently indicate that spatial anisotropy is the primary driver of predictive improvement in the datasets considered, whereas cross-variable dependence plays a secondary role. This explains the similarity between models~$\mathcal{M}_3$ and~$\mathcal{M}_4$, and suggests that the main structural gains are achieved through the deformation mechanism.

From a computational perspective, the proposed framework is scalable in the temporal dimension, as inference based on FFBS grows linearly with the number of time points. This efficiency is enabled by the factorization of the dynamic regression parameters $\boldsymbol{\beta}_{0:T}$ implied by the state-space formulation. In contrast, the main computational bottleneck arises from spatial covariance operations, which scale cubically with the number of locations. Empirical results reported in Tables~\ref{tab:Sim2_model-checking} and~\ref{tab:Application_model-checking}, both based on bivariate responses, indicate that computation times remain manageable for moderate problem sizes, and that incorporating spatial deformation leads to only a moderate increase in computational cost. Complementary evidence from the first simulation study, conducted under a trivariate specification, shows that increasing the dimension of the response vector leads to longer runtimes and increased autocorrelation, as reflected in reduced effective sample sizes for larger $N$. Nevertheless, the MCMC algorithm remains stable and provides reliable inference under these configurations.

For larger spatial datasets, several approximation strategies have been proposed in the literature to alleviate the cubic computational burden of Gaussian process models. In particular, approaches based on Vecchia-type approximations, such as the hierarchical nearest-neighbor Gaussian process (NNGP) model of \citet{Datta02042016}, replace the full covariance structure with sparse conditional specifications defined over local neighborhoods. Incorporating such approximations within the proposed framework represents a promising direction for extending the methodology to large-scale spatiotemporal datasets without sacrificing its ability to capture anisotropy and multivariate dependence.

The proposed framework adopts a parsimonious multivariate specification to ensure computational tractability and identifiability. We assume that all responses share a common spatial deformation map and a common spatial range parameter, inducing a shared anisotropy structure across variables and enabling borrowing of strength that stabilizes posterior inference. However, this assumption may be restrictive in applications where different variables exhibit distinct spatial dependence scales or anisotropy patterns. In such cases, the model should be interpreted as capturing a dominant common anisotropy structure, which may lead to oversmoothing for some responses and undersmoothing for others. A natural extension would be to allow each response variable to possess its own spatial dependence structure, even in the isotropic case, providing a useful intermediate step before introducing response-specific spatial deformations.

The model also adopts a separable covariance structure across space and responses, in which spatial dependence and cross-variable dependence are modeled independently. This separability substantially reduces computational cost and enables tractable inference in multivariate settings, but it restricts the class of dependence structures that can be represented.

Regarding temporal dependence, the observation model assumes conditional independence across time given the evolving regression coefficients, so that temporal dependence in the outcomes is induced through the state evolution equation. While this is a standard and computationally efficient dynamic modeling strategy, it may be insufficient for processes exhibiting strong short-term temporal autocorrelation not captured by the latent state dynamics. Extensions incorporating additional temporal dependence structures, such as autoregressive components at the observation level or more flexible state evolution equations, could be considered in such cases.

In addition, the model assumes a common contemporaneous covariance matrix for both the observation process and the dynamic regression coefficients. This specification induces a coherent multivariate dependence structure across responses and ensures that cross-variable dependence is consistently propagated through time, while improving identifiability and stabilizing posterior inference, particularly in settings with moderate sample sizes. Allowing separate covariance matrices for the observation and state evolution components would increase modeling flexibility, but at the cost of a substantial increase in the number of parameters, potentially leading to weak identifiability and slower MCMC mixing.

The simulation settings considered in this work include relatively small spatial designs, which facilitate interpretation but may limit the detectability of anisotropy, particularly under regular grid configurations such as those used in the second simulation study. This limitation is less pronounced in the first simulation study and in the real-data application, both of which involve irregular spatial layouts. The results obtained across these settings indicate that the proposed framework is able to recover meaningful deformation structures and provide reliable inference even under limited spatial resolution. At the same time, denser and more irregular spatial designs are expected to further improve the identification of anisotropic patterns and represent an important direction for future investigation.

Finally, the proposed framework remains stable across the configurations considered, providing meaningful inference and competitive predictive performance. Effective sample sizes computed from the unnormalized log-posterior indicate that, even in challenging scenarios, MCMC chains remain well behaved, with reductions primarily attributable to increased autocorrelation rather than pathological posterior behavior. Additional diagnostic analyses (see Figure~\ref{fig:Sim2_TracePlots}) support adequate exploration of the posterior distribution.

Other natural extensions of this work include allowing different anisotropy mechanisms across response variables, relaxing the Gaussianity assumption, and exploring alternative approaches to handle anisotropy and non-MCAR missingness mechanisms.

%--- Section ---%
\section*{Acknowledgments}

This article is based on the D.Sc. thesis in Statistics of the first author \citep{Bulhoes2024}, developed at the Federal University of Rio de Janeiro (UFRJ), Brazil, under the supervision of the second and third authors. The first author acknowledges financial support from the Coordenação de Aperfeiçoamento de Pessoal de Nível Superior (CAPES), Brazil, during his doctoral studies. 
His current research is supported by the Universidade Federal da Bahia (UFBA), Brazil (Edital PRPPG~010/2024 – Programa de Apoio a Jovens Professores(as)/Pesquisadores(as) Doutores(as) – JOVEMPESQ~2024).

The second and third authors acknowledge financial support from the Conselho Nacional de Desenvolvimento Científico e Tecnológico (CNPq), Brazil, through productivity research grants (\emph{Bolsas de Produtividade em Pesquisa}). 
The third author also acknowledges financial support from the Fundação de Amparo à Pesquisa do Estado de Minas Gerais (FAPEMIG), Brazil.

%--- Section ---%
\section*{Statements and Declarations}

\bmhead{Funding}

This work was supported by the Coordenação de Aperfeiçoamento de Pessoal de Nível Superior (CAPES), Brazil, supporting the first author; by the Universidade Federal da Bahia (UFBA), Brazil (Edital PRPPG~010/2024 - JOVEMPESQ~2024), supporting the first author; by the Conselho Nacional de Desenvolvimento Científico e Tecnológico (CNPq), Brazil, through productivity research grants awarded to the second and third authors; 
and by the Fundação de Amparo à Pesquisa do Estado de Minas Gerais (FAPEMIG), Brazil, supporting the third author.

\bmhead{Competing interests}

The authors declare that they have no competing interests.

\bmhead{Data availability}

All materials required to reproduce the Bayesian analyses are provided as Supplementary Material. These include (i) the Python scripts implementing Algorithms~\ref{alg:MCMC} and \ref{alg:Interpolation} for the simulation studies (Section~\ref{sec:Simulation}), namely \texttt{Simulation1.py} and \texttt{Simulation2.py}; (ii) the corresponding script for the real-data application (Section~\ref{sec:Application}) (\texttt{Application.py}); (iii) the dataset used in the application (\texttt{DS\_CentralValley.txt}); and (iv) the file containing the spatial coordinates of the monitoring stations (\texttt{Stations\_CentralValley.txt}), as shown in Figure~\ref{fig:Application_S}. All scripts were executed using Python~3.11 \citep{Python}. 

Due to their size, the full MCMC output files are not distributed, but can be fully reproduced using the provided scripts. Post-processing and diagnostic analyses were carried out in \textsf{R} \citep{RProject} and can be replicated directly from the reproduced samples. 

To provide practical guidance on storage requirements, we report representative file sizes for each study. In Simulation~1 (Section~\ref{subsec:Sim1}), storing the full posterior output requires approximately 2.05~GB for $N=10$ and 3.75~GB for $N=20$. In Simulation~2 (Section~\ref{subsec:Sim2}), file size scales primarily with the number of time points $T$: for $T=100$, each scenario (combining a model and a missing-data proportion) occupies approximately 517~MB, whereas for $T=500$ this increases to approximately 2.48~GB per scenario. In the real-data application, the output for each model occupies approximately 2.07~GB. These values correspond to storing full posterior samples of all model parameters and latent quantities.

\begin{appendices}

%--- Section ---%
\section{Mathematical derivations}

This section collects the main analytical derivations underlying the computational procedures used in posterior inference. Subsection~\ref{subsec:Sigma} derives the full conditional distributions of the contemporaneous covariance matrix $\boldsymbol{\Sigma}$ under both the unrestricted and diagonal specifications. Subsection~\ref{subsec:sigma2d} establishes the full conditional distributions of the deformation variance parameters collected in $\boldsymbol{\sigma}_{d}^{2}$. Finally, Subsection~\ref{subsec:D_gradient} derives the gradient of the log-target density with respect to the deformation matrix $\mathbf{D}$, which is required for the implementation of the No-U-Turn Sampler (NUTS).

\subsection[{Full conditional distribution of $\Sigma$}]{Full conditional distribution of $\boldsymbol{\Sigma}$}
\label{subsec:Sigma}

%In the derivations presented in Subsections~\ref{subsubsec:Sigma_Complete} and~\ref{subsubsec:Sigma_Diagonal}, we adopt the matrix representation of the observations and latent states given in \eqref{eq:Observation_Mat}, \eqref{eq:Evolution_Mat}, and \eqref{eq:InitialInformation_Mat}.

\subsubsection[{Unrestricted case}]{Unrestricted case}
\label{subsubsec:Sigma_Complete}

If $\boldsymbol{\Sigma} \sim \sfIW_{q}(a_{\boldsymbol{\Sigma}}, \mathbf{b}_{\boldsymbol{\Sigma}})$, then the full conditional density of $\boldsymbol{\Sigma}$ is given by
\begin{equation*}
\begin{array}{l}
f(\boldsymbol{\Sigma} \mid\mathbf{y}, \boldsymbol{\beta}_{0:T}, \mathbf{W}, \phi, \mathbf{D})
\propto
f(\boldsymbol{\Sigma})
f(\boldsymbol{\beta}_{0} \mid  \boldsymbol{\Sigma}) \\
\times \prod\limits_{t=1}^{T}
\Big[
f\boldsymbol{\beta}_{t} \mid \boldsymbol{\beta}_{t-1}, \mathbf{W}, \boldsymbol{\Sigma})
f(\mathbf{y}_{t} \mid \boldsymbol{\beta}_{t}, \phi, \mathbf{D}, \boldsymbol{\Sigma})
\Big]
\\[6pt]
\propto
|\boldsymbol{\Sigma}|^{-\left(q + \frac{a_{\boldsymbol{\Sigma}}}{2}\right)}
\exp\!\left\{-\tfrac{1}{2}\tr[\mathbf{b}_{\boldsymbol{\Sigma}}\boldsymbol{\Sigma}^{-1}]\right\}
\mathds{1}_{\SPD(q)}(\boldsymbol{\Sigma})
\\[6pt]
\times\;
|\boldsymbol{\Sigma}|^{-\frac{p}{2}}
\exp\!\left\{-\tfrac{1}{2}\tr[(\boldsymbol{\beta}_{0}-\mathbf{M}_{0})^{\top}\mathbf{C}_{0}^{-1}(\boldsymbol{\beta}_{0}-\mathbf{M}_{0})\boldsymbol{\Sigma}^{-1}]\right\}
\\[6pt]
\times\;
|\boldsymbol{\Sigma}|^{-\frac{Tp}{2}}
\exp\!\left\{-\tfrac{1}{2}\sum\limits_{t=1}^{T}
\tr[(\boldsymbol{\beta}_{t}-\mathbf{G}_{t}\boldsymbol{\beta}_{t-1})^{\top}\mathbf{W}^{-1}(\boldsymbol{\beta}_{t}-\mathbf{G}_{t}\boldsymbol{\beta}_{t-1})\boldsymbol{\Sigma}^{-1}]
\right\}
\\[6pt]
\times\;
|\boldsymbol{\Sigma}|^{-\frac{TN}{2}}
\exp\!\left\{-\tfrac{1}{2}\sum\limits_{t=1}^{T}
\tr[(\mathbf{y}_{t}-\mathbf{X}_{t}\boldsymbol{\beta}_{t})^{\top}\mathbf{B}^{-1}(\mathbf{y}_{t}-\mathbf{X}_{t}\boldsymbol{\beta}_{t})\boldsymbol{\Sigma}^{-1}]
\right\}
\\[6pt]
\propto
|\boldsymbol{\Sigma}|^{-\left(q + \frac{a'_{\boldsymbol{\Sigma}}}{2}\right)}
\exp\!\left\{-\tfrac{1}{2}\tr[\mathbf{b}'_{\boldsymbol{\Sigma}}\boldsymbol{\Sigma}^{-1}]\right\}
\mathds{1}_{\SPD(q)}(\boldsymbol{\Sigma}),
\end{array}
\end{equation*}
which implies that $[\boldsymbol{\Sigma} \mid \mathbf{Y} = \mathbf{y}, \boldsymbol{\beta}_{0:T}, \mathbf{W}, \phi, \mathbf{D}]
\sim \sfIW_{q}(a'_{\boldsymbol{\Sigma}}, \mathbf{b}'_{\boldsymbol{\Sigma}})$,
where $a'_{\boldsymbol{\Sigma}}$
and
$\mathbf{b}'_{\boldsymbol{\Sigma}}$ are respectively given in \eqref{eq:updated_a-Sigma} and \eqref{eq:updated_b-Sigma}.

\subsubsection[{Diagonal case}]{Diagonal case}
\label{subsubsec:Sigma_Diagonal}

Suppose $f(\boldsymbol{\Sigma}) = \prod_{i=1}^{q} f(\Sigma_{i,i})$ and assign $\Sigma_{i,i} \sim \sfIG(a_{\Sigma_{i,i}}, b_{\Sigma_{i,i}})$ for all $i \in \{1,\ldots,q\}$. Noting that $\boldsymbol{\Sigma}^{-1} = \diag\{\Sigma_{1,1}^{-1},\ldots,\Sigma_{q,q}^{-1}\}$, the full conditional density of $\boldsymbol{\Sigma}$ is given by 
\begin{equation*}
\begin{array}{l}
f(\boldsymbol{\Sigma} \mid \mathbf{y}, \boldsymbol{\beta}_{0:T}, \mathbf{W}, \phi, \mathbf{D})
\propto
\Bigl[\prod\limits_{i=1}^{q} f(\Sigma_{i,i})\Bigr]
f(\boldsymbol{\beta}_{0} \mid \boldsymbol{\Sigma}) \\
\times \prod\limits_{t=1}^{T}\!\Bigl[
f(\boldsymbol{\beta}_{t}\mid \boldsymbol{\beta}_{t-1}, \mathbf{W}, \boldsymbol{\Sigma})
f(\mathbf{y}_{t}\mid \boldsymbol{\beta}_{t}, \phi, \mathbf{D}, \boldsymbol{\Sigma})
\Bigr]
\\
\propto
\Bigl(\prod\limits_{i=1}^{q} \Sigma_{i,i}^{-a_{\Sigma_{i,i}}-1}\Bigr)
\exp\!\Bigl\{-\sum\limits_{i=1}^{q}\frac{b_{\Sigma_{i,i}}}{\Sigma_{i,i}}\Bigr\}
\prod\limits_{i=1}^{q}\mathds{1}_{(0,\infty)}(\Sigma_{i,i})
\\
\times\;
\Bigl(\prod\limits_{i=1}^{q}\Sigma_{i,i}^{-\frac{p}{2}}\Bigr)
\exp\!\left\{-\tfrac{1}{2}\tr\!\Bigl[
\boldsymbol{\Sigma}^{-1}
(\boldsymbol{\beta}_{0}-\mathbf{M}_{0})^{\top}\mathbf{C}_{0}^{-1}(\boldsymbol{\beta}_{0}-\mathbf{M}_{0})
\Bigr]\right\}
\\\times\;
\Bigl(\prod\limits_{i=1}^{q}\Sigma_{i,i}^{-\frac{Tp}{2}}\Bigr)
\exp\!\left\{-\tfrac{1}{2}\tr\!\Bigl[
\boldsymbol{\Sigma}^{-1}
\sum\limits_{t=1}^{T}(\boldsymbol{\beta}_{t}-\mathbf{G}_{t}\boldsymbol{\beta}_{t-1})^{\top}\mathbf{W}^{-1}(\boldsymbol{\beta}_{t}-\mathbf{G}_{t}\boldsymbol{\beta}_{t-1})
\Bigr]\right\}
\\
\times\;
\Bigl(\prod\limits_{i=1}^{q}\Sigma_{i,i}^{-\frac{TN}{2}}\Bigr)
\exp\!\left\{-\tfrac{1}{2}\tr\!\Bigl[
\boldsymbol{\Sigma}^{-1}
\sum\limits_{t=1}^{T}(\mathbf{y}_{t}-\mathbf{X}_{t}\boldsymbol{\beta}_{t})^{\top}\mathbf{B}^{-1}(\mathbf{y}_{t}-\mathbf{X}_{t}\boldsymbol{\beta}_{t})
\Bigr]\right\}
\\
\propto \prod\limits_{i=1}^{q} f(\Sigma_{i,i} \mid \mathbf{y}, \boldsymbol{\beta}_{0:T}, \mathbf{W}, \phi, \mathbf{D}),
\end{array}
\end{equation*}
where
\begin{equation}
f(\Sigma_{i,i} \mid \mathbf{y}, \boldsymbol{\beta}_{0:T}, \mathbf{W}, \phi, \mathbf{D})
\propto
\Sigma_{i,i}^{-(a'_{\Sigma_{i,i}}+1)}
\exp\!\left\{-\frac{b'_{\Sigma_{i,i}}}{\Sigma_{i,i}}\right\}
\mathds{1}_{(0,\infty)}(\Sigma_{i,i}),
\end{equation}
which implies that $[\Sigma_{i,i} \mid \mathbf{Y} = \mathbf{y}, \boldsymbol{\beta}_{0:T}, \mathbf{W}, \phi, \mathbf{D}]
\sim \sfIG(a'_{\Sigma_{i,i}}, b'_{\Sigma_{i,i}})$, with $a'_{\Sigma_{i,i}}$ and $b'_{\Sigma_{i,i}}$ given in \eqref{eq:updated_a-Sigma_i} and \eqref{eq:updated_b-Sigma_i}, respectively.

\subsection[{Full conditional distribution of $\sigma_{d}^{2}$}]
{Full conditional distribution of $\boldsymbol{\sigma}_{d}^{2}$}
\label{subsec:sigma2d}

Suppose $\boldsymbol{\sigma}_{d}^{2} = \diag\{\sigma_{d_{1,1}}^{2}, \sigma_{d_{2,2}}^{2}\}$. This diagonal specification in \eqref{eq:D_Prior} implies that the rows of the deformation matrix are conditionally independent. Let $\uvec{\mathbf{d}}_{m,\cdot}^{\top}$ and $\uvec{\mathbf{s}}_{m,\cdot}^{\top}$ denote, respectively, the $m$th rows of $\mathbf{D}$ and $\mathbf{S}$, viewed as vectors in $\mathds{R}^{N}$. Then, for $m \in \{1,2\}$,
\[
\left[\uvec{\mathbf{d}}_{m,\cdot} \mid \sigma_{d_{m,m}}^{2}\right]
\sim
\sfNor_{N}\!\left(
\uvec{\mathbf{s}}_{m,\cdot}, \;
\sigma_{d_{m,m}}^{2}\mathbf{R}_{d}
\right).
\]

It follows that the joint density factorizes as
\[
f(\mathbf{D} \mid \boldsymbol{\sigma}_{d}^{2})
=
\prod_{m=1}^{2}
f(\uvec{\mathbf{d}}_{m,\cdot} \mid \sigma_{d_{m,m}}^{2}),
\]
and each term in the product is proportional to
\[
f(\uvec{\mathbf{d}}_{m,\cdot} \mid \sigma_{d_{m,m}}^{2})
\propto
(\sigma_{d_{m,m}}^{2})^{-N/2}
\exp\!\left\{
-\frac{1}{2\sigma_{d_{m,m}}^{2}}
(\uvec{\mathbf{d}}_{m,\cdot} - \uvec{\mathbf{s}}_{m,\cdot})^{\top}
\mathbf{R}_{d}^{-1}
(\uvec{\mathbf{d}}_{m,\cdot} - \uvec{\mathbf{s}}_{m,\cdot})
\right\}.
\]

Assume independent inverse-gamma priors for $\sigma_{d_{1,1}}^{2}$ and $\sigma_{d_{2,2}}^{2}$. Then,
\[
f(\boldsymbol{\sigma}_{d}^{2}) = \prod_{m = 1}^{2} f(\sigma_{d_{m,m}}^{2}).
\]

Since the prior distribution \eqref{eq:sigma2d_Prior} has kernel
\[
f(\sigma_{d_{m,m}}^{2}) \propto (\sigma_{d_{m,m}}^{2})^{-a_{\sigma_{d,m}}-1}
\exp\left\{-\frac{b_{\sigma_{d,m}}}{\sigma_{d_{m,m}}^{2}}\right\}
\mathds{1}_{(0,\infty)}(\sigma_{d_{m,m}}^{2}),
\]
it follows by conjugacy that the full conditional distributions are inverse-gamma, given by
\begin{equation*}
[\sigma_{d_{m,m}}^2\mid \mathbf{D}]
\sim
\sfIG\!\left(
a_{\sigma_{d,m}}+\frac{N}{2},
\;
b_{\sigma_{d,m}}+\frac{1}{2}\mathbf{\Delta}_{m,\cdot}\mathbf{R}_d^{-1}\mathbf{\Delta}_{m,\cdot}^{\top}
\right),
\end{equation*}
where $\mathbf{\Delta}_{m,\cdot} = (\uvec{\mathbf{d}}_{m,\cdot} - \uvec{\mathbf{s}}_{m,\cdot})^{\top}$.

\subsection[{Gradient of the log-target with respect to $\mathbf{D}$}]
{Gradient of the log-target with respect to $\mathbf{D}$}
\label{subsec:D_gradient}

This section provides the analytical gradient of the log-target density \eqref{eq:D_FCD} with respect to the deformation matrix $\mathbf{D}$. This gradient is required for the implementation of the No-U-Turn Sampler (NUTS).

We write $\ell(\mathbf{D}; \mathbf{y})$ to emphasize the dependence of the likelihood \eqref{eq:Likelihood} on $\mathbf{D}$. Recall that the log-target density can be written as
\[
\ln f(\mathbf{D} \mid \mathbf{y}, \boldsymbol{\beta}_{0:T}, \phi, \boldsymbol{\sigma}_{d}^{2}, \boldsymbol{\Sigma})
=
\ln f(\mathbf{D} \mid \boldsymbol{\sigma}_{d}^{2})
+
\ln \ell(\mathbf{D}; \mathbf{y})
+\textnormal{Constant},
\]
where
\[
\ln f(\mathbf{D} \mid \boldsymbol{\sigma}_{d}^{2})
=
-\frac{1}{2}
\tr\!\left[
(\mathbf{D}-\mathbf{S})^\top
\boldsymbol{\sigma}_d^{-2}
(\mathbf{D}-\mathbf{S})
\mathbf{R}_d^{-1}
\right] + \textnormal{Constant}
\]
and
\[
\ln \ell(\mathbf{D}; \mathbf{y})
=
-\frac{Tq}{2}\ln\det\mathbf{B}
-\frac{1}{2}
\tr\!\left(\mathbf{B}^{-1}\mathbf{Q}\right) + \textnormal{Constant}.
\]

\subsubsection*{Gradient of the prior term}

Using standard matrix calculus results, the gradient of the prior term with respect to $\mathbf{D}$ is given by
\[
\nabla_{\mathbf{D}} \ln f(\mathbf{D} \mid \boldsymbol{\sigma}_{d}^{2})
=
-\boldsymbol{\sigma}_d^{-2}
(\mathbf{D}-\mathbf{S})
\mathbf{R}_d^{-1}.
\]

\subsubsection*{Derivative of the spatial correlation matrix}

Using the notation $\uvec{\mathbf{d}}_{n} = d(\uvec{\mathbf{s}}_{n})$, we rewrite \eqref{eq:B} as
$B_{n,n'} = \exp\{-\phi\|\uvec{\mathbf{d}}_n - \uvec{\mathbf{d}}_{n'}\|\}$.

The gradient $\nabla_{\mathbf{D}} B_{n,n'}$ is a $2 \times N$ matrix, with nonzero entries only in the columns corresponding to indices $n$ and $n'$.

For each pair $(n,n')$ with $n \neq n'$ and for $m \in \{1,2\}$,
\[
\frac{\partial B_{n,n'}}{\partial D_{m,n}}
=
-\phi B_{n,n'}
\frac{(D_{m,n} - D_{m,n'})}
{\|\uvec{\mathbf{d}}_n - \uvec{\mathbf{d}}_{n'}\|},
\]
\[
\frac{\partial B_{n,n'}}{\partial D_{m,n'}}
=
-\phi B_{n,n'}
\frac{(D_{m,n'} - D_{m,n})}
{\|\uvec{\mathbf{d}}_n - \uvec{\mathbf{d}}_{n'}\|},
\]
and all other partial derivatives are equal to zero. The diagonal elements satisfy $\partial B_{n,n}/\partial D_{m,n} = 0$.

\subsubsection*{Gradient of the likelihood term}

The gradient of the likelihood term with respect to $\mathbf{D}$ is obtained using matrix differential identities. Specifically,
\[
\nabla_{\mathbf{D}} \ln \ell(\mathbf{D}; \mathbf{y})
=
-\frac{Tq}{2} \, \nabla_{\mathbf{D}} \ln\det\mathbf{B}
-\frac{1}{2} \, \nabla_{\mathbf{D}} \tr(\mathbf{B}^{-1}\mathbf{Q}).
\]

Using the identities
\[
\nabla_{\mathbf{B}} \ln\det\mathbf{B} = \mathbf{B}^{-1},
\qquad
\nabla_{\mathbf{B}} \tr(\mathbf{B}^{-1}\mathbf{Q})
=
- \mathbf{B}^{-1}\mathbf{Q}\mathbf{B}^{-1},
\]
it follows that
\[
\nabla_{\mathbf{D}} \ln \ell(\mathbf{D}; \mathbf{y})
=
\sum_{n=1}^{N}
\sum_{n'=1}^{N}
H_{n,n'} \, \nabla_{\mathbf{D}} B_{n,n'},
\]
where
\[
\mathbf{H}
=
-\frac{Tq}{2}\mathbf{B}^{-1}
+\frac{1}{2}\mathbf{B}^{-1}\mathbf{Q}\mathbf{B}^{-1}.
\]

\subsubsection*{Full gradient}

Combining the previous results, the full gradient of the log-target density is
\[
\nabla_{\mathbf{D}} \ln f(\mathbf{D} \mid \mathbf{y}, \boldsymbol{\beta}_{0:T}, \phi, \boldsymbol{\sigma}_{d}^{2}, \boldsymbol{\Sigma})
=
\nabla_{\mathbf{D}} \ln f(\mathbf{D} \mid \boldsymbol{\sigma}_{d}^{2})
+
\nabla_{\mathbf{D}} \ln \ell(\mathbf{D}; \mathbf{y}).
\]

Define $\uvec{\mathbf{D}}_{\free} = \vect(\mathbf{D}_{\free})$. In practice, the gradient with respect to the free deformation coordinates $\uvec{\mathbf{D}}_{\free}$ is obtained by extracting the corresponding components of $\nabla_{\mathbf{D}} \ln f(\mathbf{D} \mid \mathbf{y}, \boldsymbol{\beta}_{0:T}, \phi, \boldsymbol{\sigma}_{d}^{2}, \boldsymbol{\Sigma})$ and stacking them into a vector, which is then used within the leapfrog integrator in the NUTS algorithm.

%--- Section ---%
\section{Algorithms}\label{app:Algorithms}

This section summarizes the computational procedures used for posterior inference under the proposed model. The overall strategy is based on a hybrid MCMC scheme targeting the joint posterior distribution 
$f(\uvec{\mathbf{y}}_{\mis}, \boldsymbol{\theta} \mid \uvec{\mathbf{y}}_{\obs})$, alternating between data imputation and parameter updates within an iterative loop.

Algorithm~\ref{alg:Missing} describes a single Gibbs step for sampling the missing responses from their full conditional distributions, returning a completed dataset to be used in subsequent updates. Algorithm~\ref{alg:Update_W} presents a single Metropolis--Hastings update for the state-evolution covariance matrix $\mathbf{W}$, based on the marginal likelihood obtained by integrating out the latent state trajectory. These components are embedded within the full MCMC scheme described in Algorithm~\ref{alg:MCMC}, which iterates until convergence.

Finally, Algorithm~\ref{alg:Interpolation} shows how posterior samples are used to construct a Monte Carlo approximation to the predictive distribution $f(\uvec{\mathbf{y}}_{\interp} \mid \uvec{\mathbf{y}}_{\obs})$ for interpolation at ungauged locations.

\begin{algorithm}[H]
  \caption{Sampling the missing responses from their full conditional distribution.}
  \label{alg:Missing}
  \begin{algorithmic}[1]

    \Require Current state $\boldsymbol{\theta}^{(k)} \in \boldsymbol{\Theta}$, 
      observed vectorized responses $\uvec{\mathbf{y}}_{\obs} = \{\uvec{\mathbf{y}}_{t,\obs}\}_{t=1}^T$, 
      index sets $\mathcal{T}_{\OMV}$ and $\mathcal{T}_{\SMV}$,
      and matrices $\{\mathbf{P}_t,\mathbf{L}_{t,\obs},\mathbf{L}_{t,\mis}\}_{t=1}^T$.
    \Ensure Completed vectorized dataset 
      $\uvec{\mathbf{y}}^{(k+1)} 
      = \big\{\uvec{\mathbf{y}}_{t,\obs}, \uvec{\mathbf{y}}_{t,\mis}^{(k+1)}\big\}_{t=1}^{T}$.

    \For{$t = 1,\ldots,T$}

       \If{$t \in \mathcal{T}_{\OMV}$}
         \State Sample $\uvec{\mathbf{y}}_{t,\mis}^{(k+1)}$ 
                from the marginal distribution in~\eqref{eq:Dist_Mis}.
         \State Set 
                $\uvec{\mathbf{y}}_{t}^{(k+1)}
                \gets \uvec{\mathbf{y}}_{t,\mis}^{(k+1)}$.

       \ElsIf{$t \in \mathcal{T}_{\SMV}$}
         \State Sample $\uvec{\mathbf{y}}_{t,\mis}^{(k+1)}$ 
                from the conditional distribution in~\eqref{eq:Dist_Mis_Obs}.
         \State Construct 
         $\uvec{\mathbf{y}}_{t}^{(k+1)} 
             = \mathbf{P}_{t}^{-1}
               \begin{bmatrix}
                 \uvec{\mathbf{y}}_{t,\obs} \\
                 \uvec{\mathbf{y}}_{t,\mis}^{(k+1)}
               \end{bmatrix}$.

       \Else
         \State Set 
         $\uvec{\mathbf{y}}_{t}^{(k+1)} \gets \uvec{\mathbf{y}}_{t,\obs}$.
       \EndIf

    \EndFor

    \State \Return 
      $\uvec{\mathbf{y}}^{(k+1)}$.

  \end{algorithmic}
\end{algorithm}

\begin{algorithm}[H]
\caption{Marginal Metropolis--Hastings update for $\mathbf{W}$.}
\label{alg:Update_W}
\begin{algorithmic}[1]

\Require Current state $\mathbf{W}^{(k)}$, other model parameters, initial state moments $\{\mathbf{M}_0,\mathbf{C}_0\}$, tuning variances $\{\delta_j^2\}_{j=1}^{p}$, hyperparameters $\{\tau_{j}^{2}\}_{j=1}^{p}$ and $\lambda$, and data $\{\mathbf{Y}_{t}, \mathbf{X}_t, \mathbf{G}_{t}\}_{t=1}^T$.
\Ensure Updated state $\mathbf{W}^{(k+1)}$, together with $\ln L^{(k+1)}$ and $\ln P^{(k+1)}$.

\For{$j = 1,\ldots,p$}
    \State Set $\eta_j^{(k)} = \ln W_{j,j}^{(k)}$.
    \State Sample $\eta_j^\star \sim \sfNor(\eta_j^{(k)}, \delta_j^2)$.
\EndFor

\State Set $\mathbf{W}^\star = \diag\{\exp(\eta_1^\star), \ldots, \exp(\eta_p^\star)\}$.
\State Initialize $\ln L^\star \gets 0$ and set $\{\mathbf{M}_0^\star,\mathbf{C}_0^\star\} = \{\mathbf{M}_0,\mathbf{C}_0\}$.

\For{$t = 1,\ldots,T$}
    \State Compute:
    \begin{eqnarray*}
    \mathbf{a}_t^\star &=& \mathbf{G}_t \mathbf{M}_{t-1}^\star, \\
    \mathbf{E}_t^\star &=& \mathbf{G}_t \mathbf{C}_{t-1}^\star \mathbf{G}_t^\top + \mathbf{W}^\star, \\
    \mathbf{f}_t^\star &=& \mathbf{X}_t \mathbf{a}_t^\star, \\
    \mathbf{Q}_t^\star &=& \mathbf{X}_t \mathbf{E}_t^\star \mathbf{X}_t^\top + \mathbf{B}, \\
    \ln L^\star &\gets& \ln L^\star -\frac{q}{2}\ln\det\mathbf{Q}_t^\star
    -\frac{1}{2}\tr\!\left(
    \boldsymbol{\Sigma}^{-1}
    (\mathbf{Y}_{t} - \mathbf{f}_t^\star)^\top
    (\mathbf{Q}_t^\star)^{-1}
    (\mathbf{Y}_{t} - \mathbf{f}_t^\star)
    \right), \\
    \mathbf{A}_t^\star &=& \mathbf{E}_t^\star \mathbf{X}_t^\top (\mathbf{Q}_t^\star)^{-1}, \\
    \mathbf{M}_t^\star &=& \mathbf{a}_t^\star + \mathbf{A}_t^\star (\mathbf{Y}_{t} - \mathbf{f}_t^\star), \\
    \mathbf{C}_t^\star &=& \mathbf{E}_t^\star - \mathbf{A}_t^\star \mathbf{X}_t \mathbf{E}_t^\star.
    \end{eqnarray*}
\EndFor

\State Compute:
\begin{align*}
\ln P^\star
&=
\sum_{j=1}^{p}
\left[
-(\lambda+1)\ln\left(1+\frac{\exp(\eta_j^\star)}{\tau_j^2}\right)
+ \eta_j^\star
\right], \\
\alpha_{\text{MH}}
&=
\min\left\{
1,\,
\exp\!\Big[
(\ln L^\star + \ln P^\star)
-
(\ln L^{(k)} + \ln P^{(k)})
\Big]
\right\}.
\end{align*}

\State Sample $u \sim \sfUnif(0,1)$.
\If{$u < \alpha_{\text{MH}}$}
    \State Set $\mathbf{W}^{(k+1)} \gets \mathbf{W}^\star$, $\ln L^{(k+1)} \gets \ln L^\star$, and $\ln P^{(k+1)} \gets \ln P^\star$.
\Else
    \State Set $\mathbf{W}^{(k+1)} \gets \mathbf{W}^{(k)}$, $\ln L^{(k+1)} \gets \ln L^{(k)}$, and $\ln P^{(k+1)} \gets \ln P^{(k)}$.
\EndIf

\If{$k$ is in burn-in period}
    \State Adapt $\{\delta_j\}_{j=1}^{p}$ to target a desired acceptance rate.
\EndIf

\State \Return $\{\mathbf{W}^{(k+1)}, \ln L^{(k+1)}, \ln P^{(k+1)}\}$.

\end{algorithmic}
\end{algorithm}

\begin{algorithm}[H]
\caption{Hybrid MCMC sampler targeting 
$f(\uvec{\mathbf{y}}_{\mis}, \boldsymbol{\theta} \mid \uvec{\mathbf{y}}_{\obs})$.}
\label{alg:MCMC}
\begin{algorithmic}[1]

\Require Initial values $\boldsymbol{\theta}^{(0)} \in \boldsymbol{\Theta}$, initial imputed values $\uvec{\mathbf{y}}_{\mis}^{(0)} \in \mathcal{Y}_{\mis}$, observed vectorized data $\uvec{\mathbf{y}}_{\obs}$, fixed matrices $\{\mathbf{X}_t, \mathbf{G}_t\}_{t=1}^T$, model indicator $\mathcal{M}_{h}$ ($h \in \{1,2,3,4\}$), and hyperparameters. \Comment{See Section~\ref{sec:Model_Comparison-Checking} for a description of $h$.}

\State Set $k \gets 0$.
\State Form the initial completed vectorized data $\uvec{\mathbf{y}}^{(0)} = \{\uvec{\mathbf{y}}_{\obs}, \uvec{\mathbf{y}}_{\mis}^{(0)}\}$.

\Repeat

    \Statex \textsc{Imputation step}
    \State Work with the vector parameterization of the latent states, writing them as $\uvec{\boldsymbol{\beta}}_{0:T}^{(k)}$ for the imputation update.
    \State Apply Algorithm~\ref{alg:Missing} given $\{\boldsymbol{\theta}^{(k)}, \uvec{\mathbf{y}}_{\obs}\}$ to obtain the completed vectorized data 
    $\uvec{\mathbf{y}}^{(k+1)} = \{\uvec{\mathbf{y}}_{\obs}, \uvec{\mathbf{y}}_{\mis}^{(k+1)}\}$.
    \State Form the collection $\mathbf{y}^{(k+1)} = \{\mathbf{y}_{t}^{(k+1)}\}_{t=1}^{T}$, where each $N \times q$ matrix $\mathbf{y}_{t}^{(k+1)}$ is reconstructed from $\uvec{\mathbf{y}}_{t}^{(k+1)}$.

    \Statex \textsc{Parameter update step}
    \State Work with the matrix parameterization of the latent states, writing them as $\boldsymbol{\beta}_{0:T}^{(k)}$ for the subsequent updates.

    \If{$h \in \{2,4\}$}
        \State Sample $\boldsymbol{\Sigma}^{(k+1)}$ from \eqref{eq:Sigma_Full}, given $\{\mathbf{y}^{(k+1)}, \mathbf{W}^{(k)}, \boldsymbol{\beta}_{0:T}^{(k)}, \phi^{(k)}, \mathbf{D}^{(k)}\}$.
    \Else
        \For{$i = 1,\ldots,q$}
            \State Sample $\Sigma_{i,i}^{(k+1)}$ from \eqref{eq:Sigma_Diag}, given $\{\mathbf{y}^{(k+1)}, \mathbf{W}^{(k)}, \boldsymbol{\beta}_{0:T}^{(k)}, \phi^{(k)}, \mathbf{D}^{(k)}\}$.
        \EndFor
        \State Set $\boldsymbol{\Sigma}^{(k+1)} = \diag\left\{\Sigma_{1,1}^{(k+1)}, \ldots, \Sigma_{q,q}^{(k+1)}\right\}$.
    \EndIf

    \State Apply Algorithm~\ref{alg:Update_W} to obtain $\mathbf{W}^{(k+1)}$, given $\{\mathbf{y}^{(k+1)}, \boldsymbol{\Sigma}^{(k+1)}, \phi^{(k)}, \mathbf{D}^{(k)}\}$.

    \State Sample $\boldsymbol{\beta}_{0:T}^{(k+1)}$ via FFBS from \eqref{eq:beta_FCD}, given $\{\mathbf{y}^{(k+1)}, \mathbf{W}^{(k+1)}, \boldsymbol{\Sigma}^{(k+1)}, \phi^{(k)}, \mathbf{D}^{(k)}\}$.

    \State Update $\phi^{(k+1)}$ via Metropolis--Hastings from \eqref{eq:phi_FCD}, conditional on $\{\mathbf{y}^{(k+1)}, \mathbf{W}^{(k+1)}, \boldsymbol{\Sigma}^{(k+1)}, \boldsymbol{\beta}_{0:T}^{(k+1)}, \mathbf{D}^{(k)}\}$.

    \If{$h \in \{3,4\}$}
        \For{$m \in \{1, 2\}$}
            \State Sample $\sigma_{d_{m,m}}^{2(k+1)}$ from \eqref{eq:sigma2d_m-FCD}, given $\mathbf{D}^{(k)}$.
        \EndFor
        \State Set $\boldsymbol{\sigma}_{d}^{2(k+1)} = \diag\left\{\sigma_{d_{1,1}}^{2(k+1)}, \sigma_{d_{2,2}}^{2(k+1)}\right\}$.
        \State Update $\mathbf{D}^{(k+1)}$ via NUTS targeting \eqref{eq:D_FCD}, conditional on $\{\mathbf{y}^{(k+1)}, \mathbf{W}^{(k+1)}, \boldsymbol{\Sigma}^{(k+1)}, \boldsymbol{\beta}_{0:T}^{(k+1)}, \phi^{(k+1)}, \boldsymbol{\sigma}_{d}^{2(k+1)}\}$.
    \Else
        \State Set $\mathbf{D}^{(k+1)} = \mathbf{S}$.
    \EndIf

    \State Set $k \gets k+1$.

\Until{a convergence criterion is satisfied.}

\end{algorithmic}
\end{algorithm}

\begin{algorithm}[H]
  \caption{Monte Carlo approximation of 
  $f(\uvec{\mathbf{y}}_{\interp} \mid \uvec{\mathbf{y}}_{\obs})$.}
  \label{alg:Interpolation}
  \begin{algorithmic}[1]

    \Require Posterior samples 
      $\big\{\mathbf{y}^{(k)}, \boldsymbol{\theta}^{(k)}\big\}_{k=1}^{K}$, 
      fixed matrices $\{\mathbf{X}_{t}, \mathbf{X}_{t}^{\ast}\}_{t=1}^{T}$, 
      and $\psi > 0$.
    \Ensure Monte Carlo sample 
      $\big\{\mathbf{y}_{\interp}^{(k)}\big\}_{k=1}^{K}$ 
      approximating $f(\uvec{\mathbf{y}}_{\interp} \mid \uvec{\mathbf{y}}_{\obs})$.

    \For{$k = 1,\ldots,K$}

      \State Generate the interpolated deformation 
        $\mathbf{D}^{*(k)}$ using the conditional distribution 
        in~\eqref{eq:Interpolation_D}.
      
      \For{$t = 1,\ldots,T$}
        \State Generate the interpolated response 
               $\mathbf{y}_{t}^{*(k)}$ 
               from the conditional distribution 
               in~\eqref{eq:Interpolation_Y}.
      \EndFor

      \State Set 
      $\mathbf{y}_{\interp}^{(k)}
         \gets \big\{\mathbf{y}_{t}^{*(k)}\big\}_{t=1}^{T}$.

    \EndFor

    \State \Return $\big\{\mathbf{y}_{\interp}^{(k)}\big\}_{k=1}^{K}$.

  \end{algorithmic}
\end{algorithm}

%--- Section ---%
\section{Simulation protocol details}
\label{app:SimProtocol}

This appendix provides detailed simulation protocols for the two studies described in Section~\ref{sec:Simulation}. Common elements of the data-generating mechanism are shared across studies, with study-specific configurations detailed below.

\subsection{First simulation study}
\label{app:SimProtocol_Sim1}

\begin{enumerate}
  \item \textbf{Spatial locations and deformation.} 
  Generate $N \in \{10,20\}$ spatial locations in $\mathcal{S}=[0,1]^2$ and construct $\mathbf{D}=[\uvec{\mathbf{d}}_{1}\ \cdots\ \uvec{\mathbf{d}}_{N}]$:
  \begin{enumerate}
    \item Fix $\uvec{\mathbf{s}}_{1}=(0,0)$ and $\uvec{\mathbf{s}}_{2}=(1,1)$.
    \item Sample $\uvec{\mathbf{s}}_{3},\ldots,\uvec{\mathbf{s}}_{N}$ independently from $\sfUnif([0,1]^2)$ and form $\mathbf{S}=[\uvec{\mathbf{s}}_{1}\ \cdots\ \uvec{\mathbf{s}}_{N}]$.
    \item Fix the anchor points $\uvec{\mathbf{d}}_{1}=\uvec{\mathbf{s}}_{1}$ and $\uvec{\mathbf{d}}_{2}=\uvec{\mathbf{s}}_{2}$.
    \item Let $\psi_{\simu} = 1.5$. Construct $\mathbf{R}_{d}$ with entries
    \[
    R_{n,n'} = \exp\{-\psi_{\simu}\|\uvec{\mathbf{s}}_{n} - \uvec{\mathbf{s}}_{n'}\|^{2}\}.
    \]
    \item Partition $\mathbf{S}$ and $\mathbf{R}_{d}$ as
    \[
    \mathbf{S} = [\mathbf{S}_{\anc}\ \mathbf{S}_{\free}], \qquad
    \mathbf{R}_{d} = 
    \begin{bmatrix}
      \mathbf{R}_{\anc} & \mathbf{R}_{\anc,\free} \\
      \mathbf{R}_{\anc,\free}^{\top} & \mathbf{R}_{\free}
    \end{bmatrix},
    \]
    where the subscript ``$\anc$'' refers to the anchor locations $\{1,2\}$ and ``$\free$'' to the remaining locations $\{3,\ldots,N\}$.
    \item Let $\boldsymbol{\sigma}_{d}^{2} = \diag\{0.015, 0.020\}$. Draw $\mathbf{D}_{\free} = [\uvec{\mathbf{d}}_{3}\ \cdots\ \uvec{\mathbf{d}}_{N}]$ from
    \[
    \sfNor_{2 \times (N - 2)}(\mathbf{S}_{\free}, \boldsymbol{\sigma}_{d}^{2}, \mathbf{R}_{\free}).
    \]
  \end{enumerate}

  \item \textbf{Latent states and covariance.} 
  Let $\mathbf{M}_{0,\simu} = \mathbf{0}_{2 \times 3}$ and $\mathbf{C}_{0,\simu}=0.2 \cdot \mathbf{I}_{2}$. Draw $\boldsymbol{\beta}_{0}\sim\sfNor_{2\times3}(\mathbf{M}_{0,\simu}, \mathbf{C}_{0,\simu}, \boldsymbol{\Sigma})$, where 
  \[
  \boldsymbol{\Sigma}=
  \begin{bmatrix}
  1.00 & 0.10 & 0.05\\
  0.10 & 1.00 & 0.15\\
  0.05 & 0.15 & 1.00
  \end{bmatrix}.
  \]

  \item \textbf{Time evolution, covariates, and responses.} 
  For $t=1,\ldots,T$ with $T=600$:
  \begin{enumerate}
    \item Draw $\boldsymbol{\beta}_{t}\sim\sfNor_{2\times3}(\mathbf{G}_{t}\boldsymbol{\beta}_{t-1}, \mathbf{W}, \boldsymbol{\Sigma})$, where $\mathbf{G}_{t} = \mathbf{I}_{2}$ and $\mathbf{W}=0.04 \cdot \mathbf{C}_{0,\simu}$.
    \item Generate covariate matrices 
    \[
    \mathbf{X}_{t}=
    \begin{bmatrix}
    1 & \cdots & 1\\[2pt]
    U_{1,t} & \cdots & U_{N,t}
    \end{bmatrix}^{\!\top},
    \]
    with $U_{1,t},\ldots,U_{N,t}\sim \sfUnif(0,1)$ independently.
    \item Set $\phi=0.5$. Draw $\mathbf{Y}_{t}\sim \sfNor_{N\times3}(\mathbf{X}_{t}\boldsymbol{\beta}_{t}, \mathbf{B}, \boldsymbol{\Sigma})$, where $\mathbf{B}$ has entries
    \[
    B_{n,n'}=\exp\{-\phi\|\uvec{\mathbf{d}}_{n}-\uvec{\mathbf{d}}_{n'}\|\}.
    \]
    \item Impose missingness: for each response $i\in\{1,\ldots,q\}$, randomly select $\Gamma=\gamma N$ indices from $\{1,\ldots,N\}$ (without replacement), say $n_{1}^{i},\ldots,n_{\Gamma}^{i}$, and treat $Y_{n_{1}^{i},i,t},\ldots,Y_{n_{\Gamma}^{i},i,t}$ as missing.
    \item Vectorize $\uvec{\mathbf{y}}_{t}=\vect(\mathbf{y}_{t})$ and obtain the observed subvector $\uvec{\mathbf{y}}_{t,\obs}=\mathbf{L}_{t,\obs}\mathbf{P}_{t}\uvec{\mathbf{y}}_{t}$.
  \end{enumerate}
\end{enumerate}

The resulting datasets are used to assess parameter recovery under model $\mathcal{M}_{4}$, as described in Section~\ref{subsec:Sim1}, using the MCMC algorithm detailed in Section~\ref{subsec:MCMC}.

\subsection{Second simulation study}
\label{app:SimProtocol_Sim2}

\begin{enumerate}
  \item \textbf{Spatial locations and deformation.} 
  Using the spatial configurations $\mathbf{S}$ and $\mathbf{S}^{\ast}$ shown in Figure~\ref{fig:Sim2_S}, construct $\mathbf{D}=[\uvec{\mathbf{d}}_{1}\ \cdots \ \uvec{\mathbf{d}}_{16}]$ and $\mathbf{D}^{\ast}=[\uvec{\mathbf{d}}_{17}\ \uvec{\mathbf{d}}_{18}\ \uvec{\mathbf{d}}_{19}]$:
  \begin{enumerate}
    \item Fix the two anchors: $\uvec{\mathbf{d}}_{1} = \uvec{\mathbf{s}}_{1}$ and $\uvec{\mathbf{d}}_{2} = \uvec{\mathbf{s}}_{2}$.
    \item For $n\in\{3,\ldots,19\}$, set $\uvec{\mathbf{d}}_{n}=d(\uvec{\mathbf{s}}_{n})=\boldsymbol{\Lambda}\uvec{\mathbf{s}}_{n}$, inducing geometric anisotropy via $\boldsymbol{\Lambda}^{\top}\boldsymbol{\Lambda}=\mathbf{A}$, where $\boldsymbol{\Lambda}$ is given in \eqref{eq:Lambda}.
  \end{enumerate}
  The resulting deformation (Figure~\ref{fig:Sim2_D-True}) shows axis stretching and a $45^{\circ}$ rotation at non-anchor locations.
  
  \item \textbf{Latent states and covariance.} 
  Set $\phi=0.4$ and
  \[
  \boldsymbol{\Sigma}=
  \begin{bmatrix}
  1.00 & 0.85\\[2pt]
  0.85 & 1.00
  \end{bmatrix}.
  \]
  Let $\mathbf{M}_{0,\simu}= \mathbf{0}_{2 \times 2}$ and $\mathbf{C}_{0,\simu}=0.1 \cdot \mathbf{I}_{2}$. 
  Draw $\boldsymbol{\beta}_{0}\sim\sfNor_{2\times2}(\mathbf{M}_{0,\simu}, \mathbf{C}_{0,\simu},\boldsymbol{\Sigma})$.
  Construct the block matrix
  \[
  \mathbf{B}^{\mathrm{aug}}=
  \begin{bmatrix}
  \mathbf{B} & \mathbf{B}_{\mathrm{g,u}}\\[2pt]
  \mathbf{B}_{\mathrm{u,g}} & \mathbf{B}^{\ast}
  \end{bmatrix}
  \]
  of dimension $19\times19$, where $\mathbf{B}$ ($16\times16$), $\mathbf{B}_{\mathrm{g,u}}$ ($16\times3$), $\mathbf{B}_{\mathrm{u,g}}$ ($3\times16$), and $\mathbf{B}^{\ast}$ ($3\times3$) have generic element
  \[
  B^{\mathrm{aug}}_{n,n'}=\exp\{-\phi\|d(\uvec{\mathbf{s}}_{n})-d(\uvec{\mathbf{s}}_{n'})\|\}.
  \]
  
  \item \textbf{Time evolution, covariates, and responses.} 
  For each combination $(T,\gamma)$ with $T\in\{100,500\}$ and $\gamma\in\{0.0625, 0.25\}$, iterate for $t=1,\ldots,T$:
  \begin{enumerate}
    \item Draw $\boldsymbol{\beta}_{t}\sim\sfNor_{2\times2}(\mathbf{G}_{t}\boldsymbol{\beta}_{t-1}, \mathbf{W}, \boldsymbol{\Sigma})$, where $\mathbf{G}_{t} = \mathbf{I}_{2}$ and $\mathbf{W} = 0.05 \cdot \mathbf{C}_{0,\simu}$.
    \item Generate covariate matrices 
    \[
    \mathbf{X}_{t}=
    \begin{bmatrix}
    1 & \cdots & 1\\[2pt]
    U_{1,t} & \cdots & U_{16,t}
    \end{bmatrix}^{\!\top}
    \ \ (16\times2), 
    \qquad
    \mathbf{X}_{t}^{\ast}=
    \begin{bmatrix}
    1 & 1 & 1\\[2pt]
    U_{17,t} & U_{18,t} & U_{19,t}
    \end{bmatrix}^{\!\top}
    \ \ (3\times2),
    \]
    with $U_{1,t},\ldots,U_{19,t}\sim \sfUnif(0,1)$ independently.
    \item Jointly draw
    \[
    \begin{bmatrix}\mathbf{y}_{t}\\[3pt] \mathbf{y}_{t}^{\ast}\end{bmatrix}\sim
    \sfNor_{19\times2}\!\left(
    \begin{bmatrix}\mathbf{X}_{t}\\[3pt]\mathbf{X}_{t}^{\ast}\end{bmatrix} \boldsymbol{\beta}_{t},
    \mathbf{B}^{\mathrm{aug}}, \boldsymbol{\Sigma}\right).
    \]
    \item Impose missingness: for each response $i\in\{1,\ldots,q\}$, randomly select $\Gamma=\gamma N$ indices from $\{1,\ldots,N\}$ (without replacement), say $n_{1}^{i},\ldots,n_{\Gamma}^{i}$, and treat $Y_{n_{1}^{i},i,t},\ldots,Y_{n_{\Gamma}^{i},i,t}$ as missing.
    \item Vectorize $\uvec{\mathbf{y}}_{t}=\vect(\mathbf{y}_{t})$ and obtain the observed subvector $\uvec{\mathbf{y}}_{t,\obs}=\mathbf{L}_{t,\obs}\mathbf{P}_{t}\uvec{\mathbf{y}}_{t}$.
  \end{enumerate}
\end{enumerate}

The generated datasets serve as the basis for fitting models $\mathcal{M}_{1}$–$\mathcal{M}_{4}$ under the prior and MCMC configurations described in Section~\ref{subsec:Sim2}. 
These settings ensure full reproducibility of the simulation results reported in the main text.

\end{appendices}

\bibliography{sn-bibliography}% common bib file
%% if required, the content of .bbl file can be included here once bbl is generated
%%\input sn-article.bbl

\end{document}